\documentclass[twocolumn,notitlepage,aps,prb,10pt,superscriptaddress,floatfix,letterpaper,reprint
]{revtex4-2}

\usepackage{hyperref}
\usepackage[usenames,dvipsnames]{color}
\usepackage{amsmath}
\usepackage[english]{babel}
\usepackage{amssymb}
\usepackage{graphicx}
\usepackage{epstopdf}
\usepackage[normalem]{ulem}
\usepackage{subfigure}
\usepackage{todonotes}
\usepackage{braket}
\usepackage{bbold}
\usepackage{placeins}
\usepackage[para]{threeparttable}
\usepackage{lipsum}
\usepackage{multirow}
\usepackage{bm}
\usepackage{xcolor}
\usepackage{amsfonts}
\definecolor{linkcolor}{HTML}{399B03}
\definecolor{urlcolor}{HTML}{399B03}
\hypersetup{pdfstartview=FitH, linkcolor=linkcolor,urlcolor=urlcolor,colorlinks=true}
\hypersetup{
    unicode=false,          % non-Latin characters in Acrobat’s  bookmarks
    pdftoolbar=true,        % show Acrobat’s toolbar?
    pdfmenubar=true,        % show Acrobat’s menu?
    pdffitwindow=false,     % window fit to page when opened
    pdfstartview={FitH},    % fits the width of the page to the window
    pdfauthor={},         % author
    colorlinks=true,        % false: boxed links; true: colored links
    linkcolor=blue,         % color of internal links
    citecolor=red,          % color of links to bibliography
}

\newcommand{\tk}{\tilde{\textbf{k}}}

\newcommand{\tK}{\textbf{K}}
\newcommand{\tQ}{\textbf{Q}}
\newcommand{\la}{\textbf{a}}

\begin{document}

%\preprint{APS/123-QED}

    \title{Phase transitions in partial summation methods: Results from the 3D Hubbard model.}

    \author{Sergei Iskakov}
    %\affiliation{Department of Physics, University of Michigan, Ann Arbor, MI 48109, USA}
    \author{Emanuel Gull}
    \affiliation{Department of Physics, University of Michigan, Ann Arbor, MI 48109, USA}

    \date{\today}

    \begin{abstract}
        %Motivation
        The accurate determination of magnetic phase transitions in electronic systems is an important task of solid state theory.
        %Problem
        While numerically exact results are readily available for model systems such as the half-filled 3D Hubbard model,
        the complexity of real materials requires additional approximations, such as the restriction to certain classes of diagrams in perturbation theory, that reduce the precision with which magnetic properties are described. 
        %Method
        In this work, we examine the description of magnetic properties in second order perturbation theory, GW, FLEX, and two T-Matrix approximations to numerically exact CT-QMC reference data.
        We assess finite-size effects and compare periodic lattice simulations to cluster embedding.
        %Results
        We find that embedding substantially improves finite size convergence. However, by analyzing different partial summation
        methods we find no systematic improvement in the description of magnetic properties, with most methods considered in
        this work predicting first-order instead of continuous transitions, leading us to the conclusion that 
        %Conclusion
        non-perturbative methods are necessary for the accurate determination of magnetic properties and phase transitions.
    \end{abstract}

    \maketitle
    \section{Introduction}\label{sec:introduction}
%phase transitions are important
The accurate quantum mechanical description of magnetic phase
transitions in real materials is an open problem, despite the enormous practical importance of magnetic materials, as it requires a simultaneous description of electronic structure and correlation effects, finite-temperature phenomena, and (at continuous transitions) criticality.

%prototypical magnetic phase transition is 3d hubbard. Fruit fly of fermionic PT, vry high tc.
In model systems, where the complication of electronic structure effects is absent, the prototypical  realization of a magnetic phase transition occurs in the half-filled three-dimensional version of the Hubbard model \cite{Qin21,Arovas21}. There, the transition temperature is  small for both weak and strong interactions and becomes maximal at an interaction strength $\sim 2/3$ of the bandwidth of the model. This transition has been studied extensively with numerically exact methods \cite{Staudt2000,Kent2005,Fuchs2011c,Fuchs2011,Paiva2011,Kozik2013}, motivated in particular by cold atomic gas experiments that emulate the half-filled 3D model in an optical trap \cite{Kohl05,Jordens2008,Schneider2008,Esslinger10,Duarte2015,Hart2015}.

%want numerically exact. Can only get approx. Best approx would be systematically improvable convergent
These numerically exact methods cannot easily be extended to realistic three-dimensional systems, where one is limited to approximations such as low-order self-consistent perturbation theory and partial summation methods. These approximations capture electronic structure and finite-temperature effects reasonably well, while severely approximating electron correlation. Nevertheless, one may aim to find a hierarchy or `Jacob's ladder' \cite{Sousa2007}  of approximate methods that systematically converge to the exact result as their complexity or approximation level is increased. 

\begin{figure}[tb]
        \centering
        \includegraphics[width=0.95\columnwidth]{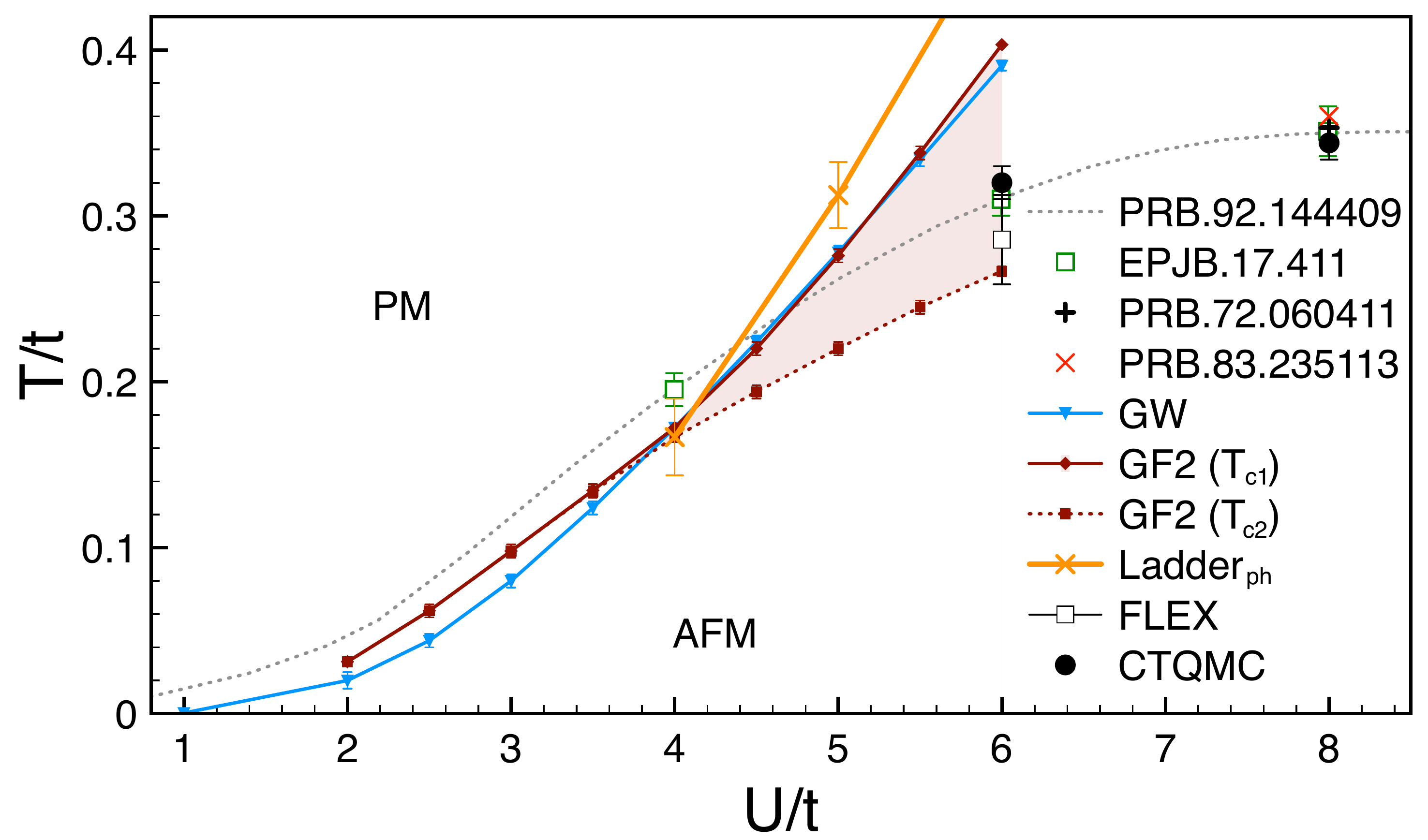}
        \caption{Overview of the weak-to-intermediate coupling phase diagram for the half-filled 3D Hubbard model on a simple cubic lattice.
        Shown are numerically exact CT-QMC results (black circles) along with GW (blue), GF2 (dark red, shaded metastable region), particle-hole ladder (orange), and FLEX (black squares). Also shown are literature results extracted from Refs.~\cite{Hirschmeier2015,Staudt2000,Kent2005,Fuchs2011}.
        }
        \label{fig:phasediagram}
    \end{figure}

%We try to do this here. Focus on methods that one could apply in difficult contexts
In this paper, we explore such a hierarchy. We  focus on semi-analytical partial summation methods suitable for realistic systems at the example of the simple and well-known half-filled 3D Hubbard model at weak-to-intermediate coupling (see Fig.~\ref{fig:phasediagram}), with special emphasis on converging the results in the critical regime to the thermodynamic limit. 
We employ self-consistent second order perturbation theory (GF2, \cite{Rusakov2016}) in the bare interaction and self-consistent first-order perturbation theory in the screened interaction (GW, \cite{Hedin1965}), and explore convergence and accuracy of these methods. We then explore two variants of T-matrix or `ladder' summation methods and a combination of `fluctuation exchange' diagrams with ladder and polarization terms (FLEX, \cite{Bickers1989a,Bickers1997,Bickers2004}). All results are then compared to numerically exact continuous-time quantum Monte Carlo calculations \cite{Gull2011b,Gull2008,Gull2011a} on the same system and in the thermodynamic limit.

%describe results
We show that finite size effects in all of these methods can be substantially reduced by employing the dynamical cluster approximation (DCA, \cite{Hettler1998,Maier2005}), even near criticality. All DCA results converge to the results on  finite size systems with periodic boundary conditions, as expected, while requiring simulations on substantially smaller systems. We also show that commonly used cluster selection criteria \cite{Betts1997,Betts1999} generally do not accelerate convergence.

We then show that despite adding `more' diagrams, no systematic convergence of the hierarchy of diagrammatic methods is observed in the weak-to-intermediate coupling regime. Instead, methods with additional geometric series tend to exhibit first-order metastability rather than second-order criticality, and the numerical values either do not improve systematically or worsen when compared to the exact results. This indicates that increasing the accuracy of simulations of magnetic transitions will require the use of other hierarchy paradigms, such as, for instance, quantum embedding \cite{Georges1996,Maier2005,Zgid2017}.

The remainder of this paper proceeds as follows. Sec.~\ref{sec:model} introduces the 3D Hubbard model and Sec.~\ref{sec:methods} the diagrammatic and finite size scaling methods. Sec.~\ref{sec:results} presents reference results, analyzes finite size effects, and discusses results from GW and GF2.  Sec.~\ref{sec:results2} shows results from ladder and FLEX approximations, and Sec.~\ref{sec:conclusion} discusses our conclusions. The appendix provides additional derivations for general spin-dependent ladder and FLEX methods and discusses the issue of causality violations.

\section{Model}\label{sec:model}
The Hamiltonian of the Hubbard model \cite{Qin21,Arovas21} is
    \begin{align}
        H = -t \sum_{\langle i,j\rangle,\sigma} c^{\dagger}_{i,\sigma}c_{j,\sigma} + U \sum_{i} n_{i\uparrow}n_{i,\downarrow},
        \label{eqn:Hamiltonian}
    \end{align}
    where $t$ denotes the hopping term, $U$ the onsite Coulomb interaction, and $\langle i, j\rangle$ nearest
    neighbors on a simple cubic lattice.
 In order to study $(\pi, \pi, \pi)$ antiferromagnetic order,
    we use a two-site supercell formalism with lattice vectors $\la_1 = (0,1,1)^T$, $\la_2 = (1,1,0)^T$ and $\la_3 = (1,0,1)^T$.
    The antiferromagnetic order parameter is $M = \left|\rho_{0,0,\uparrow} - \rho_{0,0,\downarrow} \right|$,
    where $\rho_{0,0,\sigma}$ is the local density matrix for spin $\sigma$ at site $0$.

    The imaginary time Green's function is
    \begin{align}
        G_{\tK,ij,\sigma}(\tau)  = -\frac{1}{\mathcal{Z}}
        \text{Tr}\left[e^{-(\beta-\tau)\mathcal{H}} c_{\tK,i,\sigma} e^{-\tau\mathcal{H}} c^{\dagger}_{\tK,j,\sigma}\right].
    \end{align}
    Here $\mathcal{Z} = \text{Tr}\left[e^{-\beta\mathcal{H}}\right]$ is the grand partition function, $\beta$ the
    inverse temperature, $\mathcal{H} = H - \mu N$ with $\mu$ the chemical potential and $N$ the particle number, and $\tK$ a reciprocal lattice vector. $i$ and $j$ denote site indices in the supercell.
    In this paper, we limit ourselves to the half-filled case of $\mu = \frac{U}{2}$.
    
 The Dyson equation  relates the non-interacting ($U=0$) Green's function $G^0_{\tK,ij,\sigma}(\tau)$ to the interacting Green's function
    $G_{\tK,ij,\sigma}(\tau)$ via the self-energy $\Sigma_{\tK,ij,\sigma}(\tau)$, 
    \begin{align}
        G_{\tK,ij,\sigma}(\omega_n) &= G^{0}_{\tK,ij,\sigma}(\omega_n) + \\&\sum_{kl}G^{0}_{\tK,ik,\sigma}(\omega_n)&
          \Sigma_{\tK,kl,\sigma}(\omega_n) G_{\tK,lj,\sigma}(\omega_n),\nonumber
    \end{align}
    where $\omega_n = (2n+1)\pi/\beta$ are fermionic Matsubara frequencies related to imaginary time $\tau$ via Fourier transform \cite{Mahan2000}. 
    
    The Schwinger-Dyson equation \cite{Rohringer2018} relates the dynamical part of the self-energy to a `fully reducible' vertex function $F$,
    \begin{align}
        \Sigma_{\tK,ij,\sigma}(\omega_n) &= \frac{1}{(\beta N_c)^2}\sum_{\substack{n'n'' ;\ pml \\\tK'\tQ,\sigma'}} U
        G_{\tK',ip,\sigma'}(\omega_{n'}) \times  \nonumber \\
        &G_{(\tK'+\tQ),mi,\sigma'}(\omega_{n'+n''}) G_{(\tK+\tQ),il,\sigma}(\omega_{n+n''}) \times  \nonumber \\
        &F^{\substack{\sigma\sigma'\sigma'\sigma\\pmlj}}_{\tK'\tK\tQ} (\omega_{n'},\omega_n,\Omega_{n''}),
        \label{eqn:schwinger}
    \end{align}
    with $N_c$ the number of lattice sites and $\Omega_m = 2m\pi/\beta$ bosonic Matsubara
    frequencies.
    
    The Bethe-Salpeter equation relates the vertex function to the generalized susceptibility $\chi$,
    \begin{align}
        \chi^{\alpha\beta\gamma\kappa}_{\tK\tK'\tQ} (\omega_n,\omega_{n'},\Omega_m) &= \nonumber \\
        \chi^{(0),\alpha\beta\gamma\kappa}_{\tK\tK'\tQ}(\omega_n,\omega_{n'},\Omega_m)\delta_{\substack{n,n'\\\tK,\tK'}}&-
        \tilde{\chi}^{,\alpha\beta\gamma\kappa}_{\tK\tK'\tQ} (\omega_n,\omega_{n'},\Omega_m) \\
        \tilde{\chi}^{\alpha\beta\gamma\kappa}_{\tK\tK'\tQ} (\omega_n,\omega_{n'},\Omega_m) &=
        \nonumber \\
        \sum_{\psi\xi\eta\theta}\chi^{(0),\alpha\beta\psi\xi}_{\tK\tK\tQ}(\omega_n,\omega_{n},\Omega_m)
        &F^{\psi\xi\eta\theta}_{\tK\tK'\tQ} (\omega_n,\omega_{n'},\Omega_m) \times  \nonumber \\
        \chi^{(0),\eta\theta\gamma\kappa}_{\tK'\tK'\tQ}& (\omega_{n'},\omega_{n'},\Omega_m).
    \end{align}
    Here
    \begin{align}
        \chi^{(0),\alpha\beta\gamma\kappa}_{\tK\tK'\tQ} (\omega_n,\omega_n,\Omega_m) &= \nonumber \\
    - \beta N_c G_{\tK,\alpha\beta}&(\omega_n)G_{(\tK+\tQ),\gamma\kappa}(\omega_n+\Omega_m)
    \end{align}
    is the unconnected and $\tilde{\chi}^{\alpha\beta\gamma\kappa}_{\tK\tK'\tQ}(\omega_n,\omega_{n'},\Omega_m)$
    the connected part of the susceptibility~\cite{Rohringer2018}, and Greek indices correspond to combined spin and orbital
    indices.

    The Galitzkii-Migdal formula expresses the energetics in terms of Green's functions and self-energies ~\cite{Galitskii1958,Holm2000,LeBlanc2013,Welden2016,Neuhauser2017} as
    \begin{align}
        \langle V \rangle &= \frac{1}{\beta N_c}\sum_{n,\tK,ij,\sigma} \Sigma_{\tK,ij,\sigma}(\omega_n)G_{\tK,ji,\sigma}(\omega_n),
        \\
        \langle T \rangle &= \frac{1}{\beta N_c}\sum_{n,\tK,ij,\sigma} \text{Tr}\left[\left( \epsilon_{\tK,ij} - \mu\delta_{ij}
        \right)
        \mathbf{G}_{\tK,ji,\sigma}(\omega_n)\right].
        \label{eqn:energy}
    \end{align}

    \section{Methods}\label{sec:methods}
    \begin{figure}[tb]
        \centering
        \includegraphics[width=0.47\textwidth]{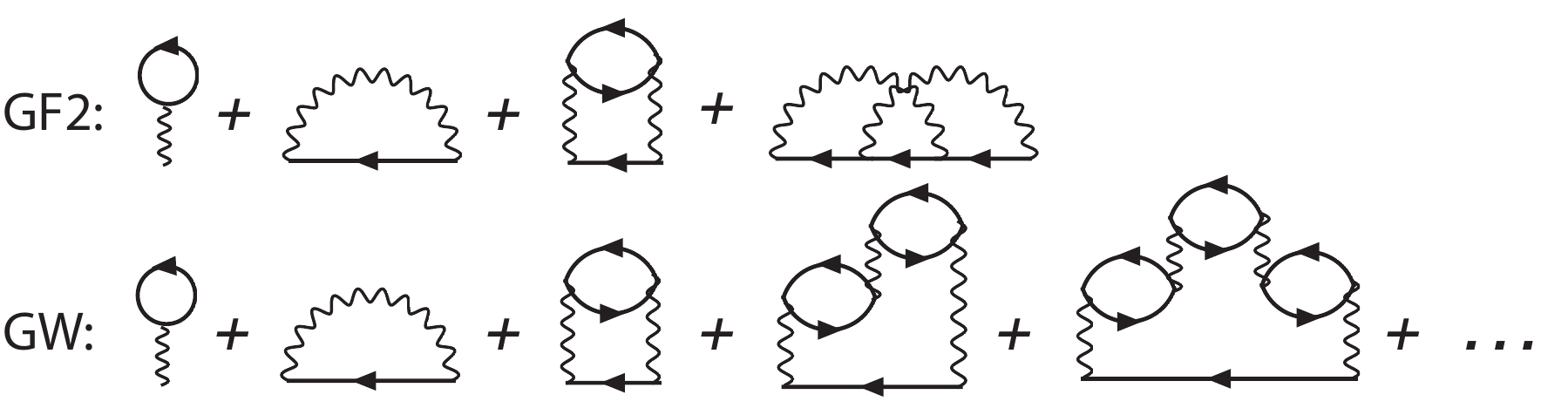}
        \caption{Self-energy diagrams for GF2 (top) and GW (bottom). Straight lines denote interacting Green's functions.
        Wiggly lines denote Coulomb interaction.}
        \label{fig:selfenergy}
    \end{figure}

    \subsection{GF2}\label{subsec:gf2}
    In self-consistent second order perturbation theory, the dynamical part of the self-energy in the presence of local interactions is (see Fig.~\ref{fig:selfenergy})
    \begin{align}
        \Sigma^{(2)}_{\tK,ij,\sigma}(\omega_n) &= - \frac{1}{(\beta N_c)^2}\sum_{\substack{n'm,\sigma'\\ \tK'\tQ}} U^2
        G_{\tK',ij,\sigma'}(\omega_{n'}) \times  \nonumber \\
        G_{(\tK'+\tQ),ji,\sigma'}&(\omega_{n'+m}) G_{(\tK+\tQ),ij,\sigma}(\omega_{n+m}).
        \label{eqn:sigma_gf2}
    \end{align}
    Comparison to Eq.~\ref{eqn:schwinger} implies that GF2 corresponds to approximating $F$
    by the bare interaction $U$~\cite{Rohringer2018}. The connected part of the susceptibility in case of the Hubbard model then
    reduces to:
    \begin{align}
        \tilde{\chi}^{(2)\alpha\beta\gamma\kappa}_{\tK\tK'\tQ} (\omega_n,\omega_{n'},\Omega_m) &=
        \sum_{\psi\xi\eta\theta}\chi^{(0),\alpha\beta\psi\xi}_{\tK\tK\tQ}(\omega_n,\omega_{n},\Omega_m)\times \nonumber \\
        U
        \chi^{(0),\eta\theta\gamma\kappa}_{\tK'\tK'\tQ}& (\omega_{n'},\omega_{n'},\Omega_m).
        \label{eqn:chigf2}
    \end{align}
    The static analogue of this expression has been presented in Refs.~\cite{Pokhilko2021a,Pokhilko2021b}.
    Because the GF2 is not a two particle self-consistent method, the susceptibility can also be evaluated from the second order
    approximation to the vertex function obtained through the functional derivative of the Luttinger-Ward functional.
    In this approach the GF2 susceptibility is~\cite{Hotta1996}
    \begin{align}
        \mathbf{\tilde{\chi}}^{(2_2)}&= \tilde{\chi}^{(2)} + \mathbf{\chi}^{(0)} \mathbf{U} \mathbf{\chi}^{(0)} \mathbf{U} \mathbf{\chi}^{(0)}.
        \label{eqn:chigf2_2}
    \end{align}
    Neither Eq.~\ref{eqn:chigf2} nor Eq.~\ref{eqn:chigf2_2} can generate a divergent susceptibility, as would be expected
    at a continuous phase transition. Transitions in this method are therefore necessarily first order.

    The direct evaluation of the self-energy diagram in GF2, Eq.~\ref{eqn:sigma_gf2}, requires integration over two
    momentum indices. The overall scaling with system size is therefore $O(N_c^3)$. Due to the locality of the interaction in the Hubbard model, this scaling can be reduced to
    $O(N_c\log(N_c))$ by evaluating diagrams in real space.

    \subsection{GW}\label{subsec:gw}
    In GW~\cite{Hedin1965}, the dynamical part of the self-energy is expressed in terms of the renormalized
    screened
    interaction $W$:
    \begin{align}
        W_{\tQ,ij}(\Omega_m) &= U - \frac{1}{\beta N_c}\sum_{\tK n l\sigma} U \times \nonumber \\
        G_{\tK,il,\sigma}(\omega_n) &G_{\tK+\tQ,li,\sigma}(\omega_{n+m}) W_{\tQ,lj}(\Omega_m).
    \end{align}
    $W$ can also be expressed as a geometric series,
    \begin{align}
        \mathbf{W}_{\tQ}(\Omega_m) = \mathbf{U} \left[\mathbb{1} - \mathbf{U} \mathbf{\Pi}^{(GW)}_{\tQ}(\Omega_m)\right]^{-1},
    \end{align}
    where
    \begin{align}
    \left[\mathbf{\Pi}^{(GW)}_{\tQ}(\Omega_m)\right]_{il} &=  -\frac{1}{\beta N_c} \sum_{\tK,n',\sigma'} \nonumber \\
    &G_{\tK',il,\sigma'}(\omega_{n'}) G_{\tK+\tQ,li,\sigma'}(\omega_{n'+m})
    \end{align}
    is the GW approximation of the polarization operator~\cite{Hedin1965} and bold symbols correspond to a tensor representation in
    orbital space.
    The dynamical part of the self energy is (see  Fig.~\ref{fig:selfenergy})
    \begin{align}
        \Sigma^{(GW)}_{\tK,ij,\sigma}(\omega_n) &= \frac{1}{\beta N_c}
        \sum_{\substack{\tQ,m,l}} U G_{\tK+\tQ,ij,\sigma}(\omega_{n+m}) \times \nonumber \\
        &\Pi_{\tQ,il}(\Omega_m) W_{\tQ,lj}(\Omega_m).
    \end{align}

    The connected part of the susceptibility in GW is
    \begin{align}
        \tilde{\chi}^{\alpha\beta\gamma\kappa}_{\tK\tK'\tQ} (\omega_n,\omega_{n'},\Omega_m) &=
        \sum_{\psi\xi\eta\theta}\chi^{(0),\alpha\beta\psi\xi}_{\tK\tK\tQ}(\omega_n,\omega_{n},\Omega_m)\times \nonumber \\
        W_{\tQ,\psi\xi\eta\theta}(\Omega_m)
        \chi^{(0),\eta\theta\gamma\kappa}_{\tK'\tK'\tQ}& (\omega_{n'},\omega_{n'},\Omega_m).
        \label{eqn:chigw}
    \end{align}

    Like GF2, GW is not a two-particle self-consistent approach~\cite{Hedin1965,Vilk1997}.
    Unlike Eqs.~\ref{eqn:chigf2} and~\ref{eqn:chigf2_2}, Eq.~\ref{eqn:chigw} will diverge when $\mathbb{1} - \mathbf{U\Pi}^{(GW)}_{\tQ}(\Omega_m)$
    becomes singular or, equivalently, when the leading eigenvalue of $\mathbf{U}\mathbf{\Pi}^{(GW)}_{\tQ}(\Omega_m)$ passes through $1$.
    The complexity of evaluating the self-energy diagram with local interactions is only $O(N_c^2)$.
    
    \subsection{CT-QMC}\label{subsec:ct-qmc}
    To obtain unbiased results for the model, we validate our results with continuous time~\cite{Rubtsov2004,Gull2011b} auxiliary
    field quantum Monte Carlo (CT-AUX)~\cite{Gull2008} with sub-matrix updates~\cite{Gull2011a}. This algorithm obtains
    numerically exact results (within statistical error) for finite-size lattice and impurity models. In the half-filled case
    the sign problem is absent and the algorithm scales as $O(N_c^3)$. In the 3D Hubbard model, the method has been successfully applied to the 
    study of thermodynamic properties in the continuum limit with
    systematic finite-size extrapolations \cite{Fuchs2011c}, as well as to spectral properties~\cite{Fuchs2011,Fuchs2011b}.

In order to obtain results in the continuum limit, finite system size results need to be extrapolated in the system size \cite{Fuchs2011}. We extrapolate our reference results from a sequence of clusters with up to $90$ cluster sites, using a linear extrapolation of the critical temperature in the inverse cluster size.

    \subsection{Treatment of the finite-size effects.}\label{subsec:finitesize}

    The thermodynamic limit of Eq.~\ref{eqn:Hamiltonian} is typically approximated by restricting the system to a
    finite size lattice with periodic boundary conditions. By gradually enlarging the lattice, convergence to the
    thermodynamic limit is observed.

    \subsubsection{Finite lattice selection}
    The traditional approach on simple cubic lattices uses periodic systems defined by the orthogonal translational vectors
    $2 n \{(1 0 0)^T,(0 1 0)^T,(0 0 1)^T\}$, with $n$ integer~\cite{Staudt2000}.
    With this choice, few lattices ($2\times2\times2$, $4\times4\times4$, \ldots, $10\times10\times10$)
are readily accessible, complicating the extrapolation to the thermodynamic limit. This choice of lattices is equivalent in reciprocal
    space to the standard choice of $\Gamma$-centered Monkhorst-Pack
    grids~\cite{MonkhorstPack1976} used in real-material calculations, and fully respects all lattice symmetries.

    If  symmetries of the infinite lattice are allowed to be broken by the finite system, many more ``cluster'' geometries become
    available. Refs.~\cite{Betts1997} and \cite{Betts1999} classified the suitability of these clusters for finite system
    simulations at the example of the 2D and 3D Heisenberg model and developed empirical criteria for cluster selection based
    on geometric properties. These so-called `Betts' clusters are popular in DCA~\cite{Hettler1998,Maier2005b} (see Refs.~\cite{Kent2005,Fuchs2011} for
    larger 3D geometries), as they allow to perform thermodynamic limit extrapolations from comparatively small system sizes
    with many
    more points. We emphasize that selecting clusters according to these criteria does not guarantee accelerated
    convergence for small clusters, and that finite size effects need to be assessed for each simulation.

    Other approaches to select finite size clusters exist. For instance, Ref.~\cite{Morgan2018} (in a real
    materials context)  suggests using different $\tk-$grids, reducing grids to their symmetrically distinct points
    and thereby substantially reducing the number of sampling points by better symmetry reduction~\cite{Hart2019}. Methods using
    non-equidistant grids could also improve finite size convergence by increasing grid density near the high symmetry
    points~\cite{Jorgensen2021}.

    \subsubsection{Quantum embedding methods}
    Finite size convergence may be drastically accelerated with quantum embedding techniques~\cite{Maier2005b}. Here, the original
    lattice is replaced by an impurity embedded into a self-consistently adjusted bath of non-interacting
    fermions~\cite{Georges1996,Lichtenstein2000,Kotliar2001,Zgid2017}. This allows to treat correlations within the size of the
    cluster explicitly while long-range correlations are approximated, and corresponds to an approximation on the level of the self-energy, rather than the Green's function. Similarly to lattice calculations, embedding
    calculations converge to the continuum limit as the system size is increased, such that the inverse system size is a control
    parameter for finite size corrections~\cite{Biroli2002,Jarrell2005,Biroli2005,Fuchs2011c,Gull2011a}.

    Numerous embedding methods have been proposed~\cite{Hettler1998,Lichtenstein2000,Kotliar2001,Biermann2003,Zgid2017}. Here we
    use the dynamical cluster approximation (DCA)~\cite{Hettler1998,Maier2005b}, which is based on a
    translationally invariant embedding where local quantities such as the order parameter or the energy converge
    quadratically with the linear system size~\cite{Hettler1998,Maier2005}. DCA can also be understood as the approximation of the continuous
    self-energy by a few momentum-dependent basis functions~\cite{Fuhrmann2007}, see also~\cite{Staar2013}.

    DCA can be formulated in conjunction with any impurity solver including the numerically exact CT-QMC methods and diagrammatic
    methods such as GF2 or GW and on any finite-size cluster, including the Betts clusters discussed above. As with lattice
    methods, the cluster selection criteria are not rigorous and convergence to the thermodynamic limit must be assessed for any
    simulation~\cite{Maier2005b,Kent2005,Gull2013,LeBlanc2013,LeBlanc2015}.

    \subsubsection{Twisted boundary conditions}
    Twisted boundary conditions, and twist averaging,~\cite{Wilkens2000,Ceperley2001} are powerful techniques commonly used in
    quantum Monte Carlo simulations~\cite{Karakuzu2018,Leclercq2011} for accelerating finite size convergence and eliminating
    shell effects. Twist averaging has been applied to the Hubbard model, e.g., in Refs.~\cite{Shiwei2016,Karakuzu2018}. The main
    idea of this method is to shift the momentum of cluster states. This provides information about additional momentum points  and, in the absence of the long range correlations, to simulate systems of effectively much larger size. The
    averaging over different twisting phase factors then accelerates the convergence to the thermodynamic limit. In the
    context of embedding, the method has been explored in Ref.~\cite{Paki2019}. 
    
    \section{Results for CT-QMC, GW and GF2}\label{sec:results}
    Because of its importance as a  paradigmatic model for phase transitions in periodic systems \cite{Qin21}, the 3D Hubbard model has been extensively studied
    over the years with  methods including mean-field theory~\cite{Dongen1991,Dongen1994}, quantum Monte Carlo~\cite{Scalettar1989,Staudt2000,Kozik2013,Paiva2011},
    dynamical mean field theory and its cluster extensions~\cite{Kent2005,Fuchs2011}, the two-particle self-consistent
    approximation~\cite{Dare2000} and diagrammatic extensions to the DMFT~\cite{Rohringer2011,Hirschmeier2015,Schafer2015}.
    Several benchmark results for this model in the weak-to-intermediate coupling regime exist, making it an ideal testbed for
    numerical methods. Fig.~\ref{fig:phasediagram} contains an overview of literature results and their respective uncertainties.

    \subsection{Reference results}\label{subsec:reference-results}
    \textcite{Staudt2000} studied the model with lattice QMC based on a discrete Hubbard-Stratonovich
    transformation~\cite{Blankenbecler1981,Hirsch1982} on finite cubic lattices with linear dimension up to $L=10$.
    Results from this work have been established as the main reference for $T_N$ in the intermediate coupling
    regime. Lattice studies~\cite{Kozik2013} with continuous time Monte-Carlo methods~\cite{Rubtsov2004,Burovski2006,Gull2011b}
    that eliminate potential time discretization errors~\cite{Gull2007} are in good agreement with these results.

    Approaches based on the dynamical cluster approximation have been used in conjunction with
    both discrete~\cite{Kent2005} and continuous~\cite{Fuchs2011} time Monte Carlo cluster solvers and have
    shown a good agreement with reference results~\cite{Staudt2000} while using much smaller system sizes.

    Motivated by cold atom experiments~\cite{Schneider2008,Jordens2008,Kondov2015} the problem was revisited in the context of
    energetic and thermodynamic properties~\cite{Paiva2011,Fuchs2011c}. Studies were mostly performed in the
    region near $U\approx8t$, where the entropy of the ordered phase is maximal~\cite{Fuchs2011c}, and for trap
    geometries~\cite{Tang2012,Paiva2015,Hart2015,Paiva2020}.

%    \begin{table}[tbh]
%        \begin{tabular}{p{1cm}|c|c|c|c|c|c|c|c}
%            U             & 3.0 & 3.5 & 4.0 & 4.5 & 5.0 & 6.0 & 7.0 & 8.0 \\
%            \hline
%            T$_c$         & 0.11& --  &0.196& --  & --  & --  & --  & 0.33  \\
%            \hline
%            $\Delta$T$_c$ & 0.01& --  &0.007& --  & --  & --  & --  & 0.01  \\
%        \end{tabular}
%        \caption{Numerically exact CTQMC estimates for the transition temperature obtained using DCA. $\Delta$T$_c$ estimated
%        from the temperature grid resolution. \textcolor{red}{THIS NEEDS UPDATING. Also put finite cluster values and their errors.}}
%        \label{tab:exact}
%    \end{table}

    \subsection{Order parameter and energetics}\label{subsec:orderparameter}
    Our focus is on the weaker correlation regime ($2\leq U/t \leq 6$), where perturbative partial summation methods  have a
    higher likelihood of success than at $U=8$.% Table~\ref{tab:exact} shows numerically exact CTQMC reference values for the transition
 %   temperature. The transition temperature on each finite lattice is {\color{red} DISCUSS: obtained between the highest temperature point where order is present and the lowest temperature point where order is absent, with an uncertainty given by the difference between those two points. These values are then extrapolated to the infinite system size limit according to $XXX$. The  uncertainties in the infinite system are estimated from a regression error of the extrapolation.}  
    
\begin{figure}[tb]
        \centering
        \includegraphics[width=0.95\columnwidth]{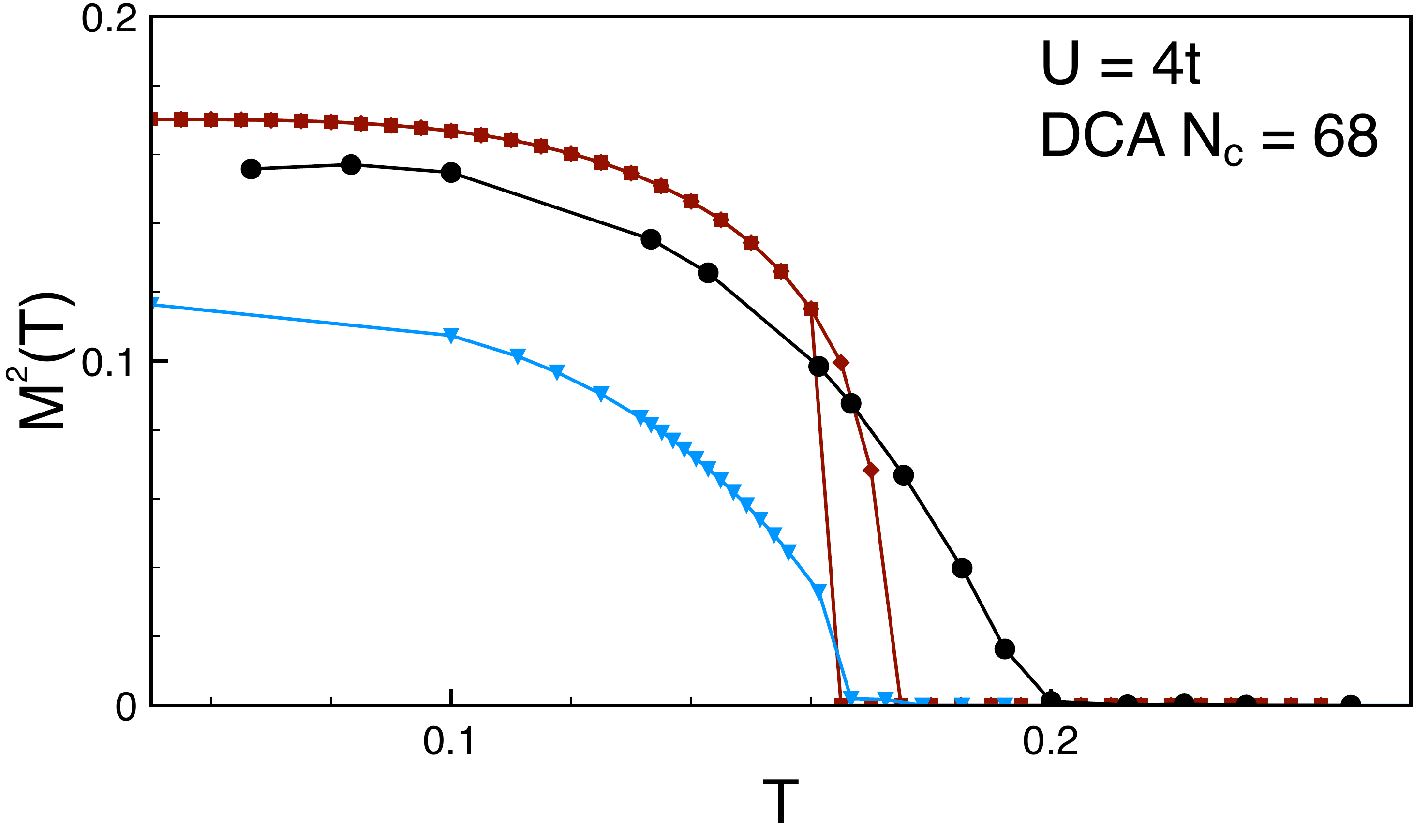}
        \includegraphics[width=0.95\columnwidth]{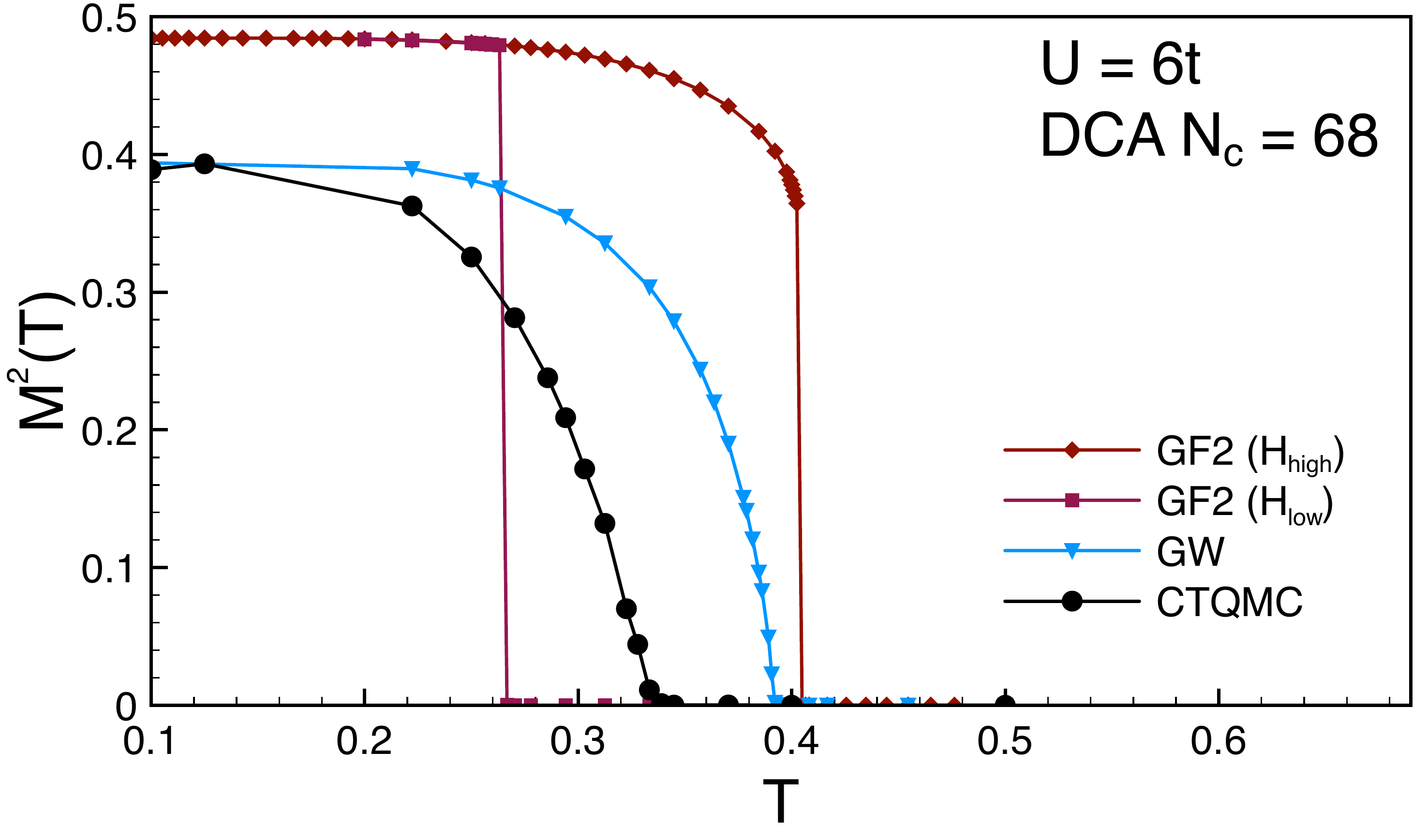}
        \caption{Temperature dependence of the antiferromagnetic order parameter for $U=4t$ (top) and $U=6t$ (bottom)
            obtained from DCA for $N_c=68$. Blue: GW. Red: GF2 for high (diamonds) and low (squares) initial field.
            Black circles correspond to numerically exact CT-QMC results.
        }
        \label{fig:M_T}
    \end{figure}

    We start the discussion from an analysis of the antiferromagnetic order parameter on a lattice with $N_c = 68$
    sites characterized by the lattice vectors $\mathbf{a}_1 = (1, 2, 3)^T$, $\mathbf{a}_2 = (3, 3, -2)^T$,
    $\mathbf{a}_3 = (2, -3, 3)^T$, using DCA self-consistency to reduce finite size effects. This lattice is large enough that finite size effects are small, but small enough that calculations can readily be performed.
    Fig.~\ref{fig:M_T} shows the temperature dependence of the order parameter for numerically exact (CT-QMC) and perturbative (GW
    and GF2) methods at $U=4t$ and $U=6t$. For both values of the interaction, perturbative methods find a phase transition
    within 20\% of the exact transition temperature.

    Numerically exact CT-QMC results for both $U=4t$ and $U=6t$ show a continuous phase transition, in agreement with previously results
    by other Monte Carlo simulations, and the values of $T_N$ coincide (within statistical error) with
    previous studies~\cite{Staudt2000,Kent2005,Fuchs2011c,Kozik2013}.

    In GF2, the order parameter is $\sim 10\% -20\%$ larger than the numerically exact value for all values of interaction.
    We performed simulation with high and low initial staggered magnetic field. These initial conditions led to different self-consistent solutions and two transition temperatures, $T_{c1}$ (where the initially polarized solution becomes isotropic) and $T_{c2}$ (where the initially unpolarized solution develops order). These metastable solutions
    indicate a first order phase transition, which we attribute to the lack of divergence in the GF2
    susceptibility (Eqs.~\ref{eqn:chigf2}~and~\ref{eqn:chigf2_2}). While the metastable region becomes smaller for
    smaller values of interaction~(see red shaded area in Fig.~\ref{fig:phasediagram}) and becomes unobservable for $U\lesssim2t$, it will likely
    persist for any non-zero value $U$.

    GW, on the other hand, predicts a continuous phase transition. The order parameter is smaller than in GF2, and
    the exact value is bracketed between GF2 and GW results. Like in GF2, the transition temperature grows almost linearly for
    $U\gtrsim 4t$ and we find no indication of a $\sim 1/U$ behavior~\cite{Staudt2000,Hirschmeier2015} at
    large $U$. The order parameter in all three methods  saturates around $T\approx T_c/3$.

    An analysis of the energetics of the model using Eq.~\ref{eqn:energy} (not shown) indicates that $T_{c1}$ corresponds to the actual
    GF2 transition temperature, whereas the isotropic solution below $T_{c1}$ belongs to a metastable state.     

    \subsection{Finite size convergence}\label{subsec:finite-size-convergence}

    Next, we perform a finite size analysis. We consider two approaches: simulations on periodic finite size clusters
    and  embedding simulations using the DCA scheme. Both approaches use the same cluster geometries, and we emphasize that in the infinite cluster size limit both approaches
    converge to the same solution, both in theory~\cite{Biroli2002,Jarrell2005,Biroli2005} and in
    practice~\cite{Fuchs2011c,LeBlanc2015}.

    \begin{figure}[tbh]
        \centering
        \includegraphics[width=0.95\columnwidth]{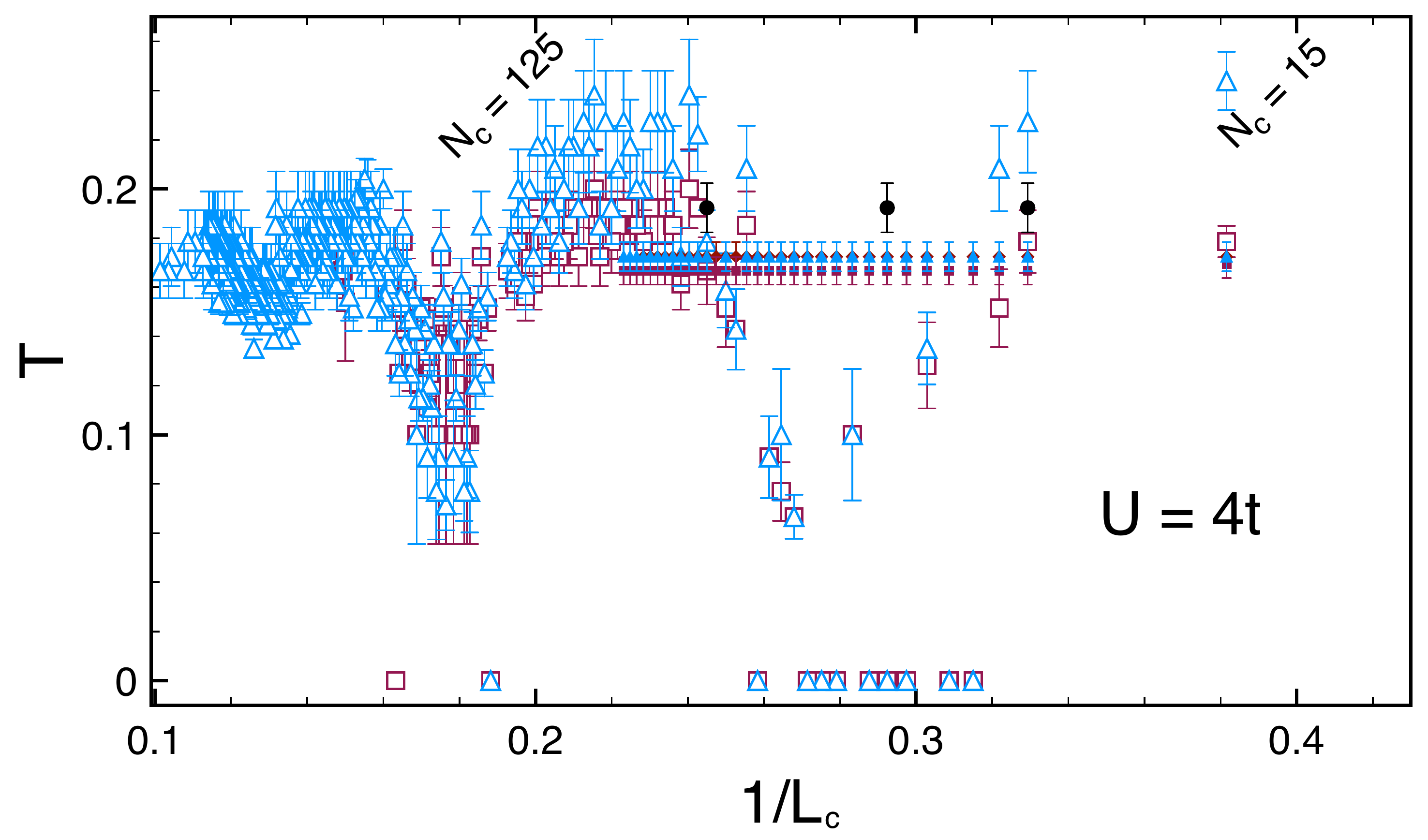}

        \includegraphics[width=0.95\columnwidth]{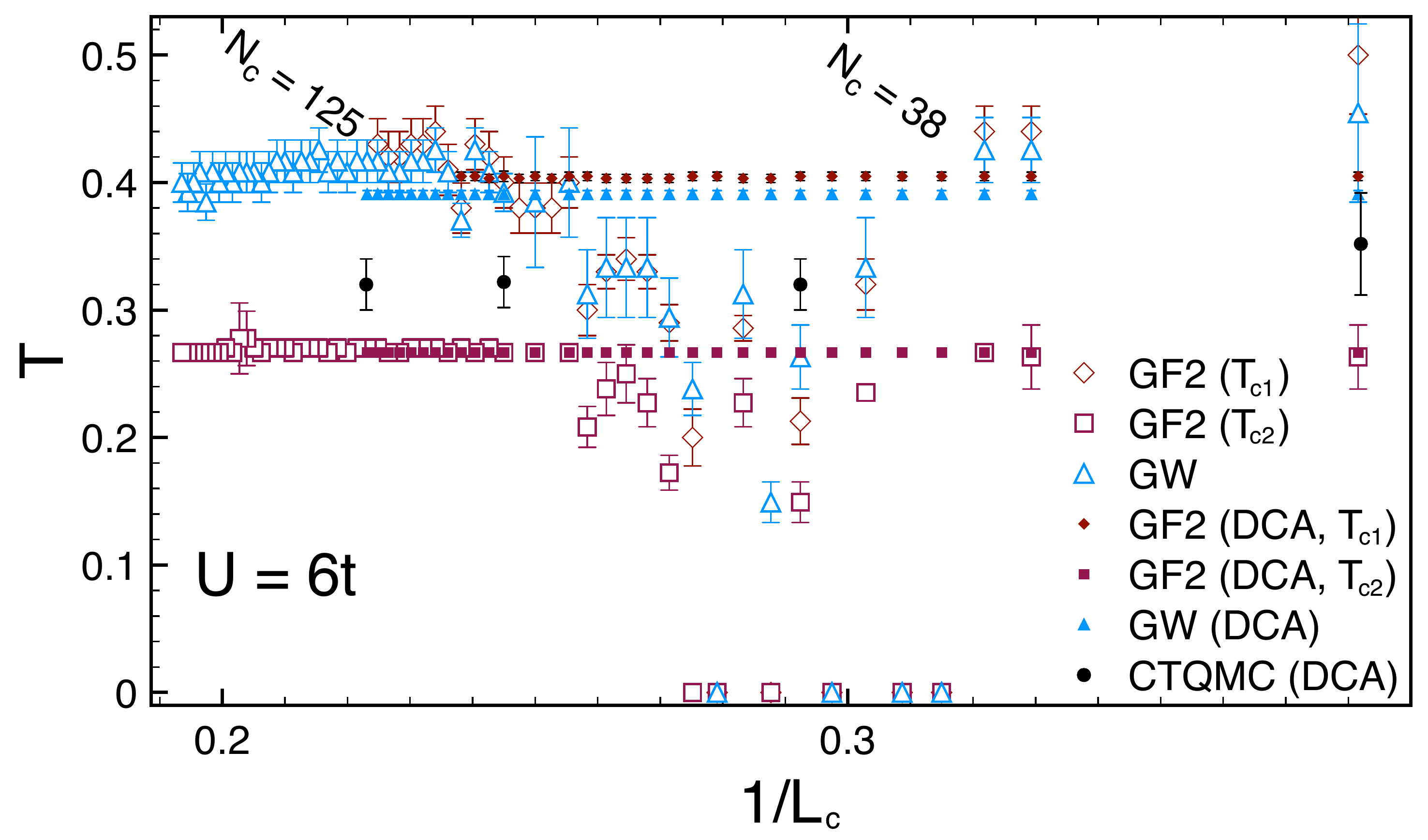}
        \caption{Dependence of the critical temperature for $U=4t$ (top) and $U=6t$ (bottom) on linear system size $L_c=N_c^{1/3}$.
        Open symbols: lattice with periodic boundary conditions. Filled symbols: DCA. Red: GF2 for high (diamonds) and
        low (squares) initial fields. Blue triangles: GW. Black circles: numerically exact CT-QMC results.}
        \label{fig:T_vs_N}
    \end{figure}

    Fig.~\ref{fig:T_vs_N} shows system size dependence of the critical temperature as obtained from GF2 (red) and GW (blue), on periodic clusters (open symbols) and with DCA embedding (filled symbols). Error bars correspond to the
     uncertainty in $T_c$, which is obtained by bracketing the critical temperature with simulations from above and below. We first consider periodic clusters without embedding. At $U=4t$ we
    see large finite size dependence and very slow convergence with system size up to $N_c = 1000$. This indicates the
    importance of long range fluctuations for small values of $U$. For $U=6t$, the convergence of the transition temperature is
    much faster and we see a clear sign of saturation at $N_c \gtrsim 100$. For some cluster geometries we observe no sign of any
    antiferromagnetic transition, e.g. $N_c = 34$. GF2 and GW exhibit a similar finite size dependence at both interaction strengths $U=4t$ and $U=6t$.
    
    In the case of DCA, for perturbative methods, the critical temperature converges to the thermodynamic limit value
    for all cluster geometries we consider. Unlike simulations without embedding, DCA simulations for all clusters demonstrate
    clear signs of the antiferromagnetic transition. This indicates that a relatively short ranged self-energy, in conjunction
    with a fine-grained resolution of the dispersion, can efficiently capture phase transition physics even in the weak
    coupling regime. For the numerically exact CT-QMC DCA results we do not observe large deviations in the transition
    temperature for clusters larger than $N_c\sim 40$, consistent with Ref.~\cite{Kent2005}.

    \begin{figure}[tbh]
        \centering
        \includegraphics[width=0.95\columnwidth]{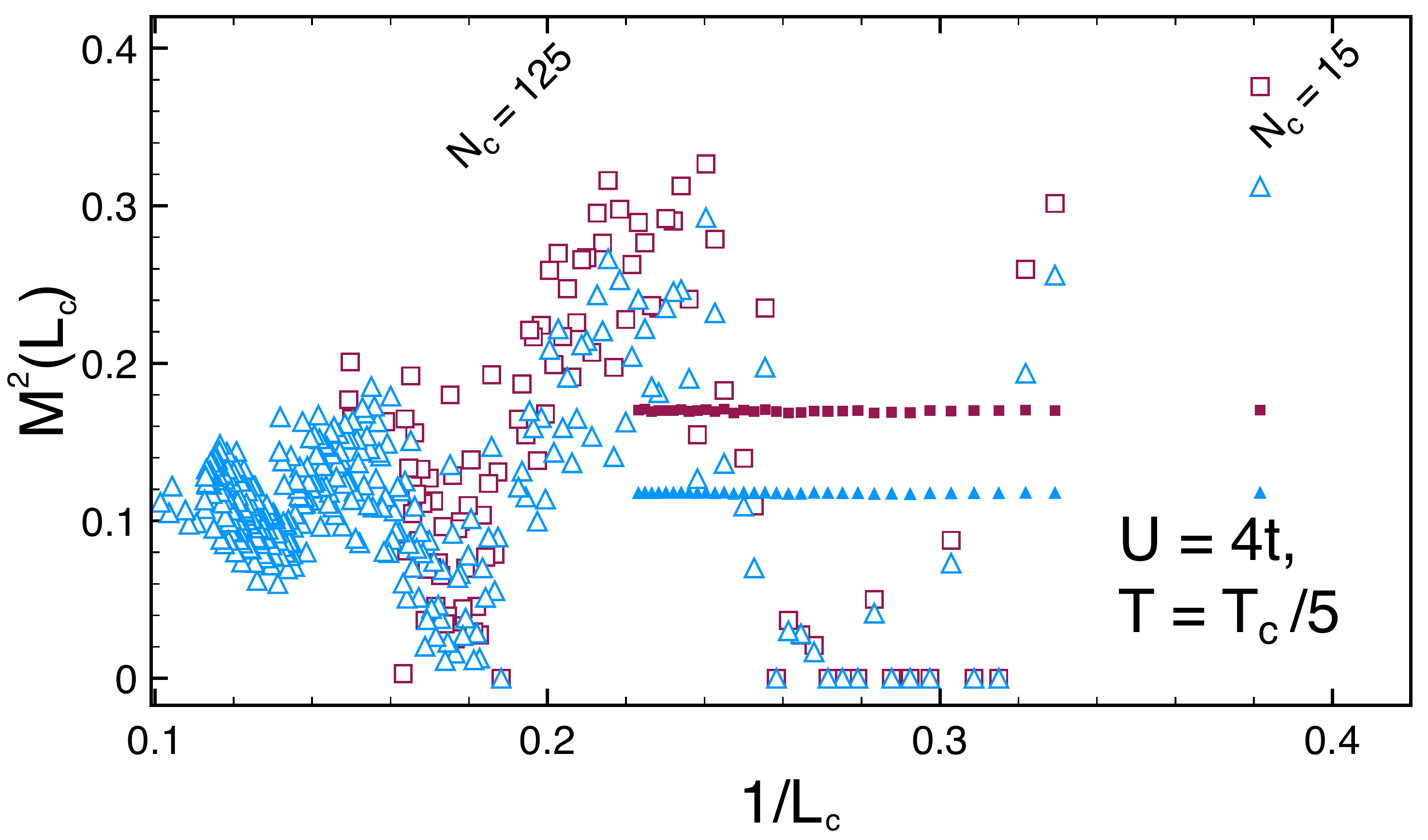}

        \includegraphics[width=0.95\columnwidth]{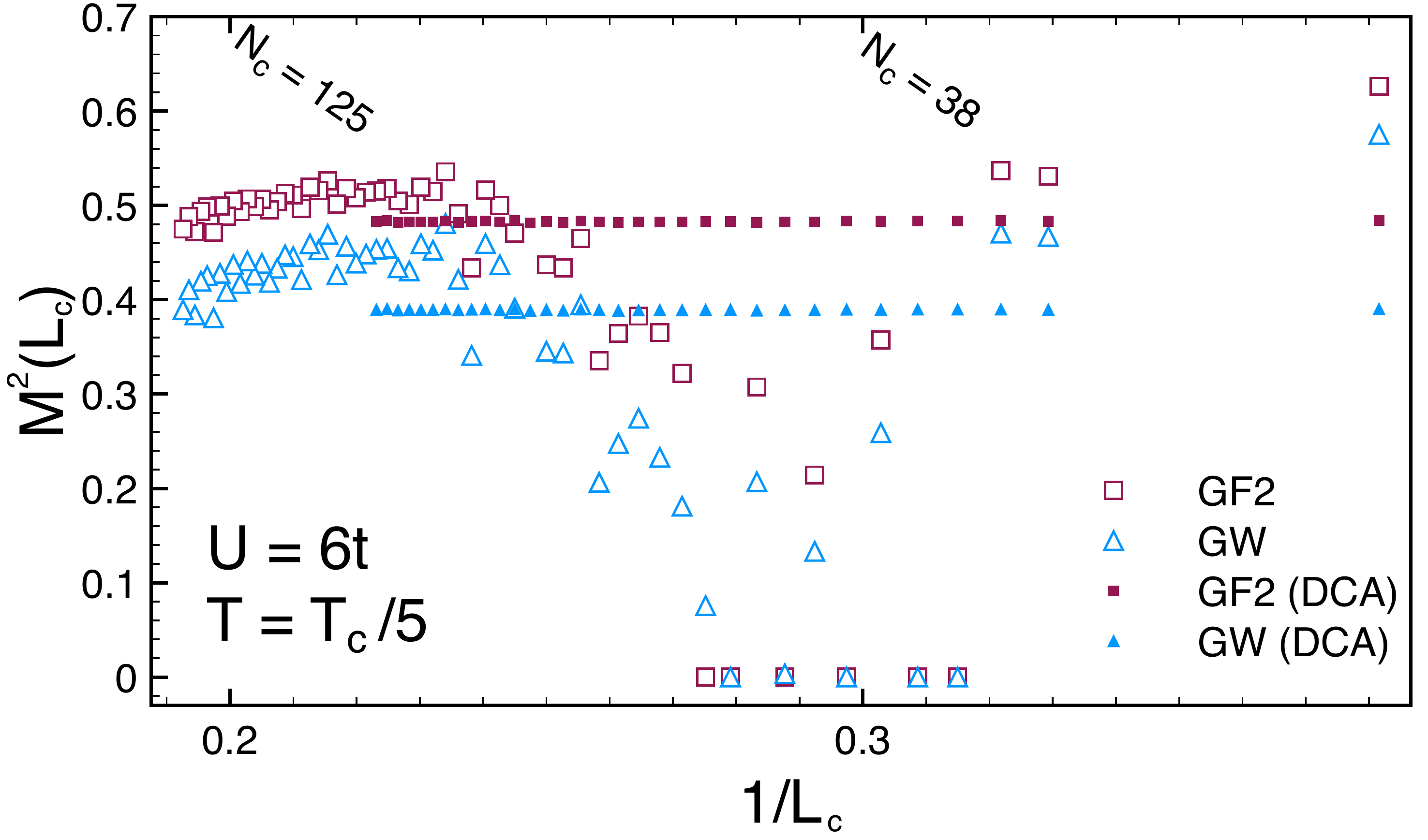}
        \caption{System size dependence of the antiferromagnetic order parameter for $U=4t$ (top) and $U=6t$ (bottom).
        Open symbols: lattice with periodic boundary conditions. Filled symbols: DCA. Red squares: GF2. Blue triangles: GW.}
        \label{fig:M_N}
    \end{figure}

    The finite size dependence of the order parameter, shown in Fig.~\ref{fig:M_N}, shows a similar trend as the critical temperature.
    For each cluster we perform simulations at a temperature $\sim 1/5 T_c$ (where the order parameter is close to saturation), as the critical temperature has a large cluster size dependence. In absence of embedding, the order parameter
    slowly saturates and even for $U=6t$ (bottom panel) does not fully converge. DCA embedding shows quick convergence and
    exhibits only small fluctuations, even for cluster with $N_c<80$.

    For interaction strength $U=4t$ and $U=6t$, we observe that in the absence of embedding some cluster geometries show no
    sign of an antiferromagnetic phase transition. Examining those clusters for their correspondence to the criteria outlined in Ref.~\cite{Betts1997}, we
    find that the cluster imperfection $I$, as the main criterion for cluster selection, is not a reliable predictor for the presence
    of a phase transition. The $N_c=38$ cluster with vectors $\mathbf{a}_1 =(1, 2, 3)^T$, $\mathbf{a}_2 =(3,-1,-2)^T$,
    $\mathbf{a}_3 =(2,-2, 2)^T$, for example, has an imperfection of $I=0$ and is therefore is classified as a `good' cluster~\cite{Kent2005}, despite having no $T_c$.

\begin{figure}[tbh]
        \centering
        \includegraphics[width=0.95\columnwidth]{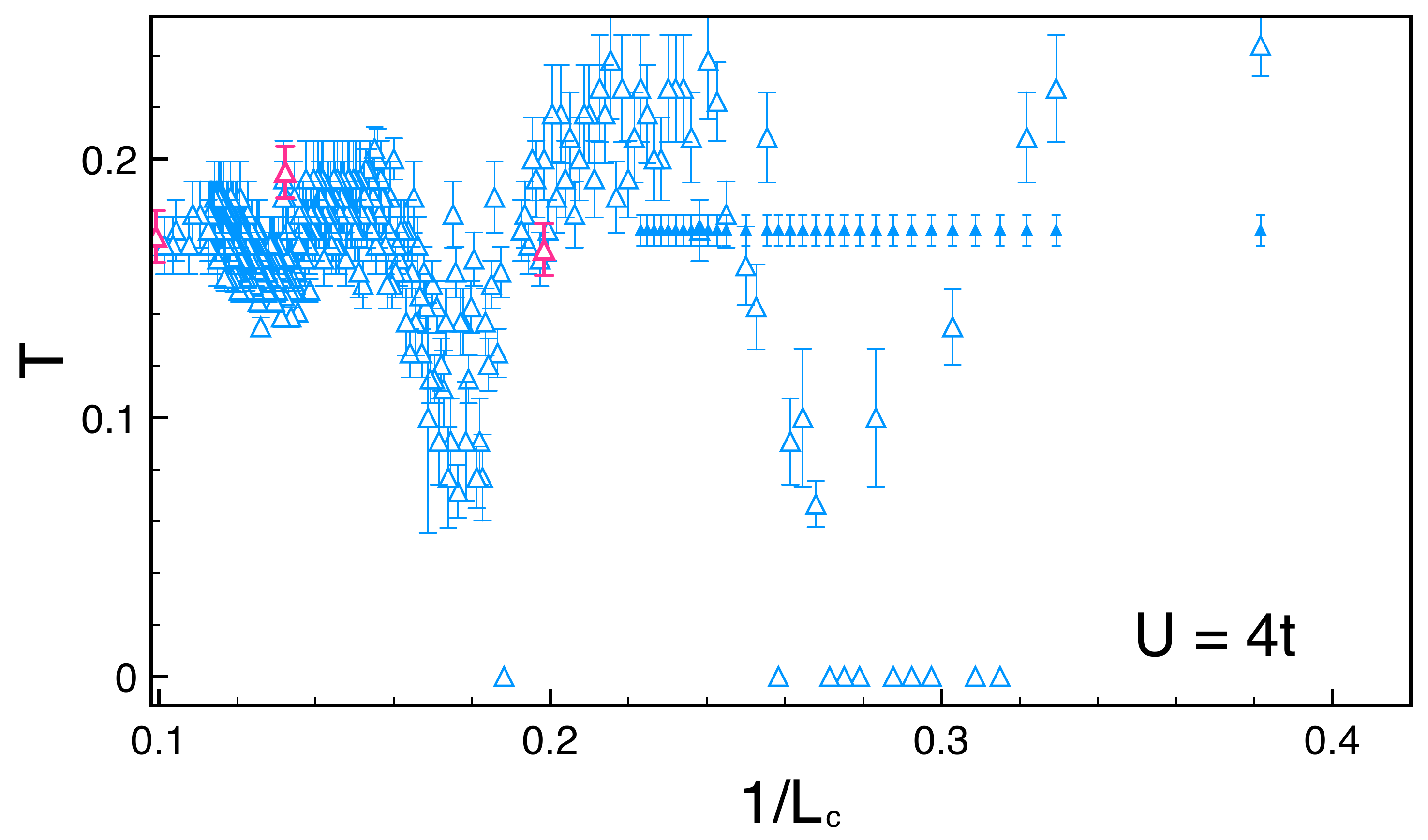}

        \includegraphics[width=0.95\columnwidth]{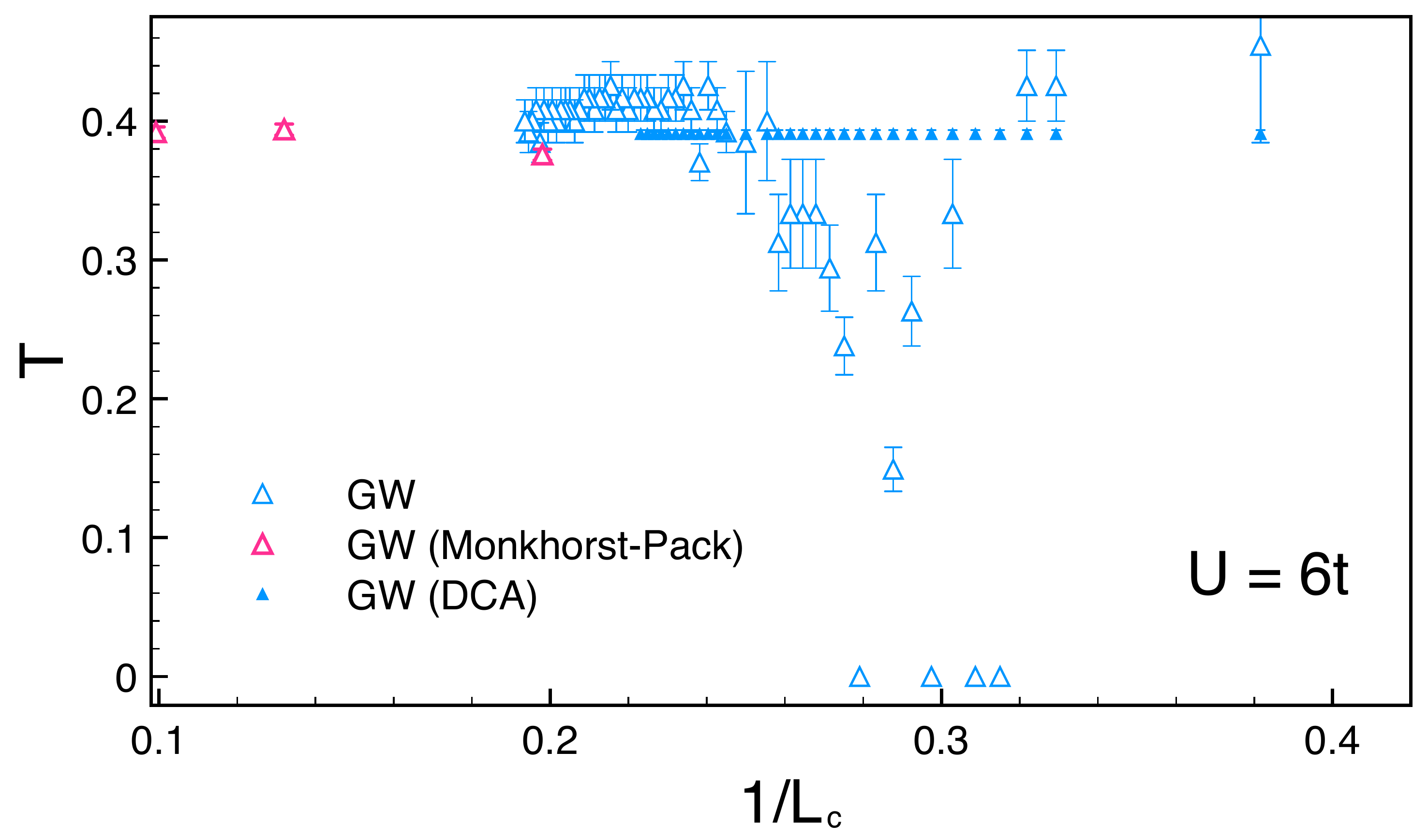}
        \caption{Comparison of system size dependence of the GW transition temperature for different type of cluster geometries
        for $U=4t$ (top) and $U=6t$ (bottom). Blue open triangle: Betts clusters with periodic boundary conditions. Purple open
        triangles: Monkhorst-Pack lattices with periodic boundary conditions. Filled blue triangles: DCA on Betts
        clusters.}
        \label{fig:Betts}
    \end{figure}

    In Fig.~\ref{fig:Betts} we compare the system size dependence of Monkhorst-Pack grids to those of Betts clusters. Shown
    are GW results for the system size dependence of transition temperature. Results without embedding for Betts clusters  are shown as blue
    open triangles, results for Monkhorst-Pack grids as open purple triangles for both values of interaction.
    DCA results (filled blue triangles) are shown as an estimate of the thermodynamic limit value. Results for Betts clusters and Monkhorst-Pack grids show similar finite size effects, leading us to conclude that the main advantage of
     Betts clusters is not an accelerated convergence, but the possibility of having many more finite-size values from which a    systematic finite size extrapolation can be performed.

    \section{Results for higher order diagrammatic methods}\label{sec:results2}

    One may hope to systematically improve perturbative methods by including additional diagrammatic topologies.
    We consider three types of diagrammatic approximations: particle-hole ladder (ph-ladder), particle-particle ladder
    (pp-ladder) and fluctuation exchange (FLEX)~\cite{Bickers1989a,Bickers1997,Bickers2004}. All of these methods are
    formulated in terms of single particle objects (such as Green's functions and self-energies), rather than two-particle
    susceptibilities~\cite{Rohringer2018}, and could therefore be extended to more complex multiorbital systems with limited
    numerical effort~\cite{Witt2021}. The ph-ladder includes the diagrams of GF2, GW and additional ``magnetic'' fluctuations.
    The pp-ladder contains charge fluctuations in addition to GF2 diagrams. FLEX contains all of the diagrams of GF2, GW,
    ph-ladder and pp-ladder combined. These methods are thermodynamically consistent and conserving~\cite{Baym1961,Baym1962,Bickers1989a}.
    Appendix~\ref{sec:ladder-and-flex} derives the formulas used.

    \subsection{Self-consistency convergence}\label{subsec:self-consistency-convergence}
    We find that all three higher order methods have a strong dependence on the starting point and may converge to a metastable
    isotropic phase instead of the ordered phase or to an unphysical solution. We attribute these convergence issues to the
    presence of sign-alternating diagrams (see Sec.~\ref{subsec:causality-properties}). Using a fixed staggered antiferromagnetic field to
    initialize the self-consistent iteration and gradually relaxing it reveals the antiferromagnetic transition in FLEX and
    in the ph-ladder. We find no indication of a phase transition in the pp-ladder.

    \begin{figure}[tbh]
        \centering
        \includegraphics[width=0.95\columnwidth]{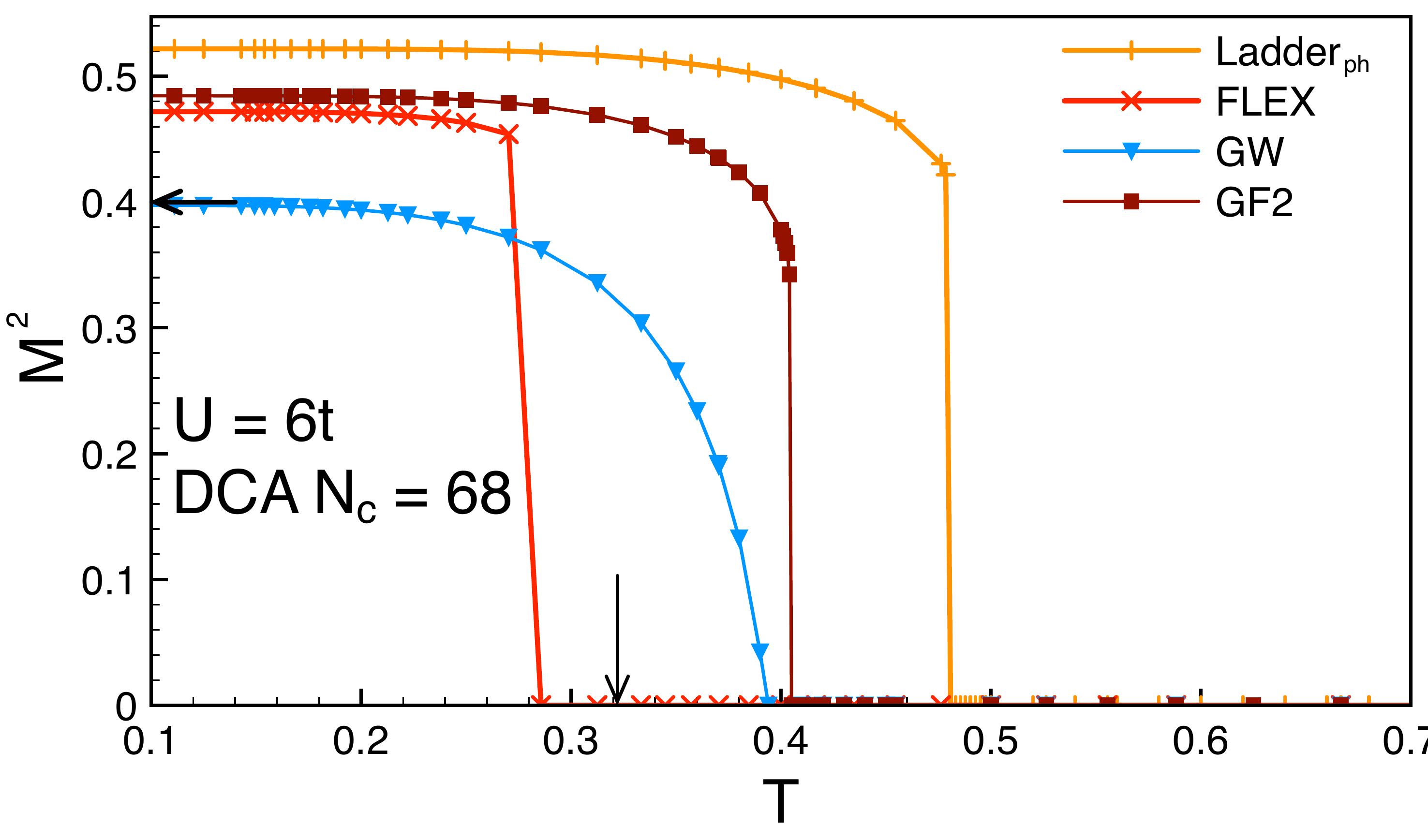}
        \caption{Temperature dependence of the order parameter at $U=6t$ for DCA cluster of size $N_c = 68$. Particle-hole
        Ladder: orange. FLEX: red. Also shown are results for GF2 (dark red) and GW (blue). Arrows indicate numerically exact
        CT-QMC solution for $T_c$ and order parameter.}
        \label{fig:Ladders_T}
    \end{figure}

    \subsection{Results for the ordered state}\label{subsec:doubled-unit-cell-results}
    As in Sec.~\ref{sec:results}, we use the double unit-cell formalism and observe the change in the order parameter. We
    present data for DCA with a fixed cluster of size $N_c=68$. The temperature dependence of the order parameter at $U=6t$ is
    shown in Fig.~\ref{fig:Ladders_T}. As shown in Fig.~\ref{fig:M_T}, GW (blue) shows a continuous phase transition and GF2
    (dark red) shows a first order transition. Both methods deviate from the numerically exact $T_c$ (black arrow) and also
    overestimate the order parameter. FLEX (red) and ph-ladder (orange) estimates bracket GF2 and GW
    values of $T_c$, as the additional renormalization in pp-channel, included in FLEX, leads to a substantial reduction of the
    transition temperature. The order parameter of FLEX and ph-ladder have a similar temperature dependence as GF2, indicating
    a first order transition.

\begin{figure}[tbh]
        \centering
        \includegraphics[width=0.95\columnwidth]{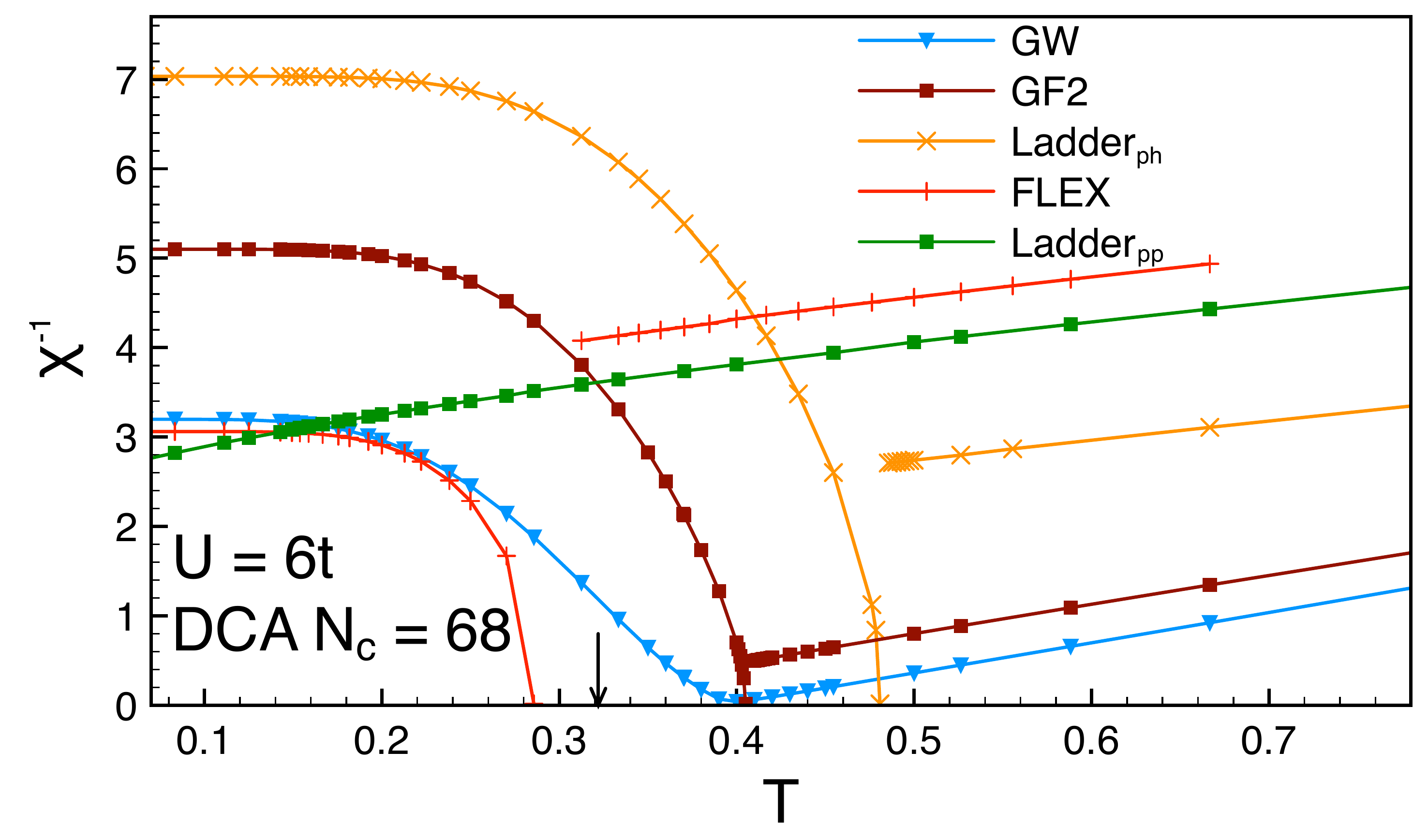}
        \caption{Inverse susceptibility in the linear response regime obtained for DCA clusters of size $N_c=68$ at $U=6t$.  GF2:
        dark red. GW: blue. ph-ladder: orange. pp-ladder: green. FLEX: red. Arrow indicates $T_c$ of numerically exact CT-QMC
        solution.}
        \label{fig:InvChi}
    \end{figure}

    In order to further analyze the transition, we evaluate the antiferromagnetic susceptibility $\chi = \frac{\partial M}{\partial H}$.
    Fig.~\ref{fig:InvChi} shows the temperature dependence of the inverse susceptibility in the linear response regime. For GW
    (blue), the susceptibility diverges from both sides of the estimated $T_c$. For GF2 (dark red), ph-ladder (orange) and
    FLEX (red) we see a divergence of the susceptibility when approaching $T_c$ from the ordered phase and a discontinuous jump to
    the isotropic phase, clearly indicating a first order transition for these methods. For pp-ladder (green) we see no
    divergence and do not observe a phase transition.

    As the results for both the order parameter (Fig.~\ref{fig:Ladders_T}) and the susceptibility (Fig.~\ref{fig:InvChi}) show,
    adding infinite series of diagrams does not lead to systematic improvement of the results, neither for the static
    low-T order parameter nor the critical temperature. Infinite geometric series, while necessary for continuous phase
    transitions, do not necessarily suppress the first order behavior in this model, and despite the ``magnetic'' particle-hole
    fluctuations contained in FLEX and ph-ladder, these methods do not capture the continuous magnetic phase transition.

    These observations are consistent with results for the weak-coupling regime of the 2d Hubbard model~\cite{Gukelberger2015},
    where systematic convergence was also not observed.

    \subsection{Comparison to $T_c$ extrapolated from high-T results}\label{subsec:comparison-to-normal-state-results}
    To evaluate the transition temperature approaching from the high-T phase we analyze the leading eigenvalue ($\lambda$) of the
    matrix $\pmb{\chi}^{0,(ph)}\pmb{\Gamma}^{0,(ph)}$ in the Eq.~\ref{eqn:phbethe} as a function of temperature. This is the
    standard method to analyze phase transitions in normal state simulations~\cite{Bulut1993,Bulut2002,Brener2008,Hafermann2009,Staar2014}
    but requires the calculation of a two particle susceptibility, rather than the simulation of the ordered state in an
    extended unit-cell. As it relies on the divergence of the susceptibility in the normal state, it is only applicable to
    continuous phase transitions.

    \begin{figure}[tbh]
        \centering
        \includegraphics[width=0.95\columnwidth]{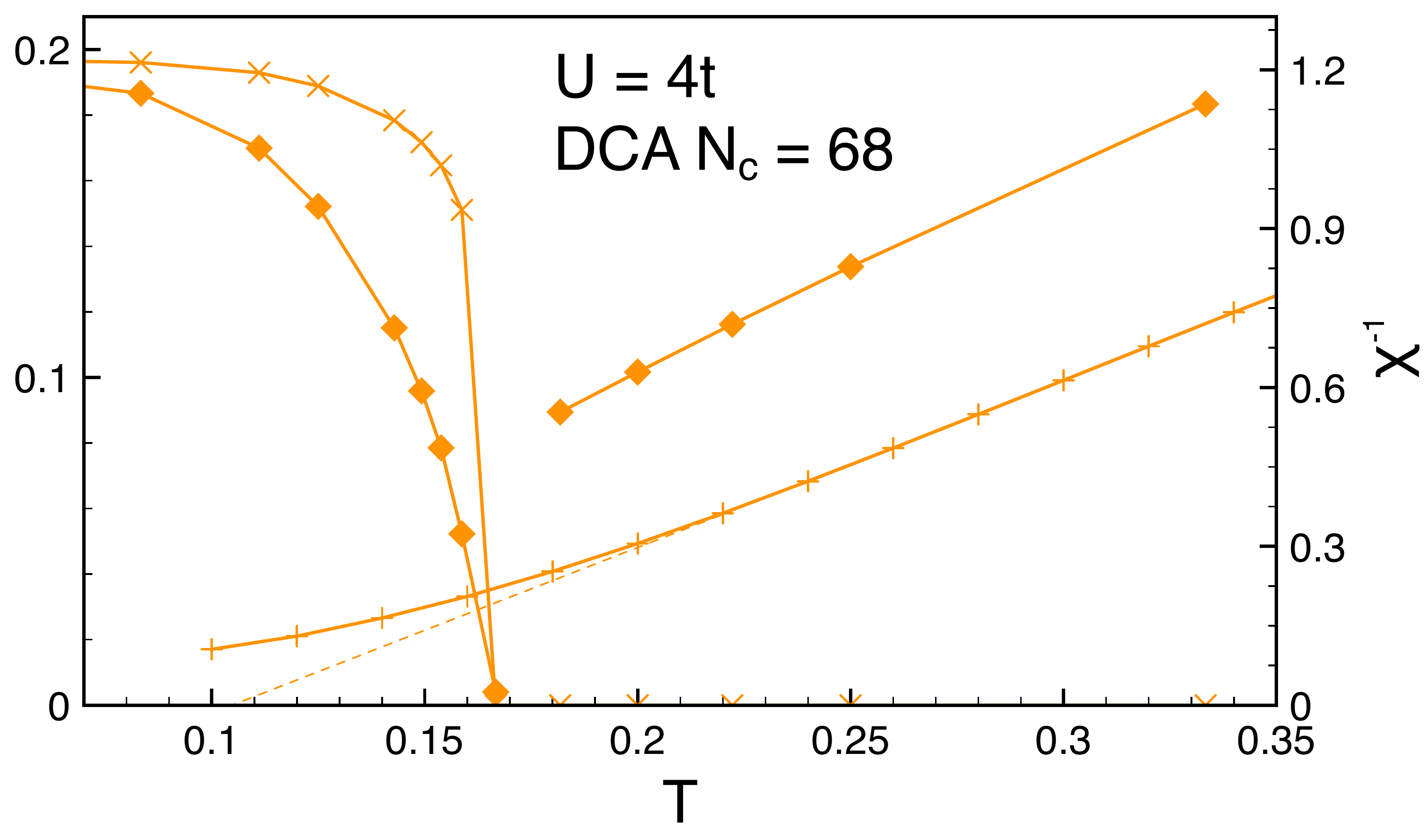}
        \includegraphics[width=0.95\columnwidth]{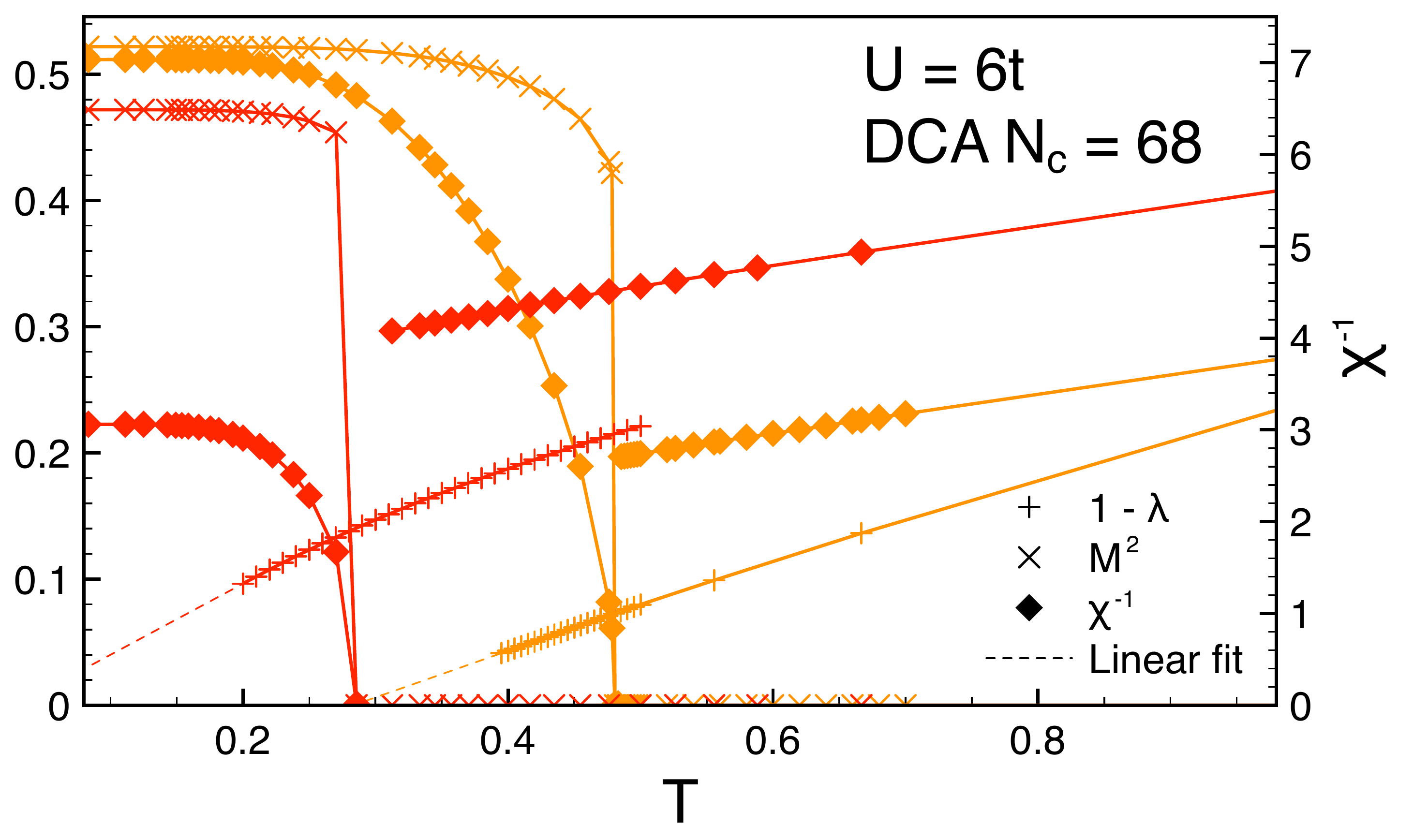}
        \caption{Methods for estimating the critical temperature at $U=4t$ (top) and $U=6t$ (bottom). Particle-hole ladder:
        orange. FLEX: red. Normal state solution: pluses. Doubled unit-cell order parameter: crosses.
        Doubled unit-cell inverse susceptibility: diamonds. Also shown is a linear extrapolation of high-T results for the leading
        eigenvalues in the ph-channel.}
        \label{fig:M_vs_l}
    \end{figure}

    For $U=4t$ (Fig.~\ref{fig:M_vs_l} top panel) we find that the $\lambda$ of the ph-ladder saturates near
    $\|\lambda\| \sim 0.96$ for temperatures down to $T \sim 0.1$, and no signature of $T_c \sim 0.166$ is visible. This
    indicates an absence of a divergence in the susceptibility when approaching from the isotropic phase and supports
    the observation of the first-order phase transition obtained in the double unit-cell formalism. Fig.~\ref{fig:M_vs_l} shows
    a comparison of the temperature dependence of the order parameter (crosses), inverse susceptibility (diamonds) and leading
    eigenvalue (pluses). For $U=6t$  (Fig.~\ref{fig:M_vs_l} bottom panel), similarly to $U=4t$, our ph-ladder
    simulations tend to either converge to unphysical solution or do not converge at all for $T<0.4t$, and no sign of $T_c\sim0.48$
    obtained in the double unit-cell formalism is visible.

    We are only able to converge FLEX to the ordered phase for $U\gtrsim5.5t$. The results for $U=6t$ are shown in the bottom
    panel of Fig.~\ref{fig:M_vs_l}. The phase transition estimates obtained in the ordered phase and in the normal state are
    substantially different due to the first order transition, similarly to what is observed for the ph-ladder.

    In addition to simulation data we show linear extrapolation of high temperature values of the leading eigenvalue
    (Fig.~\ref{fig:M_vs_l} dashed lines). Because of the first order nature of the transition, these extrapolations do not
    yield accurate transition temperatures.

    \section{Conclusion}\label{sec:conclusion}
    In this work, we examined the behavior of self-consistent diagrammatic methods at the example of the 3D Hubbard model,
    which is the prototypical fermionic model system for continuous phase transitions. Our diagrammatic methods are
    formulated in terms of single-particle quantities only, making them generalizable to real materials simulations, and compared
     to numerically exact CT-QMC data.
     
    We find that finite-size effects can be very efficiently controlled by using the DCA embedding scheme, and that embedding
    and lattice simulations constantly converge to the thermodynamic limit results, as expected.
    Betts clusters without embedding do not exhibit a faster convergence than the standard Monkhorst-Pack grids. However, in
    combination with DCA they provide many more systems from which an extrapolation with small clusters can be performed.
    An extension of the DCA scheme to realistic contexts may be able reproduce this accelerated convergence in more general systems.
 
     We also find that neither the transition temperature nor the ordered moment in GF2 or GW are accurate, and that they are not systematically improved by adding additional geometric series of diagrams. As methods that explicitly sum up  diagrams beyond those considered here are typically formulated in terms of four-operator `vertices' and are often too expensive to be applied in a realistic setup, our results imply that other paradigms, such as non-perturbative embedding methods, are needed to systematically improve results.

    The continuous magnetic phase transition examined here was found to be discontinuous in most of the methods, even in methods that
    contain infinite series of `magnetic' ladder diagrams. Therefore, the detection of phase transitions based on the
    divergence of susceptibilities (or the observation of leading eigenvalues) is not reliable.
    
    \acknowledgments{
        This work is supported by Simons Foundation via the Simons Collaboration on the Many Electron Problem. This work used the
        Extreme Science and Engineering Discovery Environment (XSEDE), which is supported by National Science Foundation grant
        number ACI-1548562, through allocation DMR130036.
    }

    \bibliographystyle{apsrev4-2}
    \bibliography{main}

\begin{thebibliography}{105}%
\makeatletter
\providecommand \@ifxundefined [1]{%
 \@ifx{#1\undefined}
}%
\providecommand \@ifnum [1]{%
 \ifnum #1\expandafter \@firstoftwo
 \else \expandafter \@secondoftwo
 \fi
}%
\providecommand \@ifx [1]{%
 \ifx #1\expandafter \@firstoftwo
 \else \expandafter \@secondoftwo
 \fi
}%
\providecommand \natexlab [1]{#1}%
\providecommand \enquote  [1]{``#1''}%
\providecommand \bibnamefont  [1]{#1}%
\providecommand \bibfnamefont [1]{#1}%
\providecommand \citenamefont [1]{#1}%
\providecommand \href@noop [0]{\@secondoftwo}%
\providecommand \href [0]{\begingroup \@sanitize@url \@href}%
\providecommand \@href[1]{\@@startlink{#1}\@@href}%
\providecommand \@@href[1]{\endgroup#1\@@endlink}%
\providecommand \@sanitize@url [0]{\catcode `\\12\catcode `\$12\catcode
  `\&12\catcode `\#12\catcode `\^12\catcode `\_12\catcode `\%12\relax}%
\providecommand \@@startlink[1]{}%
\providecommand \@@endlink[0]{}%
\providecommand \url  [0]{\begingroup\@sanitize@url \@url }%
\providecommand \@url [1]{\endgroup\@href {#1}{\urlprefix }}%
\providecommand \urlprefix  [0]{URL }%
\providecommand \Eprint [0]{\href }%
\providecommand \doibase [0]{https://doi.org/}%
\providecommand \selectlanguage [0]{\@gobble}%
\providecommand \bibinfo  [0]{\@secondoftwo}%
\providecommand \bibfield  [0]{\@secondoftwo}%
\providecommand \translation [1]{[#1]}%
\providecommand \BibitemOpen [0]{}%
\providecommand \bibitemStop [0]{}%
\providecommand \bibitemNoStop [0]{.\EOS\space}%
\providecommand \EOS [0]{\spacefactor3000\relax}%
\providecommand \BibitemShut  [1]{\csname bibitem#1\endcsname}%
\let\auto@bib@innerbib\@empty
%</preamble>
\bibitem [{\citenamefont {Qin}\ \emph {et~al.}(2021)\citenamefont {Qin},
  \citenamefont {Schäfer}, \citenamefont {Andergassen}, \citenamefont
  {Corboz},\ and\ \citenamefont {Gull}}]{Qin21}%
  \BibitemOpen
  \bibfield  {author} {\bibinfo {author} {\bibfnamefont {M.}~\bibnamefont
  {Qin}}, \bibinfo {author} {\bibfnamefont {T.}~\bibnamefont {Schäfer}},
  \bibinfo {author} {\bibfnamefont {S.}~\bibnamefont {Andergassen}}, \bibinfo
  {author} {\bibfnamefont {P.}~\bibnamefont {Corboz}},\ and\ \bibinfo {author}
  {\bibfnamefont {E.}~\bibnamefont {Gull}},\ }\href@noop {} {\bibinfo {title}
  {The hubbard model: A computational perspective}} (\bibinfo {year} {2021}),\
  \Eprint {https://arxiv.org/abs/2104.00064} {arXiv:2104.00064
  [cond-mat.str-el]} \BibitemShut {NoStop}%
\bibitem [{\citenamefont {Arovas}\ \emph {et~al.}(2021)\citenamefont {Arovas},
  \citenamefont {Berg}, \citenamefont {Kivelson},\ and\ \citenamefont
  {Raghu}}]{Arovas21}%
  \BibitemOpen
  \bibfield  {author} {\bibinfo {author} {\bibfnamefont {D.~P.}\ \bibnamefont
  {Arovas}}, \bibinfo {author} {\bibfnamefont {E.}~\bibnamefont {Berg}},
  \bibinfo {author} {\bibfnamefont {S.}~\bibnamefont {Kivelson}},\ and\
  \bibinfo {author} {\bibfnamefont {S.}~\bibnamefont {Raghu}},\ }\href@noop {}
  {\bibinfo {title} {The hubbard model}} (\bibinfo {year} {2021}),\ \Eprint
  {https://arxiv.org/abs/2103.12097} {arXiv:2103.12097 [cond-mat.str-el]}
  \BibitemShut {NoStop}%
\bibitem [{\citenamefont {{Staudt, R.}}\ \emph {et~al.}(2000)\citenamefont
  {{Staudt, R.}}, \citenamefont {{Dzierzawa, M.}},\ and\ \citenamefont
  {{Muramatsu, A.}}}]{Staudt2000}%
  \BibitemOpen
  \bibfield  {author} {\bibinfo {author} {\bibnamefont {{Staudt, R.}}},
  \bibinfo {author} {\bibnamefont {{Dzierzawa, M.}}},\ and\ \bibinfo {author}
  {\bibnamefont {{Muramatsu, A.}}},\ }\href
  {https://doi.org/10.1007/s100510070120} {\bibfield  {journal} {\bibinfo
  {journal} {Eur. Phys. J. B}\ }\textbf {\bibinfo {volume} {17}},\ \bibinfo
  {pages} {411} (\bibinfo {year} {2000})}\BibitemShut {NoStop}%
\bibitem [{\citenamefont {Kent}\ \emph {et~al.}(2005)\citenamefont {Kent},
  \citenamefont {Jarrell}, \citenamefont {Maier},\ and\ \citenamefont
  {Pruschke}}]{Kent2005}%
  \BibitemOpen
  \bibfield  {author} {\bibinfo {author} {\bibfnamefont {P.~R.~C.}\
  \bibnamefont {Kent}}, \bibinfo {author} {\bibfnamefont {M.}~\bibnamefont
  {Jarrell}}, \bibinfo {author} {\bibfnamefont {T.~A.}\ \bibnamefont {Maier}},\
  and\ \bibinfo {author} {\bibfnamefont {T.}~\bibnamefont {Pruschke}},\ }\href
  {https://doi.org/10.1103/PhysRevB.72.060411} {\bibfield  {journal} {\bibinfo
  {journal} {Phys. Rev. B}\ }\textbf {\bibinfo {volume} {72}},\ \bibinfo
  {pages} {060411} (\bibinfo {year} {2005})}\BibitemShut {NoStop}%
\bibitem [{\citenamefont {Fuchs}\ \emph
  {et~al.}(2011{\natexlab{a}})\citenamefont {Fuchs}, \citenamefont {Gull},
  \citenamefont {Pollet}, \citenamefont {Burovski}, \citenamefont {Kozik},
  \citenamefont {Pruschke},\ and\ \citenamefont {Troyer}}]{Fuchs2011c}%
  \BibitemOpen
  \bibfield  {author} {\bibinfo {author} {\bibfnamefont {S.}~\bibnamefont
  {Fuchs}}, \bibinfo {author} {\bibfnamefont {E.}~\bibnamefont {Gull}},
  \bibinfo {author} {\bibfnamefont {L.}~\bibnamefont {Pollet}}, \bibinfo
  {author} {\bibfnamefont {E.}~\bibnamefont {Burovski}}, \bibinfo {author}
  {\bibfnamefont {E.}~\bibnamefont {Kozik}}, \bibinfo {author} {\bibfnamefont
  {T.}~\bibnamefont {Pruschke}},\ and\ \bibinfo {author} {\bibfnamefont
  {M.}~\bibnamefont {Troyer}},\ }\href
  {https://doi.org/10.1103/PhysRevLett.106.030401} {\bibfield  {journal}
  {\bibinfo  {journal} {Phys. Rev. Lett.}\ }\textbf {\bibinfo {volume} {106}},\
  \bibinfo {pages} {030401} (\bibinfo {year} {2011}{\natexlab{a}})}\BibitemShut
  {NoStop}%
\bibitem [{\citenamefont {Fuchs}\ \emph
  {et~al.}(2011{\natexlab{b}})\citenamefont {Fuchs}, \citenamefont {Gull},
  \citenamefont {Troyer}, \citenamefont {Jarrell},\ and\ \citenamefont
  {Pruschke}}]{Fuchs2011}%
  \BibitemOpen
  \bibfield  {author} {\bibinfo {author} {\bibfnamefont {S.}~\bibnamefont
  {Fuchs}}, \bibinfo {author} {\bibfnamefont {E.}~\bibnamefont {Gull}},
  \bibinfo {author} {\bibfnamefont {M.}~\bibnamefont {Troyer}}, \bibinfo
  {author} {\bibfnamefont {M.}~\bibnamefont {Jarrell}},\ and\ \bibinfo {author}
  {\bibfnamefont {T.}~\bibnamefont {Pruschke}},\ }\href
  {https://doi.org/10.1103/PhysRevB.83.235113} {\bibfield  {journal} {\bibinfo
  {journal} {Phys. Rev. B}\ }\textbf {\bibinfo {volume} {83}},\ \bibinfo
  {pages} {235113} (\bibinfo {year} {2011}{\natexlab{b}})}\BibitemShut
  {NoStop}%
\bibitem [{\citenamefont {Paiva}\ \emph {et~al.}(2011)\citenamefont {Paiva},
  \citenamefont {Loh}, \citenamefont {Randeria}, \citenamefont {Scalettar},\
  and\ \citenamefont {Trivedi}}]{Paiva2011}%
  \BibitemOpen
  \bibfield  {author} {\bibinfo {author} {\bibfnamefont {T.}~\bibnamefont
  {Paiva}}, \bibinfo {author} {\bibfnamefont {Y.~L.}\ \bibnamefont {Loh}},
  \bibinfo {author} {\bibfnamefont {M.}~\bibnamefont {Randeria}}, \bibinfo
  {author} {\bibfnamefont {R.~T.}\ \bibnamefont {Scalettar}},\ and\ \bibinfo
  {author} {\bibfnamefont {N.}~\bibnamefont {Trivedi}},\ }\href
  {https://doi.org/10.1103/PhysRevLett.107.086401} {\bibfield  {journal}
  {\bibinfo  {journal} {Phys. Rev. Lett.}\ }\textbf {\bibinfo {volume} {107}},\
  \bibinfo {pages} {086401} (\bibinfo {year} {2011})}\BibitemShut {NoStop}%
\bibitem [{\citenamefont {Kozik}\ \emph {et~al.}(2013)\citenamefont {Kozik},
  \citenamefont {Burovski}, \citenamefont {Scarola},\ and\ \citenamefont
  {Troyer}}]{Kozik2013}%
  \BibitemOpen
  \bibfield  {author} {\bibinfo {author} {\bibfnamefont {E.}~\bibnamefont
  {Kozik}}, \bibinfo {author} {\bibfnamefont {E.}~\bibnamefont {Burovski}},
  \bibinfo {author} {\bibfnamefont {V.~W.}\ \bibnamefont {Scarola}},\ and\
  \bibinfo {author} {\bibfnamefont {M.}~\bibnamefont {Troyer}},\ }\href
  {https://doi.org/10.1103/PhysRevB.87.205102} {\bibfield  {journal} {\bibinfo
  {journal} {Phys. Rev. B}\ }\textbf {\bibinfo {volume} {87}},\ \bibinfo
  {pages} {205102} (\bibinfo {year} {2013})}\BibitemShut {NoStop}%
\bibitem [{\citenamefont {K\"ohl}\ \emph {et~al.}(2005)\citenamefont {K\"ohl},
  \citenamefont {Moritz}, \citenamefont {St\"oferle}, \citenamefont
  {G\"unter},\ and\ \citenamefont {Esslinger}}]{Kohl05}%
  \BibitemOpen
  \bibfield  {author} {\bibinfo {author} {\bibfnamefont {M.}~\bibnamefont
  {K\"ohl}}, \bibinfo {author} {\bibfnamefont {H.}~\bibnamefont {Moritz}},
  \bibinfo {author} {\bibfnamefont {T.}~\bibnamefont {St\"oferle}}, \bibinfo
  {author} {\bibfnamefont {K.}~\bibnamefont {G\"unter}},\ and\ \bibinfo
  {author} {\bibfnamefont {T.}~\bibnamefont {Esslinger}},\ }\href
  {https://doi.org/10.1103/PhysRevLett.94.080403} {\bibfield  {journal}
  {\bibinfo  {journal} {Phys. Rev. Lett.}\ }\textbf {\bibinfo {volume} {94}},\
  \bibinfo {pages} {080403} (\bibinfo {year} {2005})}\BibitemShut {NoStop}%
\bibitem [{\citenamefont {J{\"o}rdens}\ \emph {et~al.}(2008)\citenamefont
  {J{\"o}rdens}, \citenamefont {Strohmaier}, \citenamefont {G{\"u}nter},
  \citenamefont {Moritz},\ and\ \citenamefont {Esslinger}}]{Jordens2008}%
  \BibitemOpen
  \bibfield  {author} {\bibinfo {author} {\bibfnamefont {R.}~\bibnamefont
  {J{\"o}rdens}}, \bibinfo {author} {\bibfnamefont {N.}~\bibnamefont
  {Strohmaier}}, \bibinfo {author} {\bibfnamefont {K.}~\bibnamefont
  {G{\"u}nter}}, \bibinfo {author} {\bibfnamefont {H.}~\bibnamefont {Moritz}},\
  and\ \bibinfo {author} {\bibfnamefont {T.}~\bibnamefont {Esslinger}},\ }\href
  {https://doi.org/10.1038/nature07244} {\bibfield  {journal} {\bibinfo
  {journal} {Nature}\ }\textbf {\bibinfo {volume} {455}},\ \bibinfo {pages}
  {204} (\bibinfo {year} {2008})}\BibitemShut {NoStop}%
\bibitem [{\citenamefont {Schneider}\ \emph {et~al.}(2008)\citenamefont
  {Schneider}, \citenamefont {Hackerm{\"u}ller}, \citenamefont {Will},
  \citenamefont {Best}, \citenamefont {Bloch}, \citenamefont {Costi},
  \citenamefont {Helmes}, \citenamefont {Rasch},\ and\ \citenamefont
  {Rosch}}]{Schneider2008}%
  \BibitemOpen
  \bibfield  {author} {\bibinfo {author} {\bibfnamefont {U.}~\bibnamefont
  {Schneider}}, \bibinfo {author} {\bibfnamefont {L.}~\bibnamefont
  {Hackerm{\"u}ller}}, \bibinfo {author} {\bibfnamefont {S.}~\bibnamefont
  {Will}}, \bibinfo {author} {\bibfnamefont {T.}~\bibnamefont {Best}}, \bibinfo
  {author} {\bibfnamefont {I.}~\bibnamefont {Bloch}}, \bibinfo {author}
  {\bibfnamefont {T.~A.}\ \bibnamefont {Costi}}, \bibinfo {author}
  {\bibfnamefont {R.~W.}\ \bibnamefont {Helmes}}, \bibinfo {author}
  {\bibfnamefont {D.}~\bibnamefont {Rasch}},\ and\ \bibinfo {author}
  {\bibfnamefont {A.}~\bibnamefont {Rosch}},\ }\href
  {https://doi.org/10.1126/science.1165449} {\bibfield  {journal} {\bibinfo
  {journal} {Science}\ }\textbf {\bibinfo {volume} {322}},\ \bibinfo {pages}
  {1520} (\bibinfo {year} {2008})}\BibitemShut {NoStop}%
\bibitem [{\citenamefont {Esslinger}(2010)}]{Esslinger10}%
  \BibitemOpen
  \bibfield  {author} {\bibinfo {author} {\bibfnamefont {T.}~\bibnamefont
  {Esslinger}},\ }\href
  {https://doi.org/10.1146/annurev-conmatphys-070909-104059} {\bibfield
  {journal} {\bibinfo  {journal} {Annual Review of Condensed Matter Physics}\
  }\textbf {\bibinfo {volume} {1}},\ \bibinfo {pages} {129} (\bibinfo {year}
  {2010})},\ \Eprint
  {https://arxiv.org/abs/https://doi.org/10.1146/annurev-conmatphys-070909-104059}
  {https://doi.org/10.1146/annurev-conmatphys-070909-104059} \BibitemShut
  {NoStop}%
\bibitem [{\citenamefont {Duarte}\ \emph {et~al.}(2015)\citenamefont {Duarte},
  \citenamefont {Hart}, \citenamefont {Yang}, \citenamefont {Liu},
  \citenamefont {Paiva}, \citenamefont {Khatami}, \citenamefont {Scalettar},
  \citenamefont {Trivedi},\ and\ \citenamefont {Hulet}}]{Duarte2015}%
  \BibitemOpen
  \bibfield  {author} {\bibinfo {author} {\bibfnamefont {P.~M.}\ \bibnamefont
  {Duarte}}, \bibinfo {author} {\bibfnamefont {R.~A.}\ \bibnamefont {Hart}},
  \bibinfo {author} {\bibfnamefont {T.-L.}\ \bibnamefont {Yang}}, \bibinfo
  {author} {\bibfnamefont {X.}~\bibnamefont {Liu}}, \bibinfo {author}
  {\bibfnamefont {T.}~\bibnamefont {Paiva}}, \bibinfo {author} {\bibfnamefont
  {E.}~\bibnamefont {Khatami}}, \bibinfo {author} {\bibfnamefont {R.~T.}\
  \bibnamefont {Scalettar}}, \bibinfo {author} {\bibfnamefont {N.}~\bibnamefont
  {Trivedi}},\ and\ \bibinfo {author} {\bibfnamefont {R.~G.}\ \bibnamefont
  {Hulet}},\ }\href {https://doi.org/10.1103/PhysRevLett.114.070403} {\bibfield
   {journal} {\bibinfo  {journal} {Phys. Rev. Lett.}\ }\textbf {\bibinfo
  {volume} {114}},\ \bibinfo {pages} {070403} (\bibinfo {year}
  {2015})}\BibitemShut {NoStop}%
\bibitem [{\citenamefont {Hart}\ \emph {et~al.}(2015)\citenamefont {Hart},
  \citenamefont {Duarte}, \citenamefont {Yang}, \citenamefont {Liu},
  \citenamefont {Paiva}, \citenamefont {Khatami}, \citenamefont {Scalettar},
  \citenamefont {Trivedi}, \citenamefont {Huse},\ and\ \citenamefont
  {Hulet}}]{Hart2015}%
  \BibitemOpen
  \bibfield  {author} {\bibinfo {author} {\bibfnamefont {R.~A.}\ \bibnamefont
  {Hart}}, \bibinfo {author} {\bibfnamefont {P.~M.}\ \bibnamefont {Duarte}},
  \bibinfo {author} {\bibfnamefont {T.-L.}\ \bibnamefont {Yang}}, \bibinfo
  {author} {\bibfnamefont {X.}~\bibnamefont {Liu}}, \bibinfo {author}
  {\bibfnamefont {T.}~\bibnamefont {Paiva}}, \bibinfo {author} {\bibfnamefont
  {E.}~\bibnamefont {Khatami}}, \bibinfo {author} {\bibfnamefont {R.~T.}\
  \bibnamefont {Scalettar}}, \bibinfo {author} {\bibfnamefont {N.}~\bibnamefont
  {Trivedi}}, \bibinfo {author} {\bibfnamefont {D.~A.}\ \bibnamefont {Huse}},\
  and\ \bibinfo {author} {\bibfnamefont {R.~G.}\ \bibnamefont {Hulet}},\ }\href
  {https://doi.org/10.1038/nature14223} {\bibfield  {journal} {\bibinfo
  {journal} {Nature}\ }\textbf {\bibinfo {volume} {519}},\ \bibinfo {pages}
  {211} (\bibinfo {year} {2015})}\BibitemShut {NoStop}%
\bibitem [{\citenamefont {Sousa}\ \emph {et~al.}(2007)\citenamefont {Sousa},
  \citenamefont {Fernandes},\ and\ \citenamefont {Ramos}}]{Sousa2007}%
  \BibitemOpen
  \bibfield  {author} {\bibinfo {author} {\bibfnamefont {S.~F.}\ \bibnamefont
  {Sousa}}, \bibinfo {author} {\bibfnamefont {P.~A.}\ \bibnamefont
  {Fernandes}},\ and\ \bibinfo {author} {\bibfnamefont {M.~J.}\ \bibnamefont
  {Ramos}},\ }\href {https://doi.org/10.1021/jp0734474} {\bibfield  {journal}
  {\bibinfo  {journal} {The Journal of Physical Chemistry A}\ }\textbf
  {\bibinfo {volume} {111}},\ \bibinfo {pages} {10439} (\bibinfo {year}
  {2007})}\BibitemShut {NoStop}%
\bibitem [{\citenamefont {Rusakov}\ and\ \citenamefont
  {Zgid}(2016)}]{Rusakov2016}%
  \BibitemOpen
  \bibfield  {author} {\bibinfo {author} {\bibfnamefont {A.~A.}\ \bibnamefont
  {Rusakov}}\ and\ \bibinfo {author} {\bibfnamefont {D.}~\bibnamefont {Zgid}},\
  }\href {https://doi.org/10.1063/1.4940900} {\bibfield  {journal} {\bibinfo
  {journal} {The Journal of Chemical Physics}\ }\textbf {\bibinfo {volume}
  {144}},\ \bibinfo {pages} {054106} (\bibinfo {year} {2016})}\BibitemShut
  {NoStop}%
\bibitem [{\citenamefont {Hedin}(1965)}]{Hedin1965}%
  \BibitemOpen
  \bibfield  {author} {\bibinfo {author} {\bibfnamefont {L.}~\bibnamefont
  {Hedin}},\ }\href {https://doi.org/10.1103/PhysRev.139.A796} {\bibfield
  {journal} {\bibinfo  {journal} {Phys. Rev.}\ }\textbf {\bibinfo {volume}
  {139}},\ \bibinfo {pages} {A796} (\bibinfo {year} {1965})}\BibitemShut
  {NoStop}%
\bibitem [{\citenamefont {Bickers}\ \emph {et~al.}(1989)\citenamefont
  {Bickers}, \citenamefont {Scalapino},\ and\ \citenamefont
  {White}}]{Bickers1989a}%
  \BibitemOpen
  \bibfield  {author} {\bibinfo {author} {\bibfnamefont {N.~E.}\ \bibnamefont
  {Bickers}}, \bibinfo {author} {\bibfnamefont {D.~J.}\ \bibnamefont
  {Scalapino}},\ and\ \bibinfo {author} {\bibfnamefont {S.~R.}\ \bibnamefont
  {White}},\ }\href {https://doi.org/10.1103/PhysRevLett.62.961} {\bibfield
  {journal} {\bibinfo  {journal} {Phys. Rev. Lett.}\ }\textbf {\bibinfo
  {volume} {62}},\ \bibinfo {pages} {961} (\bibinfo {year} {1989})}\BibitemShut
  {NoStop}%
\bibitem [{\citenamefont {Esirgen}\ and\ \citenamefont
  {Bickers}(1997)}]{Bickers1997}%
  \BibitemOpen
  \bibfield  {author} {\bibinfo {author} {\bibfnamefont {G.}~\bibnamefont
  {Esirgen}}\ and\ \bibinfo {author} {\bibfnamefont {N.~E.}\ \bibnamefont
  {Bickers}},\ }\href {https://doi.org/10.1103/PhysRevB.55.2122} {\bibfield
  {journal} {\bibinfo  {journal} {Phys. Rev. B}\ }\textbf {\bibinfo {volume}
  {55}},\ \bibinfo {pages} {2122} (\bibinfo {year} {1997})}\BibitemShut
  {NoStop}%
\bibitem [{\citenamefont {Bickers}(2004)}]{Bickers2004}%
  \BibitemOpen
  \bibfield  {author} {\bibinfo {author} {\bibfnamefont {N.~E.}\ \bibnamefont
  {Bickers}},\ }\bibinfo {title} {Self-consistent many-body theory for
  condensed matter systems},\ in\ \href
  {https://doi.org/10.1007/0-387-21717-7_6} {\emph {\bibinfo {booktitle}
  {Theoretical Methods for Strongly Correlated Electrons}}},\ \bibinfo {editor}
  {edited by\ \bibinfo {editor} {\bibfnamefont {D.}~\bibnamefont
  {S{\'e}n{\'e}chal}}, \bibinfo {editor} {\bibfnamefont {A.-M.}\ \bibnamefont
  {Tremblay}},\ and\ \bibinfo {editor} {\bibfnamefont {C.}~\bibnamefont
  {Bourbonnais}}}\ (\bibinfo  {publisher} {Springer New York},\ \bibinfo
  {address} {New York, NY},\ \bibinfo {year} {2004})\ pp.\ \bibinfo {pages}
  {237--296}\BibitemShut {NoStop}%
\bibitem [{\citenamefont {Gull}\ \emph
  {et~al.}(2011{\natexlab{a}})\citenamefont {Gull}, \citenamefont {Millis},
  \citenamefont {Lichtenstein}, \citenamefont {Rubtsov}, \citenamefont
  {Troyer},\ and\ \citenamefont {Werner}}]{Gull2011b}%
  \BibitemOpen
  \bibfield  {author} {\bibinfo {author} {\bibfnamefont {E.}~\bibnamefont
  {Gull}}, \bibinfo {author} {\bibfnamefont {A.~J.}\ \bibnamefont {Millis}},
  \bibinfo {author} {\bibfnamefont {A.~I.}\ \bibnamefont {Lichtenstein}},
  \bibinfo {author} {\bibfnamefont {A.~N.}\ \bibnamefont {Rubtsov}}, \bibinfo
  {author} {\bibfnamefont {M.}~\bibnamefont {Troyer}},\ and\ \bibinfo {author}
  {\bibfnamefont {P.}~\bibnamefont {Werner}},\ }\href
  {https://doi.org/10.1103/RevModPhys.83.349} {\bibfield  {journal} {\bibinfo
  {journal} {Rev. Mod. Phys.}\ }\textbf {\bibinfo {volume} {83}},\ \bibinfo
  {pages} {349} (\bibinfo {year} {2011}{\natexlab{a}})}\BibitemShut {NoStop}%
\bibitem [{\citenamefont {Gull}\ \emph {et~al.}(2008)\citenamefont {Gull},
  \citenamefont {Werner}, \citenamefont {Parcollet},\ and\ \citenamefont
  {Troyer}}]{Gull2008}%
  \BibitemOpen
  \bibfield  {author} {\bibinfo {author} {\bibfnamefont {E.}~\bibnamefont
  {Gull}}, \bibinfo {author} {\bibfnamefont {P.}~\bibnamefont {Werner}},
  \bibinfo {author} {\bibfnamefont {O.}~\bibnamefont {Parcollet}},\ and\
  \bibinfo {author} {\bibfnamefont {M.}~\bibnamefont {Troyer}},\ }\href
  {https://doi.org/10.1209/0295-5075/82/57003} {\bibfield  {journal} {\bibinfo
  {journal} {{EPL} (Europhysics Letters)}\ }\textbf {\bibinfo {volume} {82}},\
  \bibinfo {pages} {57003} (\bibinfo {year} {2008})}\BibitemShut {NoStop}%
\bibitem [{\citenamefont {Gull}\ \emph
  {et~al.}(2011{\natexlab{b}})\citenamefont {Gull}, \citenamefont {Staar},
  \citenamefont {Fuchs}, \citenamefont {Nukala}, \citenamefont {Summers},
  \citenamefont {Pruschke}, \citenamefont {Schulthess},\ and\ \citenamefont
  {Maier}}]{Gull2011a}%
  \BibitemOpen
  \bibfield  {author} {\bibinfo {author} {\bibfnamefont {E.}~\bibnamefont
  {Gull}}, \bibinfo {author} {\bibfnamefont {P.}~\bibnamefont {Staar}},
  \bibinfo {author} {\bibfnamefont {S.}~\bibnamefont {Fuchs}}, \bibinfo
  {author} {\bibfnamefont {P.}~\bibnamefont {Nukala}}, \bibinfo {author}
  {\bibfnamefont {M.~S.}\ \bibnamefont {Summers}}, \bibinfo {author}
  {\bibfnamefont {T.}~\bibnamefont {Pruschke}}, \bibinfo {author}
  {\bibfnamefont {T.~C.}\ \bibnamefont {Schulthess}},\ and\ \bibinfo {author}
  {\bibfnamefont {T.}~\bibnamefont {Maier}},\ }\href
  {https://doi.org/10.1103/PhysRevB.83.075122} {\bibfield  {journal} {\bibinfo
  {journal} {Phys. Rev. B}\ }\textbf {\bibinfo {volume} {83}},\ \bibinfo
  {pages} {075122} (\bibinfo {year} {2011}{\natexlab{b}})}\BibitemShut
  {NoStop}%
\bibitem [{\citenamefont {Hirschmeier}\ \emph {et~al.}(2015)\citenamefont
  {Hirschmeier}, \citenamefont {Hafermann}, \citenamefont {Gull}, \citenamefont
  {Lichtenstein},\ and\ \citenamefont {Antipov}}]{Hirschmeier2015}%
  \BibitemOpen
  \bibfield  {author} {\bibinfo {author} {\bibfnamefont {D.}~\bibnamefont
  {Hirschmeier}}, \bibinfo {author} {\bibfnamefont {H.}~\bibnamefont
  {Hafermann}}, \bibinfo {author} {\bibfnamefont {E.}~\bibnamefont {Gull}},
  \bibinfo {author} {\bibfnamefont {A.~I.}\ \bibnamefont {Lichtenstein}},\ and\
  \bibinfo {author} {\bibfnamefont {A.~E.}\ \bibnamefont {Antipov}},\ }\href
  {https://doi.org/10.1103/PhysRevB.92.144409} {\bibfield  {journal} {\bibinfo
  {journal} {Phys. Rev. B}\ }\textbf {\bibinfo {volume} {92}},\ \bibinfo
  {pages} {144409} (\bibinfo {year} {2015})}\BibitemShut {NoStop}%
\bibitem [{\citenamefont {Hettler}\ \emph {et~al.}(1998)\citenamefont
  {Hettler}, \citenamefont {Tahvildar-Zadeh}, \citenamefont {Jarrell},
  \citenamefont {Pruschke},\ and\ \citenamefont {Krishnamurthy}}]{Hettler1998}%
  \BibitemOpen
  \bibfield  {author} {\bibinfo {author} {\bibfnamefont {M.~H.}\ \bibnamefont
  {Hettler}}, \bibinfo {author} {\bibfnamefont {A.~N.}\ \bibnamefont
  {Tahvildar-Zadeh}}, \bibinfo {author} {\bibfnamefont {M.}~\bibnamefont
  {Jarrell}}, \bibinfo {author} {\bibfnamefont {T.}~\bibnamefont {Pruschke}},\
  and\ \bibinfo {author} {\bibfnamefont {H.~R.}\ \bibnamefont
  {Krishnamurthy}},\ }\href {https://doi.org/10.1103/PhysRevB.58.R7475}
  {\bibfield  {journal} {\bibinfo  {journal} {Phys. Rev. B}\ }\textbf {\bibinfo
  {volume} {58}},\ \bibinfo {pages} {R7475} (\bibinfo {year}
  {1998})}\BibitemShut {NoStop}%
\bibitem [{\citenamefont {Maier}\ \emph
  {et~al.}(2005{\natexlab{a}})\citenamefont {Maier}, \citenamefont {Jarrell},
  \citenamefont {Pruschke},\ and\ \citenamefont {Hettler}}]{Maier2005}%
  \BibitemOpen
  \bibfield  {author} {\bibinfo {author} {\bibfnamefont {T.}~\bibnamefont
  {Maier}}, \bibinfo {author} {\bibfnamefont {M.}~\bibnamefont {Jarrell}},
  \bibinfo {author} {\bibfnamefont {T.}~\bibnamefont {Pruschke}},\ and\
  \bibinfo {author} {\bibfnamefont {M.~H.}\ \bibnamefont {Hettler}},\ }\href
  {https://doi.org/10.1103/RevModPhys.77.1027} {\bibfield  {journal} {\bibinfo
  {journal} {Rev. Mod. Phys.}\ }\textbf {\bibinfo {volume} {77}},\ \bibinfo
  {pages} {1027} (\bibinfo {year} {2005}{\natexlab{a}})}\BibitemShut {NoStop}%
\bibitem [{\citenamefont {Betts}\ and\ \citenamefont
  {Stewart}(1997)}]{Betts1997}%
  \BibitemOpen
  \bibfield  {author} {\bibinfo {author} {\bibfnamefont {D.~D.}\ \bibnamefont
  {Betts}}\ and\ \bibinfo {author} {\bibfnamefont {G.~E.}\ \bibnamefont
  {Stewart}},\ }\href {https://doi.org/10.1139/p96-129} {\bibfield  {journal}
  {\bibinfo  {journal} {Canadian Journal of Physics}\ }\textbf {\bibinfo
  {volume} {75}},\ \bibinfo {pages} {47} (\bibinfo {year} {1997})}\BibitemShut
  {NoStop}%
\bibitem [{\citenamefont {Betts}\ \emph {et~al.}(1999)\citenamefont {Betts},
  \citenamefont {Lin},\ and\ \citenamefont {Flynn}}]{Betts1999}%
  \BibitemOpen
  \bibfield  {author} {\bibinfo {author} {\bibfnamefont {D.~D.}\ \bibnamefont
  {Betts}}, \bibinfo {author} {\bibfnamefont {H.~Q.}\ \bibnamefont {Lin}},\
  and\ \bibinfo {author} {\bibfnamefont {J.~S.}\ \bibnamefont {Flynn}},\ }\href
  {https://doi.org/10.1139/p99-041} {\bibfield  {journal} {\bibinfo  {journal}
  {Canadian Journal of Physics}\ }\textbf {\bibinfo {volume} {77}},\ \bibinfo
  {pages} {353} (\bibinfo {year} {1999})}\BibitemShut {NoStop}%
\bibitem [{\citenamefont {Georges}\ \emph {et~al.}(1996)\citenamefont
  {Georges}, \citenamefont {Kotliar}, \citenamefont {Krauth},\ and\
  \citenamefont {Rozenberg}}]{Georges1996}%
  \BibitemOpen
  \bibfield  {author} {\bibinfo {author} {\bibfnamefont {A.}~\bibnamefont
  {Georges}}, \bibinfo {author} {\bibfnamefont {G.}~\bibnamefont {Kotliar}},
  \bibinfo {author} {\bibfnamefont {W.}~\bibnamefont {Krauth}},\ and\ \bibinfo
  {author} {\bibfnamefont {M.~J.}\ \bibnamefont {Rozenberg}},\ }\href
  {https://doi.org/10.1103/RevModPhys.68.13} {\bibfield  {journal} {\bibinfo
  {journal} {Rev. Mod. Phys.}\ }\textbf {\bibinfo {volume} {68}},\ \bibinfo
  {pages} {13} (\bibinfo {year} {1996})}\BibitemShut {NoStop}%
\bibitem [{\citenamefont {Zgid}\ and\ \citenamefont {Gull}(2017)}]{Zgid2017}%
  \BibitemOpen
  \bibfield  {author} {\bibinfo {author} {\bibfnamefont {D.}~\bibnamefont
  {Zgid}}\ and\ \bibinfo {author} {\bibfnamefont {E.}~\bibnamefont {Gull}},\
  }\href {https://doi.org/10.1088/1367-2630/aa5d34} {\bibfield  {journal}
  {\bibinfo  {journal} {New Journal of Physics}\ }\textbf {\bibinfo {volume}
  {19}},\ \bibinfo {pages} {023047} (\bibinfo {year} {2017})}\BibitemShut
  {NoStop}%
\bibitem [{\citenamefont {Mahan}(2000)}]{Mahan2000}%
  \BibitemOpen
  \bibfield  {author} {\bibinfo {author} {\bibfnamefont {G.}~\bibnamefont
  {Mahan}},\ }\href {https://books.google.com/books?id=xzSgZ4-yyMEC} {\emph
  {\bibinfo {title} {Many-Particle Physics}}},\ Physics of Solids and Liquids\
  (\bibinfo  {publisher} {Springer US},\ \bibinfo {year} {2000})\BibitemShut
  {NoStop}%
\bibitem [{\citenamefont {Rohringer}\ \emph {et~al.}(2018)\citenamefont
  {Rohringer}, \citenamefont {Hafermann}, \citenamefont {Toschi}, \citenamefont
  {Katanin}, \citenamefont {Antipov}, \citenamefont {Katsnelson}, \citenamefont
  {Lichtenstein}, \citenamefont {Rubtsov},\ and\ \citenamefont
  {Held}}]{Rohringer2018}%
  \BibitemOpen
  \bibfield  {author} {\bibinfo {author} {\bibfnamefont {G.}~\bibnamefont
  {Rohringer}}, \bibinfo {author} {\bibfnamefont {H.}~\bibnamefont
  {Hafermann}}, \bibinfo {author} {\bibfnamefont {A.}~\bibnamefont {Toschi}},
  \bibinfo {author} {\bibfnamefont {A.~A.}\ \bibnamefont {Katanin}}, \bibinfo
  {author} {\bibfnamefont {A.~E.}\ \bibnamefont {Antipov}}, \bibinfo {author}
  {\bibfnamefont {M.~I.}\ \bibnamefont {Katsnelson}}, \bibinfo {author}
  {\bibfnamefont {A.~I.}\ \bibnamefont {Lichtenstein}}, \bibinfo {author}
  {\bibfnamefont {A.~N.}\ \bibnamefont {Rubtsov}},\ and\ \bibinfo {author}
  {\bibfnamefont {K.}~\bibnamefont {Held}},\ }\href
  {https://doi.org/10.1103/RevModPhys.90.025003} {\bibfield  {journal}
  {\bibinfo  {journal} {Rev. Mod. Phys.}\ }\textbf {\bibinfo {volume} {90}},\
  \bibinfo {pages} {025003} (\bibinfo {year} {2018})}\BibitemShut {NoStop}%
\bibitem [{\citenamefont {Galitskii}\ and\ \citenamefont
  {Migdal}(1958)}]{Galitskii1958}%
  \BibitemOpen
  \bibfield  {author} {\bibinfo {author} {\bibfnamefont {V.~M.}\ \bibnamefont
  {Galitskii}}\ and\ \bibinfo {author} {\bibfnamefont {A.~B.}\ \bibnamefont
  {Migdal}},\ }\href@noop {} {\bibfield  {journal} {\bibinfo  {journal} {Sov.
  Phys. JETP}\ }\textbf {\bibinfo {volume} {7}},\ \bibinfo {pages} {96}
  (\bibinfo {year} {1958})}\BibitemShut {NoStop}%
\bibitem [{\citenamefont {Holm}\ and\ \citenamefont
  {Aryasetiawan}(2000)}]{Holm2000}%
  \BibitemOpen
  \bibfield  {author} {\bibinfo {author} {\bibfnamefont {B.}~\bibnamefont
  {Holm}}\ and\ \bibinfo {author} {\bibfnamefont {F.}~\bibnamefont
  {Aryasetiawan}},\ }\href {https://doi.org/10.1103/PhysRevB.62.4858}
  {\bibfield  {journal} {\bibinfo  {journal} {Phys. Rev. B}\ }\textbf {\bibinfo
  {volume} {62}},\ \bibinfo {pages} {4858} (\bibinfo {year}
  {2000})}\BibitemShut {NoStop}%
\bibitem [{\citenamefont {LeBlanc}\ and\ \citenamefont
  {Gull}(2013)}]{LeBlanc2013}%
  \BibitemOpen
  \bibfield  {author} {\bibinfo {author} {\bibfnamefont {J.~P.~F.}\
  \bibnamefont {LeBlanc}}\ and\ \bibinfo {author} {\bibfnamefont
  {E.}~\bibnamefont {Gull}},\ }\href
  {https://doi.org/10.1103/PhysRevB.88.155108} {\bibfield  {journal} {\bibinfo
  {journal} {Phys. Rev. B}\ }\textbf {\bibinfo {volume} {88}},\ \bibinfo
  {pages} {155108} (\bibinfo {year} {2013})}\BibitemShut {NoStop}%
\bibitem [{\citenamefont {Welden}\ \emph {et~al.}(2016)\citenamefont {Welden},
  \citenamefont {Rusakov},\ and\ \citenamefont {Zgid}}]{Welden2016}%
  \BibitemOpen
  \bibfield  {author} {\bibinfo {author} {\bibfnamefont {A.~R.}\ \bibnamefont
  {Welden}}, \bibinfo {author} {\bibfnamefont {A.~A.}\ \bibnamefont
  {Rusakov}},\ and\ \bibinfo {author} {\bibfnamefont {D.}~\bibnamefont
  {Zgid}},\ }\href {https://doi.org/10.1063/1.4967449} {\bibfield  {journal}
  {\bibinfo  {journal} {The Journal of Chemical Physics}\ }\textbf {\bibinfo
  {volume} {145}},\ \bibinfo {pages} {204106} (\bibinfo {year}
  {2016})}\BibitemShut {NoStop}%
\bibitem [{\citenamefont {Neuhauser}\ \emph {et~al.}(2017)\citenamefont
  {Neuhauser}, \citenamefont {Baer},\ and\ \citenamefont
  {Zgid}}]{Neuhauser2017}%
  \BibitemOpen
  \bibfield  {author} {\bibinfo {author} {\bibfnamefont {D.}~\bibnamefont
  {Neuhauser}}, \bibinfo {author} {\bibfnamefont {R.}~\bibnamefont {Baer}},\
  and\ \bibinfo {author} {\bibfnamefont {D.}~\bibnamefont {Zgid}},\ }\href
  {https://doi.org/10.1021/acs.jctc.7b00792} {\bibfield  {journal} {\bibinfo
  {journal} {Journal of Chemical Theory and Computation}\ }\textbf {\bibinfo
  {volume} {13}},\ \bibinfo {pages} {5396} (\bibinfo {year} {2017})},\ \bibinfo
  {note} {pMID: 28961398}\BibitemShut {NoStop}%
\bibitem [{\citenamefont {Pokhilko}\ \emph {et~al.}(2021)\citenamefont
  {Pokhilko}, \citenamefont {Iskakov}, \citenamefont {Yeh},\ and\ \citenamefont
  {Zgid}}]{Pokhilko2021a}%
  \BibitemOpen
  \bibfield  {author} {\bibinfo {author} {\bibfnamefont {P.}~\bibnamefont
  {Pokhilko}}, \bibinfo {author} {\bibfnamefont {S.}~\bibnamefont {Iskakov}},
  \bibinfo {author} {\bibfnamefont {C.-N.}\ \bibnamefont {Yeh}},\ and\ \bibinfo
  {author} {\bibfnamefont {D.}~\bibnamefont {Zgid}},\ }\href
  {https://doi.org/10.1063/5.0054661} {\bibfield  {journal} {\bibinfo
  {journal} {The Journal of Chemical Physics}\ }\textbf {\bibinfo {volume}
  {155}},\ \bibinfo {pages} {024119} (\bibinfo {year} {2021})}\BibitemShut
  {NoStop}%
\bibitem [{\citenamefont {Pokhilko}\ and\ \citenamefont
  {Zgid}(2021)}]{Pokhilko2021b}%
  \BibitemOpen
  \bibfield  {author} {\bibinfo {author} {\bibfnamefont {P.}~\bibnamefont
  {Pokhilko}}\ and\ \bibinfo {author} {\bibfnamefont {D.}~\bibnamefont
  {Zgid}},\ }\href {https://doi.org/10.1063/5.0055191} {\bibfield  {journal}
  {\bibinfo  {journal} {The Journal of Chemical Physics}\ }\textbf {\bibinfo
  {volume} {155}},\ \bibinfo {pages} {024101} (\bibinfo {year}
  {2021})}\BibitemShut {NoStop}%
\bibitem [{\citenamefont {Hotta}\ and\ \citenamefont
  {Fujimoto}(1996)}]{Hotta1996}%
  \BibitemOpen
  \bibfield  {author} {\bibinfo {author} {\bibfnamefont {T.}~\bibnamefont
  {Hotta}}\ and\ \bibinfo {author} {\bibfnamefont {S.}~\bibnamefont
  {Fujimoto}},\ }\href {https://doi.org/10.1103/PhysRevB.54.5381} {\bibfield
  {journal} {\bibinfo  {journal} {Phys. Rev. B}\ }\textbf {\bibinfo {volume}
  {54}},\ \bibinfo {pages} {5381} (\bibinfo {year} {1996})}\BibitemShut
  {NoStop}%
\bibitem [{\citenamefont {Vilk}\ and\ \citenamefont
  {Tremblay}(1997)}]{Vilk1997}%
  \BibitemOpen
  \bibfield  {author} {\bibinfo {author} {\bibfnamefont {Y.~M.}\ \bibnamefont
  {Vilk}}\ and\ \bibinfo {author} {\bibfnamefont {A.-M.}\ \bibnamefont
  {Tremblay}},\ }\href {https://doi.org/10.1051/jp1:1997135} {\bibfield
  {journal} {\bibinfo  {journal} {J. Phys. I France}\ }\textbf {\bibinfo
  {volume} {7}},\ \bibinfo {pages} {1309} (\bibinfo {year} {1997})}\BibitemShut
  {NoStop}%
\bibitem [{\citenamefont {Rubtsov}\ and\ \citenamefont
  {Lichtenstein}(2004)}]{Rubtsov2004}%
  \BibitemOpen
  \bibfield  {author} {\bibinfo {author} {\bibfnamefont {A.~N.}\ \bibnamefont
  {Rubtsov}}\ and\ \bibinfo {author} {\bibfnamefont {A.~I.}\ \bibnamefont
  {Lichtenstein}},\ }\href {https://doi.org/10.1134/1.1800216} {\bibfield
  {journal} {\bibinfo  {journal} {Journal of Experimental and Theoretical
  Physics Letters}\ }\textbf {\bibinfo {volume} {80}},\ \bibinfo {pages} {61}
  (\bibinfo {year} {2004})}\BibitemShut {NoStop}%
\bibitem [{\citenamefont {Fuchs}(2011)}]{Fuchs2011b}%
  \BibitemOpen
  \bibfield  {author} {\bibinfo {author} {\bibfnamefont {S.}~\bibnamefont
  {Fuchs}},\ }\emph {\bibinfo {title} {Thermodynamic and spectral properties of
  quantum many-particle systems}},\ \href
  {http://inis.iaea.org/search/search.aspx?orig_q=RN:47087476} {Ph.D. thesis},\
  \bibinfo  {school} {Universit\"at G\"ottingen} (\bibinfo {year} {Jan 2011}),\
  \bibinfo {note} {condesned Matter Physics, Superconductivity and
  superfluidity.}\BibitemShut {Stop}%
\bibitem [{\citenamefont {Monkhorst}\ and\ \citenamefont
  {Pack}(1976)}]{MonkhorstPack1976}%
  \BibitemOpen
  \bibfield  {author} {\bibinfo {author} {\bibfnamefont {H.~J.}\ \bibnamefont
  {Monkhorst}}\ and\ \bibinfo {author} {\bibfnamefont {J.~D.}\ \bibnamefont
  {Pack}},\ }\href {https://doi.org/10.1103/PhysRevB.13.5188} {\bibfield
  {journal} {\bibinfo  {journal} {Phys. Rev. B}\ }\textbf {\bibinfo {volume}
  {13}},\ \bibinfo {pages} {5188} (\bibinfo {year} {1976})}\BibitemShut
  {NoStop}%
\bibitem [{\citenamefont {Maier}\ \emph
  {et~al.}(2005{\natexlab{b}})\citenamefont {Maier}, \citenamefont {Jarrell},
  \citenamefont {Schulthess}, \citenamefont {Kent},\ and\ \citenamefont
  {White}}]{Maier2005b}%
  \BibitemOpen
  \bibfield  {author} {\bibinfo {author} {\bibfnamefont {T.~A.}\ \bibnamefont
  {Maier}}, \bibinfo {author} {\bibfnamefont {M.}~\bibnamefont {Jarrell}},
  \bibinfo {author} {\bibfnamefont {T.~C.}\ \bibnamefont {Schulthess}},
  \bibinfo {author} {\bibfnamefont {P.~R.~C.}\ \bibnamefont {Kent}},\ and\
  \bibinfo {author} {\bibfnamefont {J.~B.}\ \bibnamefont {White}},\ }\href
  {https://doi.org/10.1103/PhysRevLett.95.237001} {\bibfield  {journal}
  {\bibinfo  {journal} {Phys. Rev. Lett.}\ }\textbf {\bibinfo {volume} {95}},\
  \bibinfo {pages} {237001} (\bibinfo {year} {2005}{\natexlab{b}})}\BibitemShut
  {NoStop}%
\bibitem [{\citenamefont {Morgan}\ \emph {et~al.}(2018)\citenamefont {Morgan},
  \citenamefont {Jorgensen}, \citenamefont {Hess},\ and\ \citenamefont
  {Hart}}]{Morgan2018}%
  \BibitemOpen
  \bibfield  {author} {\bibinfo {author} {\bibfnamefont {W.~S.}\ \bibnamefont
  {Morgan}}, \bibinfo {author} {\bibfnamefont {J.~J.}\ \bibnamefont
  {Jorgensen}}, \bibinfo {author} {\bibfnamefont {B.~C.}\ \bibnamefont
  {Hess}},\ and\ \bibinfo {author} {\bibfnamefont {G.~L.}\ \bibnamefont
  {Hart}},\ }\href
  {https://doi.org/https://doi.org/10.1016/j.commatsci.2018.06.031} {\bibfield
  {journal} {\bibinfo  {journal} {Computational Materials Science}\ }\textbf
  {\bibinfo {volume} {153}},\ \bibinfo {pages} {424} (\bibinfo {year}
  {2018})}\BibitemShut {NoStop}%
\bibitem [{\citenamefont {Hart}\ \emph {et~al.}(2019)\citenamefont {Hart},
  \citenamefont {Jorgensen}, \citenamefont {Morgan},\ and\ \citenamefont
  {Forcade}}]{Hart2019}%
  \BibitemOpen
  \bibfield  {author} {\bibinfo {author} {\bibfnamefont {G.~L.~W.}\
  \bibnamefont {Hart}}, \bibinfo {author} {\bibfnamefont {J.~J.}\ \bibnamefont
  {Jorgensen}}, \bibinfo {author} {\bibfnamefont {W.~S.}\ \bibnamefont
  {Morgan}},\ and\ \bibinfo {author} {\bibfnamefont {R.~W.}\ \bibnamefont
  {Forcade}},\ }\href {https://doi.org/10.1088/2399-6528/ab2937} {\bibfield
  {journal} {\bibinfo  {journal} {Journal of Physics Communications}\ }\textbf
  {\bibinfo {volume} {3}},\ \bibinfo {pages} {065009} (\bibinfo {year}
  {2019})}\BibitemShut {NoStop}%
\bibitem [{\citenamefont {Jorgensen}\ \emph {et~al.}(2021)\citenamefont
  {Jorgensen}, \citenamefont {Christensen}, \citenamefont {Jarvis},\ and\
  \citenamefont {Hart}}]{Jorgensen2021}%
  \BibitemOpen
  \bibfield  {author} {\bibinfo {author} {\bibfnamefont {J.~J.}\ \bibnamefont
  {Jorgensen}}, \bibinfo {author} {\bibfnamefont {J.~E.}\ \bibnamefont
  {Christensen}}, \bibinfo {author} {\bibfnamefont {T.~J.}\ \bibnamefont
  {Jarvis}},\ and\ \bibinfo {author} {\bibfnamefont {G.~L.~W.}\ \bibnamefont
  {Hart}},\ }\href@noop {} {\bibinfo {title} {A simple, general algorithm for
  calculating the irreducible brillouin zone}} (\bibinfo {year} {2021}),\
  \Eprint {https://arxiv.org/abs/2104.05856} {arXiv:2104.05856
  [cond-mat.mtrl-sci]} \BibitemShut {NoStop}%
\bibitem [{\citenamefont {Lichtenstein}\ and\ \citenamefont
  {Katsnelson}(2000)}]{Lichtenstein2000}%
  \BibitemOpen
  \bibfield  {author} {\bibinfo {author} {\bibfnamefont {A.~I.}\ \bibnamefont
  {Lichtenstein}}\ and\ \bibinfo {author} {\bibfnamefont {M.~I.}\ \bibnamefont
  {Katsnelson}},\ }\href {https://doi.org/10.1103/PhysRevB.62.R9283} {\bibfield
   {journal} {\bibinfo  {journal} {Phys. Rev. B}\ }\textbf {\bibinfo {volume}
  {62}},\ \bibinfo {pages} {R9283} (\bibinfo {year} {2000})}\BibitemShut
  {NoStop}%
\bibitem [{\citenamefont {Kotliar}\ \emph {et~al.}(2001)\citenamefont
  {Kotliar}, \citenamefont {Savrasov}, \citenamefont {P\'alsson},\ and\
  \citenamefont {Biroli}}]{Kotliar2001}%
  \BibitemOpen
  \bibfield  {author} {\bibinfo {author} {\bibfnamefont {G.}~\bibnamefont
  {Kotliar}}, \bibinfo {author} {\bibfnamefont {S.~Y.}\ \bibnamefont
  {Savrasov}}, \bibinfo {author} {\bibfnamefont {G.}~\bibnamefont
  {P\'alsson}},\ and\ \bibinfo {author} {\bibfnamefont {G.}~\bibnamefont
  {Biroli}},\ }\href {https://doi.org/10.1103/PhysRevLett.87.186401} {\bibfield
   {journal} {\bibinfo  {journal} {Phys. Rev. Lett.}\ }\textbf {\bibinfo
  {volume} {87}},\ \bibinfo {pages} {186401} (\bibinfo {year}
  {2001})}\BibitemShut {NoStop}%
\bibitem [{\citenamefont {Biroli}\ and\ \citenamefont
  {Kotliar}(2002)}]{Biroli2002}%
  \BibitemOpen
  \bibfield  {author} {\bibinfo {author} {\bibfnamefont {G.}~\bibnamefont
  {Biroli}}\ and\ \bibinfo {author} {\bibfnamefont {G.}~\bibnamefont
  {Kotliar}},\ }\href {https://doi.org/10.1103/PhysRevB.65.155112} {\bibfield
  {journal} {\bibinfo  {journal} {Phys. Rev. B}\ }\textbf {\bibinfo {volume}
  {65}},\ \bibinfo {pages} {155112} (\bibinfo {year} {2002})}\BibitemShut
  {NoStop}%
\bibitem [{\citenamefont {Aryanpour}\ \emph {et~al.}(2005)\citenamefont
  {Aryanpour}, \citenamefont {Maier},\ and\ \citenamefont
  {Jarrell}}]{Jarrell2005}%
  \BibitemOpen
  \bibfield  {author} {\bibinfo {author} {\bibfnamefont {K.}~\bibnamefont
  {Aryanpour}}, \bibinfo {author} {\bibfnamefont {T.~A.}\ \bibnamefont
  {Maier}},\ and\ \bibinfo {author} {\bibfnamefont {M.}~\bibnamefont
  {Jarrell}},\ }\href {https://doi.org/10.1103/PhysRevB.71.037101} {\bibfield
  {journal} {\bibinfo  {journal} {Phys. Rev. B}\ }\textbf {\bibinfo {volume}
  {71}},\ \bibinfo {pages} {037101} (\bibinfo {year} {2005})}\BibitemShut
  {NoStop}%
\bibitem [{\citenamefont {Biroli}\ and\ \citenamefont
  {Kotliar}(2005)}]{Biroli2005}%
  \BibitemOpen
  \bibfield  {author} {\bibinfo {author} {\bibfnamefont {G.}~\bibnamefont
  {Biroli}}\ and\ \bibinfo {author} {\bibfnamefont {G.}~\bibnamefont
  {Kotliar}},\ }\href {https://doi.org/10.1103/PhysRevB.71.037102} {\bibfield
  {journal} {\bibinfo  {journal} {Phys. Rev. B}\ }\textbf {\bibinfo {volume}
  {71}},\ \bibinfo {pages} {037102} (\bibinfo {year} {2005})}\BibitemShut
  {NoStop}%
\bibitem [{\citenamefont {Biermann}\ \emph {et~al.}(2003)\citenamefont
  {Biermann}, \citenamefont {Aryasetiawan},\ and\ \citenamefont
  {Georges}}]{Biermann2003}%
  \BibitemOpen
  \bibfield  {author} {\bibinfo {author} {\bibfnamefont {S.}~\bibnamefont
  {Biermann}}, \bibinfo {author} {\bibfnamefont {F.}~\bibnamefont
  {Aryasetiawan}},\ and\ \bibinfo {author} {\bibfnamefont {A.}~\bibnamefont
  {Georges}},\ }\href {https://doi.org/10.1103/PhysRevLett.90.086402}
  {\bibfield  {journal} {\bibinfo  {journal} {Phys. Rev. Lett.}\ }\textbf
  {\bibinfo {volume} {90}},\ \bibinfo {pages} {086402} (\bibinfo {year}
  {2003})}\BibitemShut {NoStop}%
\bibitem [{\citenamefont {Fuhrmann}\ \emph {et~al.}(2007)\citenamefont
  {Fuhrmann}, \citenamefont {Okamoto}, \citenamefont {Monien},\ and\
  \citenamefont {Millis}}]{Fuhrmann2007}%
  \BibitemOpen
  \bibfield  {author} {\bibinfo {author} {\bibfnamefont {A.}~\bibnamefont
  {Fuhrmann}}, \bibinfo {author} {\bibfnamefont {S.}~\bibnamefont {Okamoto}},
  \bibinfo {author} {\bibfnamefont {H.}~\bibnamefont {Monien}},\ and\ \bibinfo
  {author} {\bibfnamefont {A.~J.}\ \bibnamefont {Millis}},\ }\href
  {https://doi.org/10.1103/PhysRevB.75.205118} {\bibfield  {journal} {\bibinfo
  {journal} {Phys. Rev. B}\ }\textbf {\bibinfo {volume} {75}},\ \bibinfo
  {pages} {205118} (\bibinfo {year} {2007})}\BibitemShut {NoStop}%
\bibitem [{\citenamefont {Staar}\ \emph {et~al.}(2013)\citenamefont {Staar},
  \citenamefont {Maier},\ and\ \citenamefont {Schulthess}}]{Staar2013}%
  \BibitemOpen
  \bibfield  {author} {\bibinfo {author} {\bibfnamefont {P.}~\bibnamefont
  {Staar}}, \bibinfo {author} {\bibfnamefont {T.}~\bibnamefont {Maier}},\ and\
  \bibinfo {author} {\bibfnamefont {T.~C.}\ \bibnamefont {Schulthess}},\ }\href
  {https://doi.org/10.1103/PhysRevB.88.115101} {\bibfield  {journal} {\bibinfo
  {journal} {Phys. Rev. B}\ }\textbf {\bibinfo {volume} {88}},\ \bibinfo
  {pages} {115101} (\bibinfo {year} {2013})}\BibitemShut {NoStop}%
\bibitem [{\citenamefont {Gull}\ \emph {et~al.}(2013)\citenamefont {Gull},
  \citenamefont {Parcollet},\ and\ \citenamefont {Millis}}]{Gull2013}%
  \BibitemOpen
  \bibfield  {author} {\bibinfo {author} {\bibfnamefont {E.}~\bibnamefont
  {Gull}}, \bibinfo {author} {\bibfnamefont {O.}~\bibnamefont {Parcollet}},\
  and\ \bibinfo {author} {\bibfnamefont {A.~J.}\ \bibnamefont {Millis}},\
  }\href {https://doi.org/10.1103/PhysRevLett.110.216405} {\bibfield  {journal}
  {\bibinfo  {journal} {Phys. Rev. Lett.}\ }\textbf {\bibinfo {volume} {110}},\
  \bibinfo {pages} {216405} (\bibinfo {year} {2013})}\BibitemShut {NoStop}%
\bibitem [{\citenamefont {LeBlanc}\ \emph {et~al.}(2015)\citenamefont
  {LeBlanc}, \citenamefont {Antipov}, \citenamefont {Becca}, \citenamefont
  {Bulik}, \citenamefont {Chan}, \citenamefont {Chung}, \citenamefont {Deng},
  \citenamefont {Ferrero}, \citenamefont {Henderson}, \citenamefont
  {Jim\'enez-Hoyos}, \citenamefont {Kozik}, \citenamefont {Liu}, \citenamefont
  {Millis}, \citenamefont {Prokof'ev}, \citenamefont {Qin}, \citenamefont
  {Scuseria}, \citenamefont {Shi}, \citenamefont {Svistunov}, \citenamefont
  {Tocchio}, \citenamefont {Tupitsyn}, \citenamefont {White}, \citenamefont
  {Zhang}, \citenamefont {Zheng}, \citenamefont {Zhu},\ and\ \citenamefont
  {Gull}}]{LeBlanc2015}%
  \BibitemOpen
  \bibfield  {author} {\bibinfo {author} {\bibfnamefont {J.~P.~F.}\
  \bibnamefont {LeBlanc}}, \bibinfo {author} {\bibfnamefont {A.~E.}\
  \bibnamefont {Antipov}}, \bibinfo {author} {\bibfnamefont {F.}~\bibnamefont
  {Becca}}, \bibinfo {author} {\bibfnamefont {I.~W.}\ \bibnamefont {Bulik}},
  \bibinfo {author} {\bibfnamefont {G.~K.-L.}\ \bibnamefont {Chan}}, \bibinfo
  {author} {\bibfnamefont {C.-M.}\ \bibnamefont {Chung}}, \bibinfo {author}
  {\bibfnamefont {Y.}~\bibnamefont {Deng}}, \bibinfo {author} {\bibfnamefont
  {M.}~\bibnamefont {Ferrero}}, \bibinfo {author} {\bibfnamefont {T.~M.}\
  \bibnamefont {Henderson}}, \bibinfo {author} {\bibfnamefont {C.~A.}\
  \bibnamefont {Jim\'enez-Hoyos}}, \bibinfo {author} {\bibfnamefont
  {E.}~\bibnamefont {Kozik}}, \bibinfo {author} {\bibfnamefont {X.-W.}\
  \bibnamefont {Liu}}, \bibinfo {author} {\bibfnamefont {A.~J.}\ \bibnamefont
  {Millis}}, \bibinfo {author} {\bibfnamefont {N.~V.}\ \bibnamefont
  {Prokof'ev}}, \bibinfo {author} {\bibfnamefont {M.}~\bibnamefont {Qin}},
  \bibinfo {author} {\bibfnamefont {G.~E.}\ \bibnamefont {Scuseria}}, \bibinfo
  {author} {\bibfnamefont {H.}~\bibnamefont {Shi}}, \bibinfo {author}
  {\bibfnamefont {B.~V.}\ \bibnamefont {Svistunov}}, \bibinfo {author}
  {\bibfnamefont {L.~F.}\ \bibnamefont {Tocchio}}, \bibinfo {author}
  {\bibfnamefont {I.~S.}\ \bibnamefont {Tupitsyn}}, \bibinfo {author}
  {\bibfnamefont {S.~R.}\ \bibnamefont {White}}, \bibinfo {author}
  {\bibfnamefont {S.}~\bibnamefont {Zhang}}, \bibinfo {author} {\bibfnamefont
  {B.-X.}\ \bibnamefont {Zheng}}, \bibinfo {author} {\bibfnamefont
  {Z.}~\bibnamefont {Zhu}},\ and\ \bibinfo {author} {\bibfnamefont
  {E.}~\bibnamefont {Gull}} (\bibinfo {collaboration} {Simons Collaboration on
  the Many-Electron Problem}),\ }\href
  {https://doi.org/10.1103/PhysRevX.5.041041} {\bibfield  {journal} {\bibinfo
  {journal} {Phys. Rev. X}\ }\textbf {\bibinfo {volume} {5}},\ \bibinfo {pages}
  {041041} (\bibinfo {year} {2015})}\BibitemShut {NoStop}%
\bibitem [{\citenamefont {Souza}\ \emph {et~al.}(2000)\citenamefont {Souza},
  \citenamefont {Wilkens},\ and\ \citenamefont {Martin}}]{Wilkens2000}%
  \BibitemOpen
  \bibfield  {author} {\bibinfo {author} {\bibfnamefont {I.}~\bibnamefont
  {Souza}}, \bibinfo {author} {\bibfnamefont {T.}~\bibnamefont {Wilkens}},\
  and\ \bibinfo {author} {\bibfnamefont {R.~M.}\ \bibnamefont {Martin}},\
  }\href {https://doi.org/10.1103/PhysRevB.62.1666} {\bibfield  {journal}
  {\bibinfo  {journal} {Phys. Rev. B}\ }\textbf {\bibinfo {volume} {62}},\
  \bibinfo {pages} {1666} (\bibinfo {year} {2000})}\BibitemShut {NoStop}%
\bibitem [{\citenamefont {Lin}\ \emph {et~al.}(2001)\citenamefont {Lin},
  \citenamefont {Zong},\ and\ \citenamefont {Ceperley}}]{Ceperley2001}%
  \BibitemOpen
  \bibfield  {author} {\bibinfo {author} {\bibfnamefont {C.}~\bibnamefont
  {Lin}}, \bibinfo {author} {\bibfnamefont {F.~H.}\ \bibnamefont {Zong}},\ and\
  \bibinfo {author} {\bibfnamefont {D.~M.}\ \bibnamefont {Ceperley}},\ }\href
  {https://doi.org/10.1103/PhysRevE.64.016702} {\bibfield  {journal} {\bibinfo
  {journal} {Phys. Rev. E}\ }\textbf {\bibinfo {volume} {64}},\ \bibinfo
  {pages} {016702} (\bibinfo {year} {2001})}\BibitemShut {NoStop}%
\bibitem [{\citenamefont {Karakuzu}\ \emph {et~al.}(2018)\citenamefont
  {Karakuzu}, \citenamefont {Seki},\ and\ \citenamefont
  {Sorella}}]{Karakuzu2018}%
  \BibitemOpen
  \bibfield  {author} {\bibinfo {author} {\bibfnamefont {S.}~\bibnamefont
  {Karakuzu}}, \bibinfo {author} {\bibfnamefont {K.}~\bibnamefont {Seki}},\
  and\ \bibinfo {author} {\bibfnamefont {S.}~\bibnamefont {Sorella}},\ }\href
  {https://doi.org/10.1103/PhysRevB.98.075156} {\bibfield  {journal} {\bibinfo
  {journal} {Phys. Rev. B}\ }\textbf {\bibinfo {volume} {98}},\ \bibinfo
  {pages} {075156} (\bibinfo {year} {2018})}\BibitemShut {NoStop}%
\bibitem [{\citenamefont {Gulminelli}\ \emph {et~al.}(2011)\citenamefont
  {Gulminelli}, \citenamefont {Furuta}, \citenamefont {Juillet},\ and\
  \citenamefont {Leclercq}}]{Leclercq2011}%
  \BibitemOpen
  \bibfield  {author} {\bibinfo {author} {\bibfnamefont {F.}~\bibnamefont
  {Gulminelli}}, \bibinfo {author} {\bibfnamefont {T.}~\bibnamefont {Furuta}},
  \bibinfo {author} {\bibfnamefont {O.}~\bibnamefont {Juillet}},\ and\ \bibinfo
  {author} {\bibfnamefont {C.}~\bibnamefont {Leclercq}},\ }\href
  {https://doi.org/10.1103/PhysRevC.84.065806} {\bibfield  {journal} {\bibinfo
  {journal} {Phys. Rev. C}\ }\textbf {\bibinfo {volume} {84}},\ \bibinfo
  {pages} {065806} (\bibinfo {year} {2011})}\BibitemShut {NoStop}%
\bibitem [{\citenamefont {Qin}\ \emph {et~al.}(2016)\citenamefont {Qin},
  \citenamefont {Shi},\ and\ \citenamefont {Zhang}}]{Shiwei2016}%
  \BibitemOpen
  \bibfield  {author} {\bibinfo {author} {\bibfnamefont {M.}~\bibnamefont
  {Qin}}, \bibinfo {author} {\bibfnamefont {H.}~\bibnamefont {Shi}},\ and\
  \bibinfo {author} {\bibfnamefont {S.}~\bibnamefont {Zhang}},\ }\href
  {https://doi.org/10.1103/PhysRevB.94.085103} {\bibfield  {journal} {\bibinfo
  {journal} {Phys. Rev. B}\ }\textbf {\bibinfo {volume} {94}},\ \bibinfo
  {pages} {085103} (\bibinfo {year} {2016})}\BibitemShut {NoStop}%
\bibitem [{\citenamefont {Paki}(2019)}]{Paki2019}%
  \BibitemOpen
  \bibfield  {author} {\bibinfo {author} {\bibfnamefont {J.}~\bibnamefont
  {Paki}},\ }\emph {\bibinfo {title} {Quantum Monte Carlo Methods and
  Extensions for the 2D Hubbard Model}},\ \href
  {http://hdl.handle.net/2027.42/151444} {Ph.D. thesis},\ \bibinfo  {school}
  {University of Michigan} (\bibinfo {year} {2019}),\ \bibinfo {note}
  {condensed Matter Physics}\BibitemShut {NoStop}%
\bibitem [{\citenamefont {van Dongen}(1991)}]{Dongen1991}%
  \BibitemOpen
  \bibfield  {author} {\bibinfo {author} {\bibfnamefont {P.~G.~J.}\
  \bibnamefont {van Dongen}},\ }\href
  {https://doi.org/10.1103/PhysRevLett.67.757} {\bibfield  {journal} {\bibinfo
  {journal} {Phys. Rev. Lett.}\ }\textbf {\bibinfo {volume} {67}},\ \bibinfo
  {pages} {757} (\bibinfo {year} {1991})}\BibitemShut {NoStop}%
\bibitem [{\citenamefont {van Dongen}(1994)}]{Dongen1994}%
  \BibitemOpen
  \bibfield  {author} {\bibinfo {author} {\bibfnamefont {P.~G.~J.}\
  \bibnamefont {van Dongen}},\ }\href
  {https://doi.org/10.1103/PhysRevB.50.14016} {\bibfield  {journal} {\bibinfo
  {journal} {Phys. Rev. B}\ }\textbf {\bibinfo {volume} {50}},\ \bibinfo
  {pages} {14016} (\bibinfo {year} {1994})}\BibitemShut {NoStop}%
\bibitem [{\citenamefont {Scalettar}\ \emph {et~al.}(1989)\citenamefont
  {Scalettar}, \citenamefont {Scalapino}, \citenamefont {Sugar},\ and\
  \citenamefont {Toussaint}}]{Scalettar1989}%
  \BibitemOpen
  \bibfield  {author} {\bibinfo {author} {\bibfnamefont {R.~T.}\ \bibnamefont
  {Scalettar}}, \bibinfo {author} {\bibfnamefont {D.~J.}\ \bibnamefont
  {Scalapino}}, \bibinfo {author} {\bibfnamefont {R.~L.}\ \bibnamefont
  {Sugar}},\ and\ \bibinfo {author} {\bibfnamefont {D.}~\bibnamefont
  {Toussaint}},\ }\href {https://doi.org/10.1103/PhysRevB.39.4711} {\bibfield
  {journal} {\bibinfo  {journal} {Phys. Rev. B}\ }\textbf {\bibinfo {volume}
  {39}},\ \bibinfo {pages} {4711} (\bibinfo {year} {1989})}\BibitemShut
  {NoStop}%
\bibitem [{\citenamefont {Dar\'e}\ and\ \citenamefont
  {Albinet}(2000)}]{Dare2000}%
  \BibitemOpen
  \bibfield  {author} {\bibinfo {author} {\bibfnamefont {A.-M.}\ \bibnamefont
  {Dar\'e}}\ and\ \bibinfo {author} {\bibfnamefont {G.}~\bibnamefont
  {Albinet}},\ }\href {https://doi.org/10.1103/PhysRevB.61.4567} {\bibfield
  {journal} {\bibinfo  {journal} {Phys. Rev. B}\ }\textbf {\bibinfo {volume}
  {61}},\ \bibinfo {pages} {4567} (\bibinfo {year} {2000})}\BibitemShut
  {NoStop}%
\bibitem [{\citenamefont {Rohringer}\ \emph {et~al.}(2011)\citenamefont
  {Rohringer}, \citenamefont {Toschi}, \citenamefont {Katanin},\ and\
  \citenamefont {Held}}]{Rohringer2011}%
  \BibitemOpen
  \bibfield  {author} {\bibinfo {author} {\bibfnamefont {G.}~\bibnamefont
  {Rohringer}}, \bibinfo {author} {\bibfnamefont {A.}~\bibnamefont {Toschi}},
  \bibinfo {author} {\bibfnamefont {A.}~\bibnamefont {Katanin}},\ and\ \bibinfo
  {author} {\bibfnamefont {K.}~\bibnamefont {Held}},\ }\href
  {https://doi.org/10.1103/PhysRevLett.107.256402} {\bibfield  {journal}
  {\bibinfo  {journal} {Phys. Rev. Lett.}\ }\textbf {\bibinfo {volume} {107}},\
  \bibinfo {pages} {256402} (\bibinfo {year} {2011})}\BibitemShut {NoStop}%
\bibitem [{\citenamefont {Sch\"afer}\ \emph {et~al.}(2015)\citenamefont
  {Sch\"afer}, \citenamefont {Toschi},\ and\ \citenamefont
  {Tomczak}}]{Schafer2015}%
  \BibitemOpen
  \bibfield  {author} {\bibinfo {author} {\bibfnamefont {T.}~\bibnamefont
  {Sch\"afer}}, \bibinfo {author} {\bibfnamefont {A.}~\bibnamefont {Toschi}},\
  and\ \bibinfo {author} {\bibfnamefont {J.~M.}\ \bibnamefont {Tomczak}},\
  }\href {https://doi.org/10.1103/PhysRevB.91.121107} {\bibfield  {journal}
  {\bibinfo  {journal} {Phys. Rev. B}\ }\textbf {\bibinfo {volume} {91}},\
  \bibinfo {pages} {121107} (\bibinfo {year} {2015})}\BibitemShut {NoStop}%
\bibitem [{\citenamefont {Blankenbecler}\ \emph {et~al.}(1981)\citenamefont
  {Blankenbecler}, \citenamefont {Scalapino},\ and\ \citenamefont
  {Sugar}}]{Blankenbecler1981}%
  \BibitemOpen
  \bibfield  {author} {\bibinfo {author} {\bibfnamefont {R.}~\bibnamefont
  {Blankenbecler}}, \bibinfo {author} {\bibfnamefont {D.~J.}\ \bibnamefont
  {Scalapino}},\ and\ \bibinfo {author} {\bibfnamefont {R.~L.}\ \bibnamefont
  {Sugar}},\ }\href {https://doi.org/10.1103/PhysRevD.24.2278} {\bibfield
  {journal} {\bibinfo  {journal} {Phys. Rev. D}\ }\textbf {\bibinfo {volume}
  {24}},\ \bibinfo {pages} {2278} (\bibinfo {year} {1981})}\BibitemShut
  {NoStop}%
\bibitem [{\citenamefont {Hirsch}\ \emph {et~al.}(1982)\citenamefont {Hirsch},
  \citenamefont {Sugar}, \citenamefont {Scalapino},\ and\ \citenamefont
  {Blankenbecler}}]{Hirsch1982}%
  \BibitemOpen
  \bibfield  {author} {\bibinfo {author} {\bibfnamefont {J.~E.}\ \bibnamefont
  {Hirsch}}, \bibinfo {author} {\bibfnamefont {R.~L.}\ \bibnamefont {Sugar}},
  \bibinfo {author} {\bibfnamefont {D.~J.}\ \bibnamefont {Scalapino}},\ and\
  \bibinfo {author} {\bibfnamefont {R.}~\bibnamefont {Blankenbecler}},\ }\href
  {https://doi.org/10.1103/PhysRevB.26.5033} {\bibfield  {journal} {\bibinfo
  {journal} {Phys. Rev. B}\ }\textbf {\bibinfo {volume} {26}},\ \bibinfo
  {pages} {5033} (\bibinfo {year} {1982})}\BibitemShut {NoStop}%
\bibitem [{\citenamefont {Burovski}\ \emph {et~al.}(2006)\citenamefont
  {Burovski}, \citenamefont {Prokof'ev}, \citenamefont {Svistunov},\ and\
  \citenamefont {Troyer}}]{Burovski2006}%
  \BibitemOpen
  \bibfield  {author} {\bibinfo {author} {\bibfnamefont {E.}~\bibnamefont
  {Burovski}}, \bibinfo {author} {\bibfnamefont {N.}~\bibnamefont {Prokof'ev}},
  \bibinfo {author} {\bibfnamefont {B.}~\bibnamefont {Svistunov}},\ and\
  \bibinfo {author} {\bibfnamefont {M.}~\bibnamefont {Troyer}},\ }\href
  {https://doi.org/10.1088/1367-2630/8/8/153} {\bibfield  {journal} {\bibinfo
  {journal} {New Journal of Physics}\ }\textbf {\bibinfo {volume} {8}},\
  \bibinfo {pages} {153} (\bibinfo {year} {2006})}\BibitemShut {NoStop}%
\bibitem [{\citenamefont {Gull}\ \emph {et~al.}(2007)\citenamefont {Gull},
  \citenamefont {Werner}, \citenamefont {Millis},\ and\ \citenamefont
  {Troyer}}]{Gull2007}%
  \BibitemOpen
  \bibfield  {author} {\bibinfo {author} {\bibfnamefont {E.}~\bibnamefont
  {Gull}}, \bibinfo {author} {\bibfnamefont {P.}~\bibnamefont {Werner}},
  \bibinfo {author} {\bibfnamefont {A.}~\bibnamefont {Millis}},\ and\ \bibinfo
  {author} {\bibfnamefont {M.}~\bibnamefont {Troyer}},\ }\href
  {https://doi.org/10.1103/PhysRevB.76.235123} {\bibfield  {journal} {\bibinfo
  {journal} {Phys. Rev. B}\ }\textbf {\bibinfo {volume} {76}},\ \bibinfo
  {pages} {235123} (\bibinfo {year} {2007})}\BibitemShut {NoStop}%
\bibitem [{\citenamefont {Kondov}\ \emph {et~al.}(2015)\citenamefont {Kondov},
  \citenamefont {McGehee}, \citenamefont {Xu},\ and\ \citenamefont
  {DeMarco}}]{Kondov2015}%
  \BibitemOpen
  \bibfield  {author} {\bibinfo {author} {\bibfnamefont {S.~S.}\ \bibnamefont
  {Kondov}}, \bibinfo {author} {\bibfnamefont {W.~R.}\ \bibnamefont {McGehee}},
  \bibinfo {author} {\bibfnamefont {W.}~\bibnamefont {Xu}},\ and\ \bibinfo
  {author} {\bibfnamefont {B.}~\bibnamefont {DeMarco}},\ }\href
  {https://doi.org/10.1103/PhysRevLett.114.083002} {\bibfield  {journal}
  {\bibinfo  {journal} {Phys. Rev. Lett.}\ }\textbf {\bibinfo {volume} {114}},\
  \bibinfo {pages} {083002} (\bibinfo {year} {2015})}\BibitemShut {NoStop}%
\bibitem [{\citenamefont {Tang}\ \emph {et~al.}(2012)\citenamefont {Tang},
  \citenamefont {Paiva}, \citenamefont {Khatami},\ and\ \citenamefont
  {Rigol}}]{Tang2012}%
  \BibitemOpen
  \bibfield  {author} {\bibinfo {author} {\bibfnamefont {B.}~\bibnamefont
  {Tang}}, \bibinfo {author} {\bibfnamefont {T.}~\bibnamefont {Paiva}},
  \bibinfo {author} {\bibfnamefont {E.}~\bibnamefont {Khatami}},\ and\ \bibinfo
  {author} {\bibfnamefont {M.}~\bibnamefont {Rigol}},\ }\href
  {https://doi.org/10.1103/PhysRevLett.109.205301} {\bibfield  {journal}
  {\bibinfo  {journal} {Phys. Rev. Lett.}\ }\textbf {\bibinfo {volume} {109}},\
  \bibinfo {pages} {205301} (\bibinfo {year} {2012})}\BibitemShut {NoStop}%
\bibitem [{\citenamefont {Paiva}\ \emph {et~al.}(2015)\citenamefont {Paiva},
  \citenamefont {Khatami}, \citenamefont {Yang}, \citenamefont {Rousseau},
  \citenamefont {Jarrell}, \citenamefont {Moreno}, \citenamefont {Hulet},\ and\
  \citenamefont {Scalettar}}]{Paiva2015}%
  \BibitemOpen
  \bibfield  {author} {\bibinfo {author} {\bibfnamefont {T.}~\bibnamefont
  {Paiva}}, \bibinfo {author} {\bibfnamefont {E.}~\bibnamefont {Khatami}},
  \bibinfo {author} {\bibfnamefont {S.}~\bibnamefont {Yang}}, \bibinfo {author}
  {\bibfnamefont {V.}~\bibnamefont {Rousseau}}, \bibinfo {author}
  {\bibfnamefont {M.}~\bibnamefont {Jarrell}}, \bibinfo {author} {\bibfnamefont
  {J.}~\bibnamefont {Moreno}}, \bibinfo {author} {\bibfnamefont {R.~G.}\
  \bibnamefont {Hulet}},\ and\ \bibinfo {author} {\bibfnamefont {R.~T.}\
  \bibnamefont {Scalettar}},\ }\href
  {https://doi.org/10.1103/PhysRevLett.115.240402} {\bibfield  {journal}
  {\bibinfo  {journal} {Phys. Rev. Lett.}\ }\textbf {\bibinfo {volume} {115}},\
  \bibinfo {pages} {240402} (\bibinfo {year} {2015})}\BibitemShut {NoStop}%
\bibitem [{\citenamefont {Ibarra-Garc\'{\i}a-Padilla}\ \emph
  {et~al.}(2020)\citenamefont {Ibarra-Garc\'{\i}a-Padilla}, \citenamefont
  {Mukherjee}, \citenamefont {Hulet}, \citenamefont {Hazzard}, \citenamefont
  {Paiva},\ and\ \citenamefont {Scalettar}}]{Paiva2020}%
  \BibitemOpen
  \bibfield  {author} {\bibinfo {author} {\bibfnamefont {E.}~\bibnamefont
  {Ibarra-Garc\'{\i}a-Padilla}}, \bibinfo {author} {\bibfnamefont
  {R.}~\bibnamefont {Mukherjee}}, \bibinfo {author} {\bibfnamefont {R.~G.}\
  \bibnamefont {Hulet}}, \bibinfo {author} {\bibfnamefont {K.~R.~A.}\
  \bibnamefont {Hazzard}}, \bibinfo {author} {\bibfnamefont {T.}~\bibnamefont
  {Paiva}},\ and\ \bibinfo {author} {\bibfnamefont {R.~T.}\ \bibnamefont
  {Scalettar}},\ }\href {https://doi.org/10.1103/PhysRevA.102.033340}
  {\bibfield  {journal} {\bibinfo  {journal} {Phys. Rev. A}\ }\textbf {\bibinfo
  {volume} {102}},\ \bibinfo {pages} {033340} (\bibinfo {year}
  {2020})}\BibitemShut {NoStop}%
\bibitem [{\citenamefont {Witt}\ \emph {et~al.}(2021)\citenamefont {Witt},
  \citenamefont {van Loon}, \citenamefont {Nomoto}, \citenamefont {Arita},\
  and\ \citenamefont {Wehling}}]{Witt2021}%
  \BibitemOpen
  \bibfield  {author} {\bibinfo {author} {\bibfnamefont {N.}~\bibnamefont
  {Witt}}, \bibinfo {author} {\bibfnamefont {E.~G. C.~P.}\ \bibnamefont {van
  Loon}}, \bibinfo {author} {\bibfnamefont {T.}~\bibnamefont {Nomoto}},
  \bibinfo {author} {\bibfnamefont {R.}~\bibnamefont {Arita}},\ and\ \bibinfo
  {author} {\bibfnamefont {T.~O.}\ \bibnamefont {Wehling}},\ }\href
  {https://doi.org/10.1103/PhysRevB.103.205148} {\bibfield  {journal} {\bibinfo
   {journal} {Phys. Rev. B}\ }\textbf {\bibinfo {volume} {103}},\ \bibinfo
  {pages} {205148} (\bibinfo {year} {2021})}\BibitemShut {NoStop}%
\bibitem [{\citenamefont {Baym}\ and\ \citenamefont
  {Kadanoff}(1961)}]{Baym1961}%
  \BibitemOpen
  \bibfield  {author} {\bibinfo {author} {\bibfnamefont {G.}~\bibnamefont
  {Baym}}\ and\ \bibinfo {author} {\bibfnamefont {L.~P.}\ \bibnamefont
  {Kadanoff}},\ }\href {https://doi.org/10.1103/PhysRev.124.287} {\bibfield
  {journal} {\bibinfo  {journal} {Phys. Rev.}\ }\textbf {\bibinfo {volume}
  {124}},\ \bibinfo {pages} {287} (\bibinfo {year} {1961})}\BibitemShut
  {NoStop}%
\bibitem [{\citenamefont {Baym}(1962)}]{Baym1962}%
  \BibitemOpen
  \bibfield  {author} {\bibinfo {author} {\bibfnamefont {G.}~\bibnamefont
  {Baym}},\ }\href {https://doi.org/10.1103/PhysRev.127.1391} {\bibfield
  {journal} {\bibinfo  {journal} {Phys. Rev.}\ }\textbf {\bibinfo {volume}
  {127}},\ \bibinfo {pages} {1391} (\bibinfo {year} {1962})}\BibitemShut
  {NoStop}%
\bibitem [{\citenamefont {Gukelberger}\ \emph {et~al.}(2015)\citenamefont
  {Gukelberger}, \citenamefont {Huang},\ and\ \citenamefont
  {Werner}}]{Gukelberger2015}%
  \BibitemOpen
  \bibfield  {author} {\bibinfo {author} {\bibfnamefont {J.}~\bibnamefont
  {Gukelberger}}, \bibinfo {author} {\bibfnamefont {L.}~\bibnamefont {Huang}},\
  and\ \bibinfo {author} {\bibfnamefont {P.}~\bibnamefont {Werner}},\ }\href
  {https://doi.org/10.1103/PhysRevB.91.235114} {\bibfield  {journal} {\bibinfo
  {journal} {Phys. Rev. B}\ }\textbf {\bibinfo {volume} {91}},\ \bibinfo
  {pages} {235114} (\bibinfo {year} {2015})}\BibitemShut {NoStop}%
\bibitem [{\citenamefont {Bulut}\ \emph {et~al.}(1993)\citenamefont {Bulut},
  \citenamefont {Scalapino},\ and\ \citenamefont {White}}]{Bulut1993}%
  \BibitemOpen
  \bibfield  {author} {\bibinfo {author} {\bibfnamefont {N.}~\bibnamefont
  {Bulut}}, \bibinfo {author} {\bibfnamefont {D.~J.}\ \bibnamefont
  {Scalapino}},\ and\ \bibinfo {author} {\bibfnamefont {S.~R.}\ \bibnamefont
  {White}},\ }\href {https://doi.org/10.1103/PhysRevB.47.14599} {\bibfield
  {journal} {\bibinfo  {journal} {Phys. Rev. B}\ }\textbf {\bibinfo {volume}
  {47}},\ \bibinfo {pages} {14599} (\bibinfo {year} {1993})}\BibitemShut
  {NoStop}%
\bibitem [{\citenamefont {Bulut}(2002)}]{Bulut2002}%
  \BibitemOpen
  \bibfield  {author} {\bibinfo {author} {\bibfnamefont {N.}~\bibnamefont
  {Bulut}},\ }\href {https://doi.org/10.1080/00018730210155142} {\bibfield
  {journal} {\bibinfo  {journal} {Advances in Physics}\ }\textbf {\bibinfo
  {volume} {51}},\ \bibinfo {pages} {1587} (\bibinfo {year}
  {2002})}\BibitemShut {NoStop}%
\bibitem [{\citenamefont {Brener}\ \emph {et~al.}(2008)\citenamefont {Brener},
  \citenamefont {Hafermann}, \citenamefont {Rubtsov}, \citenamefont
  {Katsnelson},\ and\ \citenamefont {Lichtenstein}}]{Brener2008}%
  \BibitemOpen
  \bibfield  {author} {\bibinfo {author} {\bibfnamefont {S.}~\bibnamefont
  {Brener}}, \bibinfo {author} {\bibfnamefont {H.}~\bibnamefont {Hafermann}},
  \bibinfo {author} {\bibfnamefont {A.~N.}\ \bibnamefont {Rubtsov}}, \bibinfo
  {author} {\bibfnamefont {M.~I.}\ \bibnamefont {Katsnelson}},\ and\ \bibinfo
  {author} {\bibfnamefont {A.~I.}\ \bibnamefont {Lichtenstein}},\ }\href
  {https://doi.org/10.1103/PhysRevB.77.195105} {\bibfield  {journal} {\bibinfo
  {journal} {Phys. Rev. B}\ }\textbf {\bibinfo {volume} {77}},\ \bibinfo
  {pages} {195105} (\bibinfo {year} {2008})}\BibitemShut {NoStop}%
\bibitem [{\citenamefont {Hafermann}\ \emph {et~al.}(2009)\citenamefont
  {Hafermann}, \citenamefont {Li}, \citenamefont {Rubtsov}, \citenamefont
  {Katsnelson}, \citenamefont {Lichtenstein},\ and\ \citenamefont
  {Monien}}]{Hafermann2009}%
  \BibitemOpen
  \bibfield  {author} {\bibinfo {author} {\bibfnamefont {H.}~\bibnamefont
  {Hafermann}}, \bibinfo {author} {\bibfnamefont {G.}~\bibnamefont {Li}},
  \bibinfo {author} {\bibfnamefont {A.~N.}\ \bibnamefont {Rubtsov}}, \bibinfo
  {author} {\bibfnamefont {M.~I.}\ \bibnamefont {Katsnelson}}, \bibinfo
  {author} {\bibfnamefont {A.~I.}\ \bibnamefont {Lichtenstein}},\ and\ \bibinfo
  {author} {\bibfnamefont {H.}~\bibnamefont {Monien}},\ }\href
  {https://doi.org/10.1103/PhysRevLett.102.206401} {\bibfield  {journal}
  {\bibinfo  {journal} {Phys. Rev. Lett.}\ }\textbf {\bibinfo {volume} {102}},\
  \bibinfo {pages} {206401} (\bibinfo {year} {2009})}\BibitemShut {NoStop}%
\bibitem [{\citenamefont {Staar}\ \emph {et~al.}(2014)\citenamefont {Staar},
  \citenamefont {Maier},\ and\ \citenamefont {Schulthess}}]{Staar2014}%
  \BibitemOpen
  \bibfield  {author} {\bibinfo {author} {\bibfnamefont {P.}~\bibnamefont
  {Staar}}, \bibinfo {author} {\bibfnamefont {T.}~\bibnamefont {Maier}},\ and\
  \bibinfo {author} {\bibfnamefont {T.~C.}\ \bibnamefont {Schulthess}},\ }\href
  {https://doi.org/10.1103/PhysRevB.89.195133} {\bibfield  {journal} {\bibinfo
  {journal} {Phys. Rev. B}\ }\textbf {\bibinfo {volume} {89}},\ \bibinfo
  {pages} {195133} (\bibinfo {year} {2014})}\BibitemShut {NoStop}%
\bibitem [{\citenamefont {Bickers}\ and\ \citenamefont
  {Scalapino}(1989)}]{Bickers1989b}%
  \BibitemOpen
  \bibfield  {author} {\bibinfo {author} {\bibfnamefont {N.~E.}\ \bibnamefont
  {Bickers}}\ and\ \bibinfo {author} {\bibfnamefont {D.~J.}\ \bibnamefont
  {Scalapino}},\ }\href
  {https://doi.org/https://doi.org/10.1016/0003-4916(89)90359-X} {\bibfield
  {journal} {\bibinfo  {journal} {Annals of Physics}\ }\textbf {\bibinfo
  {volume} {193}},\ \bibinfo {pages} {206} (\bibinfo {year}
  {1989})}\BibitemShut {NoStop}%
\bibitem [{\citenamefont {Dahm}\ and\ \citenamefont
  {Tewordt}(1995)}]{Dahm1995}%
  \BibitemOpen
  \bibfield  {author} {\bibinfo {author} {\bibfnamefont {T.}~\bibnamefont
  {Dahm}}\ and\ \bibinfo {author} {\bibfnamefont {L.}~\bibnamefont {Tewordt}},\
  }\href {https://doi.org/https://doi.org/10.1016/0921-4534(95)00139-5}
  {\bibfield  {journal} {\bibinfo  {journal} {Physica C: Superconductivity}\
  }\textbf {\bibinfo {volume} {246}},\ \bibinfo {pages} {61} (\bibinfo {year}
  {1995})}\BibitemShut {NoStop}%
\bibitem [{\citenamefont {Dahm}(1997)}]{Dahm1997}%
  \BibitemOpen
  \bibfield  {author} {\bibinfo {author} {\bibfnamefont {T.}~\bibnamefont
  {Dahm}},\ }\href
  {https://doi.org/https://doi.org/10.1016/S0038-1098(96)00650-3} {\bibfield
  {journal} {\bibinfo  {journal} {Solid State Communications}\ }\textbf
  {\bibinfo {volume} {101}},\ \bibinfo {pages} {487} (\bibinfo {year}
  {1997})}\BibitemShut {NoStop}%
\bibitem [{\citenamefont {Rodr\'iguez-N\`u\~nez}\ and\ \citenamefont
  {Schafroth}(1997)}]{Schafroth1997}%
  \BibitemOpen
  \bibfield  {author} {\bibinfo {author} {\bibfnamefont {J.~J.}\ \bibnamefont
  {Rodr\'iguez-N\`u\~nez}}\ and\ \bibinfo {author} {\bibfnamefont
  {S.}~\bibnamefont {Schafroth}},\ }\href
  {https://doi.org/10.1142/S0129183197001016} {\bibfield  {journal} {\bibinfo
  {journal} {International Journal of Modern Physics C}\ }\textbf {\bibinfo
  {volume} {08}},\ \bibinfo {pages} {1145} (\bibinfo {year}
  {1997})}\BibitemShut {NoStop}%
\bibitem [{\citenamefont {Esirgen}\ and\ \citenamefont
  {Bickers}(1998)}]{Bickers1998}%
  \BibitemOpen
  \bibfield  {author} {\bibinfo {author} {\bibfnamefont {G.}~\bibnamefont
  {Esirgen}}\ and\ \bibinfo {author} {\bibfnamefont {N.~E.}\ \bibnamefont
  {Bickers}},\ }\href {https://doi.org/10.1103/PhysRevB.57.5376} {\bibfield
  {journal} {\bibinfo  {journal} {Phys. Rev. B}\ }\textbf {\bibinfo {volume}
  {57}},\ \bibinfo {pages} {5376} (\bibinfo {year} {1998})}\BibitemShut
  {NoStop}%
\bibitem [{\citenamefont {Nakano}\ \emph {et~al.}(2007)\citenamefont {Nakano},
  \citenamefont {Kuroki},\ and\ \citenamefont {Onari}}]{Tsuguhito2007}%
  \BibitemOpen
  \bibfield  {author} {\bibinfo {author} {\bibfnamefont {T.}~\bibnamefont
  {Nakano}}, \bibinfo {author} {\bibfnamefont {K.}~\bibnamefont {Kuroki}},\
  and\ \bibinfo {author} {\bibfnamefont {S.}~\bibnamefont {Onari}},\ }\href
  {https://doi.org/10.1103/PhysRevB.76.014515} {\bibfield  {journal} {\bibinfo
  {journal} {Phys. Rev. B}\ }\textbf {\bibinfo {volume} {76}},\ \bibinfo
  {pages} {014515} (\bibinfo {year} {2007})}\BibitemShut {NoStop}%
\bibitem [{\citenamefont {Kondo}\ and\ \citenamefont
  {Moriya}(1998)}]{Hisashi1998}%
  \BibitemOpen
  \bibfield  {author} {\bibinfo {author} {\bibfnamefont {H.}~\bibnamefont
  {Kondo}}\ and\ \bibinfo {author} {\bibfnamefont {T.}~\bibnamefont {Moriya}},\
  }\href {https://doi.org/10.1143/JPSJ.67.3695} {\bibfield  {journal} {\bibinfo
   {journal} {Journal of the Physical Society of Japan}\ }\textbf {\bibinfo
  {volume} {67}},\ \bibinfo {pages} {3695} (\bibinfo {year}
  {1998})}\BibitemShut {NoStop}%
\bibitem [{\citenamefont {Kino}\ and\ \citenamefont
  {Kontani}(1998)}]{Hiori1998}%
  \BibitemOpen
  \bibfield  {author} {\bibinfo {author} {\bibfnamefont {H.}~\bibnamefont
  {Kino}}\ and\ \bibinfo {author} {\bibfnamefont {H.}~\bibnamefont {Kontani}},\
  }\href {https://doi.org/10.1143/JPSJ.67.3691} {\bibfield  {journal} {\bibinfo
   {journal} {Journal of the Physical Society of Japan}\ }\textbf {\bibinfo
  {volume} {67}},\ \bibinfo {pages} {3691} (\bibinfo {year}
  {1998})}\BibitemShut {NoStop}%
\bibitem [{\citenamefont {Schmalian}(1998)}]{Schmalian1998}%
  \BibitemOpen
  \bibfield  {author} {\bibinfo {author} {\bibfnamefont {J.}~\bibnamefont
  {Schmalian}},\ }\href {https://doi.org/10.1103/PhysRevLett.81.4232}
  {\bibfield  {journal} {\bibinfo  {journal} {Phys. Rev. Lett.}\ }\textbf
  {\bibinfo {volume} {81}},\ \bibinfo {pages} {4232} (\bibinfo {year}
  {1998})}\BibitemShut {NoStop}%
\bibitem [{\citenamefont {Kontani}\ and\ \citenamefont
  {Ueda}(1998)}]{Kontani1998}%
  \BibitemOpen
  \bibfield  {author} {\bibinfo {author} {\bibfnamefont {H.}~\bibnamefont
  {Kontani}}\ and\ \bibinfo {author} {\bibfnamefont {K.}~\bibnamefont {Ueda}},\
  }\href {https://doi.org/10.1103/PhysRevLett.80.5619} {\bibfield  {journal}
  {\bibinfo  {journal} {Phys. Rev. Lett.}\ }\textbf {\bibinfo {volume} {80}},\
  \bibinfo {pages} {5619} (\bibinfo {year} {1998})}\BibitemShut {NoStop}%
\bibitem [{\citenamefont {Takimoto}\ \emph {et~al.}(2004)\citenamefont
  {Takimoto}, \citenamefont {Hotta},\ and\ \citenamefont {Ueda}}]{Tetsuya2004}%
  \BibitemOpen
  \bibfield  {author} {\bibinfo {author} {\bibfnamefont {T.}~\bibnamefont
  {Takimoto}}, \bibinfo {author} {\bibfnamefont {T.}~\bibnamefont {Hotta}},\
  and\ \bibinfo {author} {\bibfnamefont {K.}~\bibnamefont {Ueda}},\ }\href
  {https://doi.org/10.1103/PhysRevB.69.104504} {\bibfield  {journal} {\bibinfo
  {journal} {Phys. Rev. B}\ }\textbf {\bibinfo {volume} {69}},\ \bibinfo
  {pages} {104504} (\bibinfo {year} {2004})}\BibitemShut {NoStop}%
\bibitem [{\citenamefont {Kubo}(2007)}]{Katsunori2007}%
  \BibitemOpen
  \bibfield  {author} {\bibinfo {author} {\bibfnamefont {K.}~\bibnamefont
  {Kubo}},\ }\href {https://doi.org/10.1103/PhysRevB.75.224509} {\bibfield
  {journal} {\bibinfo  {journal} {Phys. Rev. B}\ }\textbf {\bibinfo {volume}
  {75}},\ \bibinfo {pages} {224509} (\bibinfo {year} {2007})}\BibitemShut
  {NoStop}%
\bibitem [{\citenamefont {Arita}\ \emph {et~al.}(2000)\citenamefont {Arita},
  \citenamefont {Kuroki},\ and\ \citenamefont {Aoki}}]{Ryotaro2000}%
  \BibitemOpen
  \bibfield  {author} {\bibinfo {author} {\bibfnamefont {R.}~\bibnamefont
  {Arita}}, \bibinfo {author} {\bibfnamefont {K.}~\bibnamefont {Kuroki}},\ and\
  \bibinfo {author} {\bibfnamefont {H.}~\bibnamefont {Aoki}},\ }\href
  {https://doi.org/10.1143/JPSJ.69.1181} {\bibfield  {journal} {\bibinfo
  {journal} {Journal of the Physical Society of Japan}\ }\textbf {\bibinfo
  {volume} {69}},\ \bibinfo {pages} {1181} (\bibinfo {year}
  {2000})}\BibitemShut {NoStop}%
\bibitem [{\citenamefont {Link}\ \emph {et~al.}(2019)\citenamefont {Link},
  \citenamefont {Forti}, \citenamefont {St\"ohr}, \citenamefont {K\"uster},
  \citenamefont {R\"osner}, \citenamefont {Hirschmeier}, \citenamefont {Chen},
  \citenamefont {Avila}, \citenamefont {Asensio}, \citenamefont {Zakharov},
  \citenamefont {Wehling}, \citenamefont {Lichtenstein}, \citenamefont
  {Katsnelson},\ and\ \citenamefont {Starke}}]{Link2019}%
  \BibitemOpen
  \bibfield  {author} {\bibinfo {author} {\bibfnamefont {S.}~\bibnamefont
  {Link}}, \bibinfo {author} {\bibfnamefont {S.}~\bibnamefont {Forti}},
  \bibinfo {author} {\bibfnamefont {A.}~\bibnamefont {St\"ohr}}, \bibinfo
  {author} {\bibfnamefont {K.}~\bibnamefont {K\"uster}}, \bibinfo {author}
  {\bibfnamefont {M.}~\bibnamefont {R\"osner}}, \bibinfo {author}
  {\bibfnamefont {D.}~\bibnamefont {Hirschmeier}}, \bibinfo {author}
  {\bibfnamefont {C.}~\bibnamefont {Chen}}, \bibinfo {author} {\bibfnamefont
  {J.}~\bibnamefont {Avila}}, \bibinfo {author} {\bibfnamefont {M.~C.}\
  \bibnamefont {Asensio}}, \bibinfo {author} {\bibfnamefont {A.~A.}\
  \bibnamefont {Zakharov}}, \bibinfo {author} {\bibfnamefont {T.~O.}\
  \bibnamefont {Wehling}}, \bibinfo {author} {\bibfnamefont {A.~I.}\
  \bibnamefont {Lichtenstein}}, \bibinfo {author} {\bibfnamefont {M.~I.}\
  \bibnamefont {Katsnelson}},\ and\ \bibinfo {author} {\bibfnamefont
  {U.}~\bibnamefont {Starke}},\ }\href
  {https://doi.org/10.1103/PhysRevB.100.121407} {\bibfield  {journal} {\bibinfo
   {journal} {Phys. Rev. B}\ }\textbf {\bibinfo {volume} {100}},\ \bibinfo
  {pages} {121407} (\bibinfo {year} {2019})}\BibitemShut {NoStop}%
\bibitem [{\citenamefont {Katsnelson}\ and\ \citenamefont
  {Lichtenstein}(1999)}]{Katsnelson1999}%
  \BibitemOpen
  \bibfield  {author} {\bibinfo {author} {\bibfnamefont {M.~I.}\ \bibnamefont
  {Katsnelson}}\ and\ \bibinfo {author} {\bibfnamefont {A.~I.}\ \bibnamefont
  {Lichtenstein}},\ }\href {https://doi.org/10.1088/0953-8984/11/4/011}
  {\bibfield  {journal} {\bibinfo  {journal} {Journal of Physics: Condensed
  Matter}\ }\textbf {\bibinfo {volume} {11}},\ \bibinfo {pages} {1037}
  (\bibinfo {year} {1999})}\BibitemShut {NoStop}%
\bibitem [{\citenamefont {Aryanpour}\ \emph {et~al.}(2003)\citenamefont
  {Aryanpour}, \citenamefont {Hettler},\ and\ \citenamefont
  {Jarrell}}]{Jarrel2003}%
  \BibitemOpen
  \bibfield  {author} {\bibinfo {author} {\bibfnamefont {K.}~\bibnamefont
  {Aryanpour}}, \bibinfo {author} {\bibfnamefont {M.~H.}\ \bibnamefont
  {Hettler}},\ and\ \bibinfo {author} {\bibfnamefont {M.}~\bibnamefont
  {Jarrell}},\ }\href {https://doi.org/10.1103/PhysRevB.67.085101} {\bibfield
  {journal} {\bibinfo  {journal} {Phys. Rev. B}\ }\textbf {\bibinfo {volume}
  {67}},\ \bibinfo {pages} {085101} (\bibinfo {year} {2003})}\BibitemShut
  {NoStop}%
\bibitem [{\citenamefont {Abrikosov}\ \emph {et~al.}(1975)\citenamefont
  {Abrikosov}, \citenamefont {Dzyaloshinskii}, \citenamefont {Gorkov},\ and\
  \citenamefont {Silverman}}]{Abrikosov1975}%
  \BibitemOpen
  \bibfield  {author} {\bibinfo {author} {\bibfnamefont {A.~A.}\ \bibnamefont
  {Abrikosov}}, \bibinfo {author} {\bibfnamefont {I.}~\bibnamefont
  {Dzyaloshinskii}}, \bibinfo {author} {\bibfnamefont {L.~P.}\ \bibnamefont
  {Gorkov}},\ and\ \bibinfo {author} {\bibfnamefont {R.~A.}\ \bibnamefont
  {Silverman}},\ }\href {https://cds.cern.ch/record/107441} {\emph {\bibinfo
  {title} {{Methods of quantum field theory in statistical physics}}}}\
  (\bibinfo  {publisher} {Dover},\ \bibinfo {address} {New York, NY},\ \bibinfo
  {year} {1975})\BibitemShut {NoStop}%
\bibitem [{\citenamefont {Jani\ifmmode~\check{s}\else
  \v{s}\fi{}}(1999)}]{Janis1999}%
  \BibitemOpen
  \bibfield  {author} {\bibinfo {author} {\bibfnamefont {V.}~\bibnamefont
  {Jani\ifmmode~\check{s}\else \v{s}\fi{}}},\ }\href
  {https://doi.org/10.1103/PhysRevB.60.11345} {\bibfield  {journal} {\bibinfo
  {journal} {Phys. Rev. B}\ }\textbf {\bibinfo {volume} {60}},\ \bibinfo
  {pages} {11345} (\bibinfo {year} {1999})}\BibitemShut {NoStop}%
\end{thebibliography}%

    \appendix

    \section{Particle-hole ladder, particle-particle ladder and FLEX}\label{sec:ladder-and-flex}
    \begin{figure}[h]
        \centering
        \includegraphics[width=0.3\textwidth]{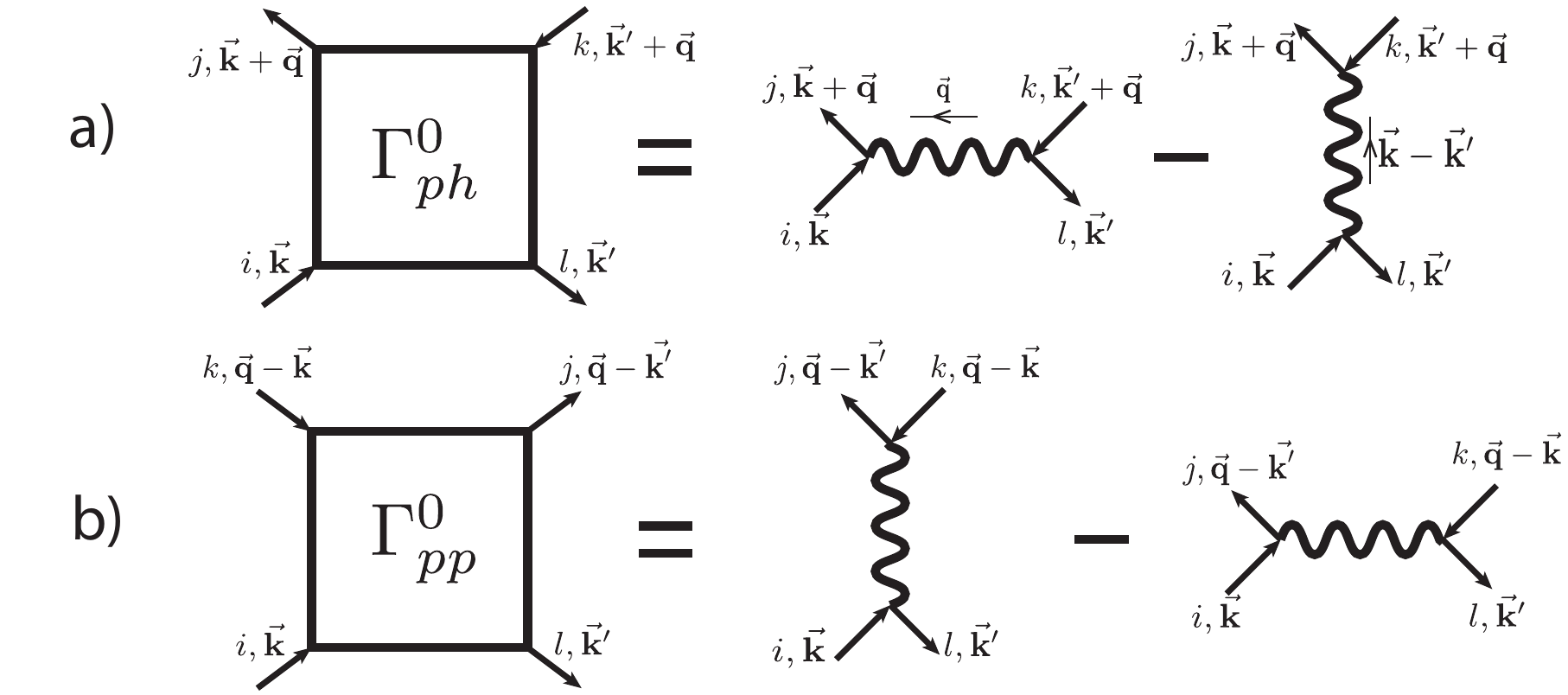}
        \caption{Antisymmetrized bare interaction in particle-hole channel (top) and in particle-particle-channel (bottom).}
        \label{fig:antisym}
    \end{figure}

    In this section we describe the equations used for solving the general spin-dependent FLEX and Ladder approximations. The
    normal state FLEX formalism was originally introduced in Refs.~\cite{Bickers1989a,Bickers1989b}. It has later been
    extended to general lattice Hamiltonians~\cite{Bickers1997}. This approach has been extensively used to study effect
    of spin and charge fluctuations on superconducting state in cuprates~\cite{Dahm1995,Dahm1997,Schafroth1997,Bickers1998,Tsuguhito2007},
    organic superconductors~\cite{Hisashi1998,Hiori1998,Schmalian1998} and spin lattice systems~\cite{Kontani1998}, due to its
    ability to treat these fluctuations on equal footing.

    Application of FLEX to multi-orbital systems is usually limited to few orbitals~\cite{Bickers1997,Tetsuya2004,Katsunori2007} due to
    its computational complexity. Often, diagrams in the particle-particle channel are
    neglected~\cite{Schmalian1998,Ryotaro2000,Link2019,Witt2021}. This additional approximation corresponds to ph-ladder in our
    notation.

    To study the ordered phase one has to consider spin-dependence of the interaction as was proposed in
    Ref.~\cite{Katsnelson1999} for a local multi-orbital problem. The extension to lattices with general non-local interactions is
    straight-forward and will be briefly discussed below.

    Ref.~\cite{Katsnelson1999} proposed to work with an interaction diagonalized in spin space. We find that this approach
    is not necessary as the susceptibilities in the ph- and pp-channels remain non-diagonal. Instead, we utilize the linear
    independence of different components in the interaction tensor. Both approaches are mathematically equivalent.

    The self-energies have the following form:
    \begin{align}
        \Sigma_{\tK_1,i_1 i_2,\sigma_1}(\omega_n) &= \Sigma^{(2)}_{\tK_1,i_1 i_2,\sigma_1}(\omega_n) + \nonumber \\
        \zeta\tilde{\Sigma}^{(ph)}_{\tK_1,i_1 i_2,\sigma_1}&(\omega_n) + \xi\tilde{\Sigma}^{(pp)}_{\tK_1,i_1 i_2,\sigma_1}(\omega_n).
    \end{align}
    Here $\Sigma^{(2)}$ is the second order self-energy (Eq.~\ref{eqn:sigma_gf2}), $\tilde{\Sigma}^{(ph)}$ and $\tilde{\Sigma}^{(pp)}$
    are higher order contribution from particle-hole and particle-particle ladders respectively.

    For FLEX both $\zeta = 1$ and
    $\xi = 1$, for ph-ladder $\zeta = 1$ and $\xi = 0$, and for pp-ladder $\zeta = 0$ and $\xi = 1$.

    The particle-hole ladder contribution to the self-energy is
    \begin{align}
        \tilde{\Sigma}^{(ph)}_{\tK_1,i_1 i_2,\sigma_1}(\omega_n) &= \nonumber \\
        \frac{1}{\beta N_c^2}\sum_{\substack{\mathbb{345}\\\mathbb{678}\\m}}\Gamma^{0,(ph)}_{\mathbb{1456}}&
        (\chi^{(ph)}_{\mathbb{6587}}(\Omega_m) - \chi^{0,(ph)}_{\mathbb{6587}}(\Omega_m))
        \Gamma^{0,(ph)}_{\mathbb{7832}}  \nonumber \\
        G_{\tK_1+\tQ,i_3 i_4,\sigma_3} &(\omega_{n+m}) \delta_{\sigma_1,\sigma_2}\delta_{\sigma_3,\sigma_4}
        \delta_{\tK_1 + \tK_5, \tK_4 + \tK_6} \\
        \delta_{\tK_7 + \tK_3, \tK_8 + \tK_2} \nonumber.
    \end{align}
    Here the indices $\mathbb{N} = (\tK_N, i_N, \sigma_N)$ are combined momentum,orbital and spin indices. The bosonic momentum
    transfer is $\tQ = \tK_4 - \tK_1$. $\Gamma^{0,(ph)}$ is the particle-hole anti-symmetrized bare Coulomb interaction, as
    shown in the top panel of Fig.~\ref{fig:antisym}. The unconnected part of the susceptibility is
    \begin{align}
        \chi^{0,(ph)}_{\mathbb{1234}}(\Omega_m) &= -\frac{1}{\beta}\sum_{n}
        G_{\tK_1,i_1 i_4,\sigma_1} (\omega_{n}) G_{\tK_2,i_3 i_2,\sigma_2} (\omega_{n+m}) \nonumber \\
        &\delta_{\sigma_1,\sigma_4}\delta_{\sigma_2,\sigma_3} \delta_{\tK_1,\tK_4} \delta_{\tK_2,\tK_3}.
    \end{align}
    The dressed particle-hole susceptibility is evaluated via solution of the Bethe-Salpeter equation in the ph-channel:
    \begin{align}
        \pmb{\chi}^{(ph)}(\Omega_m) &= \nonumber \\
              (\mathbb{1} - &\pmb{\chi}^{0,(ph)}(\Omega_m) \pmb{\Gamma}^{0,(ph)})^{-1}
        \pmb{\chi}^{0,(ph)}(\Omega_m),
        \label{eqn:phbethe}
    \end{align}
    where bold symbols $\pmb{\chi}^{(ph)}$, $\pmb{\chi}^{0,(ph)}$ and $\pmb{\Gamma}^{0,(ph)}$ correspond to matrix representations
    of the unconnected part of the susceptibility, dressed susceptibility and anti-symmetrized bare interaction respectively.

    Similarly, the particle-particle contribution
    \begin{align}
        \tilde{\Sigma}^{(pp)}_{\tK_1,i_1 i_2,\sigma_1}(\omega_n) &= \nonumber \\
        \frac{1}{\beta N_c^2}\sum_{\substack{\mathbb{345}\\\mathbb{678}\\m}}\Gamma^{0,(pp)}_{\mathbb{1456}}&
        (\psi^{(pp)}_{\mathbb{6587}}(\Omega_m) - \psi^{0,(pp)}_{\mathbb{6587}}(\Omega_m))
        \Gamma^{0,(pp)}_{\mathbb{7832}}  \nonumber \\
        G_{\tQ - \tK_1,i_3 i_4,\sigma_3} &(\omega_{m-n}) \delta_{\sigma_1,\sigma_2}\delta_{\sigma_3,\sigma_4}
        \delta_{\tK_1 + \tK_5, \tK_4 + \tK_6} \\
        \delta_{\tK_7 + \tK_3, \tK_8 + \tK_2} \nonumber.
    \end{align}
    The particle-particle bosonic momentum transfer is $\tQ = \tK_4 + \tK_1$. $\Gamma^{0,(pp)}$ is the particle-particle
    anti-symmetrized bare Coulomb interaction, as shown in bottom panel of Fig.~\ref{fig:antisym}. The bare particle-particle
    susceptibility is
    \begin{align}
        \psi^{0,(pp)}_{\mathbb{1234}}(\Omega_m) &= \frac{1}{2\beta}\sum_{n}
        G_{\tK_1,i_1 i_4,\sigma_1} (\omega_{n}) G_{\tK_2,i_2 i_3,\sigma_2} (\omega_{m-n}) \nonumber \\
        &\delta_{\sigma_1,\sigma_4}\delta_{\sigma_2,\sigma_3} \delta_{\tK_1,\tK_4} \delta_{\tK_2,\tK_3},
    \end{align}
    and the susceptibility in the pp-channel is
    \begin{align}
        \pmb{\psi}^{(pp)}(\Omega_m) &= \nonumber \\
        (\mathbb{1} - &\pmb{\psi}^{0,(pp)}(\Omega_m) \pmb{\Gamma}^{0,(pp)})^{-1}
        \pmb{\psi}^{0,(pp)}(\Omega_m).
        \label{eqn:ppbethe}
    \end{align}

    The computational complexity of Eq.~\ref{eqn:phbethe} and Eq.~\ref{eqn:ppbethe} is $O(N^6)$, where $N = n_{orb} N_c$
    is the dimension of the combined index $\mathbb{N}$. To reduce the numerical complexity of these equations it is
    advantageous to use conservation laws and symmetries. For example, due to momentum conservation these equations are
    diagonal in the bosonic momentum transfer $\tQ$ leading to overall complexity $O(n_{orb}^6 N_c^4 )$. Another important
    property is spin conservation. In the ph-channel it leads to the following structure of the two-particle quantities in spin
    space:
    \begin{align}
        X^{(ph)} =
        \left(
        \begin{matrix}
            X^{(ph)}_{\uparrow\uparrow\uparrow\uparrow} & 0 & 0 & X^{(ph)}_{\uparrow\uparrow\downarrow\downarrow}\\
            0 & X^{(ph)}_{\uparrow\downarrow\downarrow\uparrow} & 0 & 0   \\
            0 & 0 & X^{(ph)}_{\downarrow\uparrow\uparrow\downarrow} & 0   \\
            X^{(ph)}_{\downarrow\downarrow\uparrow\uparrow} & 0 & 0 & X^{(ph)}_{\downarrow\downarrow\downarrow\downarrow}
        \end{matrix}
        \right),
    \end{align}
    where $X$ is either the anti-symmetrized interaction or the susceptibility. One can see that particle-hole two-particle
    quantities in a spin space consist of three linearly independent blocks, namely longitudinal ($\parallel$), and two transverse
    $S^{\pm}$ ($\pm$) and $S^{\mp}$ ($\mp$) blocks:
    \begin{subequations}
    \begin{align}
        X^{\parallel,(ph)} &=
        \left(
        \begin{matrix}
            X^{(ph)}_{\uparrow\uparrow\uparrow\uparrow} & X^{(ph)}_{\uparrow\uparrow\downarrow\downarrow}\\
            X^{(ph)}_{\downarrow\downarrow\uparrow\uparrow} & X^{(ph)}_{\downarrow\downarrow\downarrow\downarrow}
        \end{matrix}
        \right); \label{eqn:phdm} \\
        X^{\pm,(ph)} &=
        \left(
        \begin{matrix}
            X^{(ph)}_{\uparrow\downarrow\downarrow\uparrow}
        \end{matrix}
        \right); \label{eqn:phpm} \\
        X^{\mp,(ph)} &=
        \left(
        \begin{matrix}
            X^{(ph)}_{\downarrow\uparrow\uparrow\downarrow}
        \end{matrix}
        \right). \label{eqn:phmp}
    \end{align}
    \end{subequations}
    Thus, Eq.~\ref{eqn:phbethe} can be split into set of the following linearly independent equations:
    \begin{align}
        \chi^{\gamma,(ph)}_{\tK_1 i_1 i_2, \tK_2 i_3 i_4}(\tQ,\Omega_m) &= \nonumber \\
        (\mathbb{1} - \left[\pmb{\chi}^{0,\gamma,(ph)}\Gamma^{0,\gamma,(ph)}\right]&(\tQ,\Omega_m))^{-1}_{\tK_1 i_1 i_2, \tK_3
        i_5 i_6}
        \times \nonumber \\
        \chi^{0,\gamma,(ph)}&_{\tK_3 i_7 i_8, \tK_2 i_3 i_4}(\tQ,\Omega_m),
    \end{align}
    where $\gamma \in (\parallel, \pm, \mp)$ is one of three spin channels.

    Similarly, in the particle-particle channel one can separate two-particle quantities into three spin blocks, longitudinal
    ($\parallel$), triplet up-spin ($t_{+1}$), triplet down-spin ($t_{-1}$):
    \begin{subequations}
        \begin{align}
            X^{\parallel,(pp)} &=
            \left(
            \begin{matrix}
                X^{(pp)}_{\uparrow\downarrow\uparrow\downarrow} & X^{(pp)}_{\uparrow\downarrow\downarrow\uparrow}\\
                X^{(pp)}_{\downarrow\uparrow\uparrow\downarrow} & X^{(pp)}_{\downarrow\uparrow\downarrow\uparrow}
            \end{matrix}
            \right); \label{eqn:ppst} \\
            X^{t_{+1},(pp)} &=
            \left(
            \begin{matrix}
                X^{(pp)}_{\uparrow\uparrow\uparrow\uparrow}
            \end{matrix}
            \right); \label{eqn:pptp} \\
            X^{t_{-1},(pp)} &=
            \left(
            \begin{matrix}
                X^{(pp)}_{\downarrow\downarrow\downarrow\downarrow}
            \end{matrix}
            \right). \label{eqn:pptm}
        \end{align}
    \end{subequations}
    Eq.~\ref{eqn:ppbethe}, similarly, can be simplified:
    \begin{align}
        \psi^{\gamma,(pp)}_{\tK_1 i_1 i_2, \tK_2 i_3 i_4}(\tQ,\Omega_m) &= \nonumber \\
        (\mathbb{1} - \left[\pmb{\psi}^{0,\gamma,(pp)}\Gamma^{0,\gamma,(pp)}\right]&(\tQ,\Omega_m))^{-1}_{\tK_1 i_1 i_2, \tK_3
        i_5 i_6}
        \times \nonumber \\
        \psi^{0,\gamma,(pp)}&_{\tK_3 i_7 i_8, \tK_2 i_3 i_4}(\tQ,\Omega_m),
    \end{align}
    with spin channel $\gamma \in (\parallel, t_{+1}, t_{-1})$.

    In Ref.~\cite{Katsnelson1999} the interactions are diagonal matrices in spin space, but the longitudinal susceptibilities
    are expressed as a linear combination of different spin components. For instance, the longitudinal bare particle-hole
    susceptibility that corresponds to Eq.~\ref{eqn:phdm}, in the notation of Ref.~\cite{Katsnelson1999}, is:
    \begin{subequations}
    \begin{align}
        \chi^{\parallel,0,(ph)} &=
        \left(
        \begin{matrix}
            \chi^{0,(ph)}_{\uparrow\uparrow\uparrow\uparrow} + \chi^{0,(ph)}_{\downarrow\downarrow\downarrow\downarrow} &
            \chi^{0,(ph)}_{\uparrow\uparrow\uparrow\uparrow} - \chi^{0,(ph)}_{\downarrow\downarrow\downarrow\downarrow} \\
            \chi^{0,(ph)}_{\uparrow\uparrow\uparrow\uparrow} - \chi^{0,(ph)}_{\downarrow\downarrow\downarrow\downarrow} &
            \chi^{0,(ph)}_{\uparrow\uparrow\uparrow\uparrow} + \chi^{0,(ph)}_{\downarrow\downarrow\downarrow\downarrow}
        \end{matrix}
        \right).
    \end{align}
    \end{subequations}
    The two formalisms are one-to-one equivalent.

\begin{figure}[tbh]
        \centering
        \includegraphics[width=0.3\textwidth]{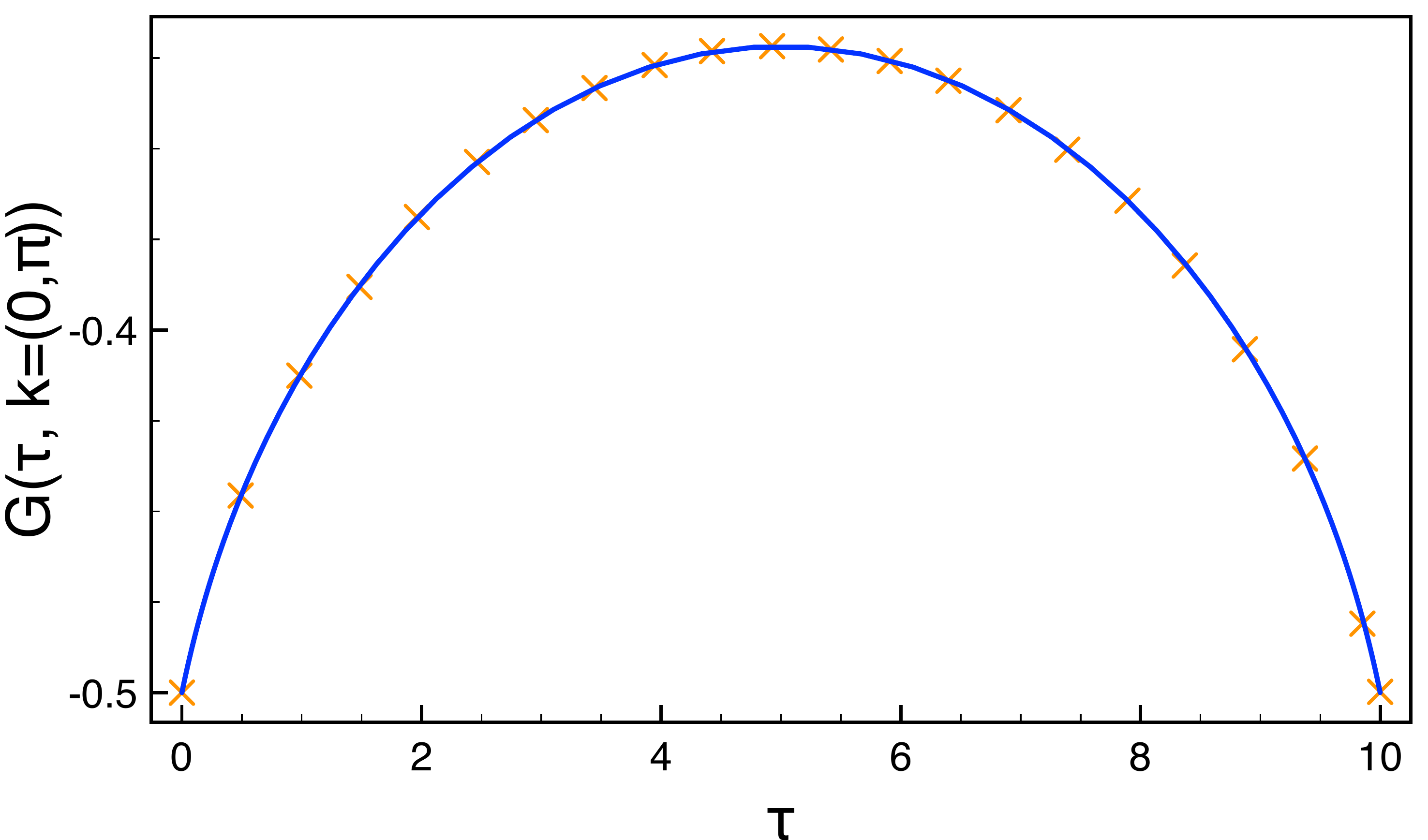}
        \caption{Self-consistent FLEX results for 2D Hubbard model at $U=1.57t$ and $T=0.1t$ in comparison to results
        from Ref.~\cite{Jarrel2003}.}
        \label{fig:jarrel}
    \end{figure}

    Results of FLEX agree with published work~\cite{Jarrel2003}, see Fig.~\ref{fig:jarrel}.

    \section{Causality violations}\label{subsec:causality-properties}
\begin{figure}[h]
        \centering
        \includegraphics[width=0.3\textwidth]{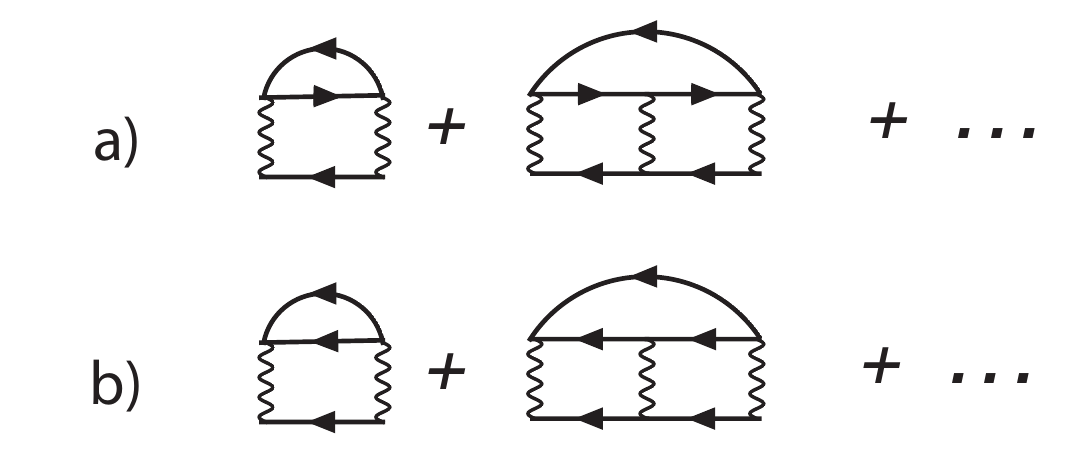}
        \caption{Particle-hole (top) and particle-particle (bottom) scattering diagrams, contained in FLEX.
        Top panel corresponds to Eqs.~\ref{eqn:phscatter}.}
        \label{fig:scatter}
    \end{figure}

    We find regimes where results become noncausal for all three infinite anti-symmetrized diagrammatic series.
    This behavior can be attributed to the presence of sign-alternating
    particle-hole (Fig.~\ref{fig:scatter} top) and particle-particle (Fig.~\ref{fig:scatter} bottom) scattering diagrams.

    We analyze this behavior at the example of the half-filled Hubbard atom. The Hamiltonian
    \begin{align}
        H = U (n_{\uparrow} - \frac{1}{2})(n_{\downarrow} - \frac{1}{2})
    \end{align}
    is defined in such a way that chemical potential $\mu=0$ corresponds to half-filled case.
    Starting from the non-interacting Green's function
    \begin{align}
        G_{0}(\omega_n) = \frac{1}{i\omega_n + \mu},
    \end{align}
    the unconnected susceptibility is $\chi^{0,(ph)}(\Omega_m) = \frac{\beta}{4}\delta_{m,0}$. The second and third order
    ph-scattering contributions to the self-energy (see Fig.~\ref{fig:scatter}a) are~\cite{Abrikosov1975}
    \begin{subequations}
    \begin{align}
        \Sigma^{2}(\omega_n) = \frac{U^2}{4} \frac{1}{i\omega_n}, \label{eqn:2ndph} \\
        \Sigma^{3}(\omega_n) = -\frac{\beta U^3}{\color{red}{16}} \frac{1}{i\omega_n}.  \label{eqn:3rdph}
    \end{align}
    \label{eqn:phscatter}
    \end{subequations}
    The contribution from the third order diagram will lead to non-causal behavior for $\beta U \geq 4$. this corresponds to
    the region where the geometric series for the ph-scattering is outside of its radius of convergence. In case of a partially
    dressed one-particle propagator the radius of convergence of the ph-scattering ladder is determined by
    $\chi^{0,(ph)}(\Omega_0) U < 1$~\cite{Janis1999}. A similar conclusion can be reached for pp-scattering diagrams
    (see Fig.~\ref{fig:scatter}b). For multi-orbital and lattice systems the radius of convergence has a more
    complicated dependence on system parameters.

\end{document}